\documentclass[11pt]{article}

\usepackage{array}
\usepackage{graphicx}
\usepackage{amsmath}
\usepackage{amsthm}
\usepackage{amssymb}
\usepackage{hhline}

\newcommand{\D}{\Delta}
\newcommand{\be}{\begin{equation}}
\newcommand{\ee}{\end{equation}}
\newcommand{\bsube}{\begin{subequations}}
\newcommand{\esube}{\end{subequations}}
\newcommand{\ba}{\begin{array}}
\newcommand{\ea}{\end{array}}
\newcommand{\To}{\rightarrow}
\newcommand{\bea}{\begin{eqnarray}}
\newcommand{\eea}{\end{eqnarray}}
\newcommand{\bc}{\begin{center}}
\newcommand{\ec}{\end{center}}
\newcommand{\so}{\Rightarrow}
\newcommand{\dst}{\displaystyle}
\newcommand{\const}{{\rm const}}
\newcommand{\tu}{\tilde{u}}
\newcommand{\tv}{\tilde{v}}
\newcommand{\tw}{\tilde{w}}
\newcommand{\bu}{\bar{u}}
\newcommand{\dt}{\Delta t}
\newcommand{\dx}{\Delta x}
\newcommand{\F}{{\mathcal F}}
\newcommand{\sech}{{\rm sech}\,}
\newcommand{\usol}{u_{\rm sol}}
\newcommand{\Usol}{U_{\rm sol}}
\newcommand{\Ub}{U_{\rm b}}

\newcommand{\omsol}{\omega_{\rm sol}}
\newcommand{\Ksol}{K_{\rm sol}}
\newcommand{\ompw}{\omega_{\rm pw}}
\newcommand{\upw}{u_{\rm pw}}
\newcommand{\ub}{u_{\rm b}}

\title{\bf Instability of the finite-difference split-step
method on the background of
localized solutions of the generalized nonlinear Schr\"odinger equation}

\author{ T.I. Lakoba\thanks{{\tt tlakoba@uvm.edu}  } \\ \\
Department of Mathematics and Statistics,  University of Vermont, \\ Burlington, VT 05401, USA }

\begin{document}

\maketitle

\begin{abstract}
We consider numerical instability that can be observed
in simulations of localized solutions of
the generalized nonlinear Schr\"odinger equation (NLS)
by a split-step method where the linear part of the evolution is solved by a
finite-difference discretization.
Properties of such an instability cannot be inferred from the
von Neumann analysis of the numerical scheme. Rather, their explanation requires
tools of stability analysis of nonlinear waves, with numerically unstable
modes exhibiting novel features not reported for ``real" unstable modes of 
nonlinear waves. For example, modes that cause numerical instability of a
standing soliton of the NLS are supported
by the sides of the soliton rather than by its core.
Furthermore, we demonstrate that both properties and analyses of the 
numerical instability may be substantially affected by specific details of the 
simulated solution; e.g., they are substantially different for standing and 
moving solitons of the NLS. 
\end{abstract}

\bigskip



\vfill 

{\bf Short title}: \ 
Instability of finite-difference split-step method for NLS


\newpage

\section{Introduction}

The split-step method (SSM), also known as the operator-splitting method, is widely used
in numerical simulations of evolutionary equations that arise in diverse areas of science:
nonlinear waves, including nonlinear optics and Bose--Einstein condensation
\cite{HardinTappert_73}--\cite{BaoWang}, atomic physics 
\cite{Bandrauk93,Bandrauk07}, studies of advection--reaction--diffusion equations
\cite{SSMReactDiff_99}--\cite{SSMConvDiff_09}, and relativistic quantum mechanics
\cite{KGE_2012}. In this paper we focus on the SSM applied to nonlinear
Schr\"odinger (NLS)-type equations
\be
i u_t - \beta u_{xx} + G(x, |u|^2)\,u =0,
\label{e_00}
\ee
where $G(x,|u|^2)$ is some smooth function such that $|G(\infty,|u|^2)| <\infty$.
(This {\em excludes}, e.g., the Gross--Pitaevskii equation, where 
$G(x,|u|^2)=\alpha x^2 + \gamma |u|^2$.) 
In fact, we will do most of the analysis for the pure NLS
\be
i u_t - \beta u_{xx} + \gamma u |u|^2 =0
\label{e_01}
\ee
and will consider differences that occur for the more general equation \eqref{e_00}, 
in Sec.~5. 
Although the real-valued constants $\beta$ and $\gamma$ in \eqref{e_01} can be scaled
out of the equation, we will keep them in order to distinguish the contributions of
the dispersive ($u_{xx}$) and nonlinear ($u|u|^2$) terms. Without loss of generality
we will consider $\gamma > 0$ in \eqref{e_01}; then solitons  exist for $\beta < 0$.
(For $\gamma <0$, one simply changes the sign of $\beta$ to obtain an equivalent equation.)

The idea of the SSM is that \eqref{e_01} can be easily solved analytically when either
the dispersive or the nonlinear term is set to zero. This alows one to seek an approximate
numerical solution of \eqref{e_01} as a sequence of steps which alternatively account
for dispersion and nonlinearity:
\be
\ba{llr}
  \mbox{for $n$ from $1$ to $n_{\max}$ do:} & & \\
    & \hspace*{-4cm} \bar{u}(x) = u_n(x)\,\exp\big(i\gamma |u_n(x)|^2 \dt \big) 
    & \mbox{(nonlinear step)} \vspace{0.2cm}\\
    & \hspace*{-4cm} u_{n+1}(x) = 
      \left\{ \ba{l} \mbox{solution of \ $iu_t=\beta u_{xx}$ at $t=\dt$} \\
              \mbox{with initial condition $u(x,0)=\bar{u}(x)$}
      \ea  \right.
    & \mbox{(dispersive step)}
\ea
\label{e_02}
\ee
where the implementation of the dispersive step will be discussed below. 
In \eqref{e_02}, $\dt$ is the time step of the numerical integration and
$u_n(x) \equiv u(x,n\dt)$. Scheme \eqref{e_02} can yield a numerical solution
of \eqref{e_01} whose accuracy is $O(\dt)$. Higher-order schemes, yielding
more accurate solutions (e.g., with accuracy $O(\dt\,^2)$, $O(\dt\,^4)$, etc.),
are known \cite{Strang,Yevick91,Bandrauk93}, but here we will restrict our
attention to the lowest-order scheme \eqref{e_02}; see also the 
paragraph after Eq.~\eqref{e_33} below.

The implementation of the dispersive step in \eqref{e_02} depends on the numerical
method by which the spatial derivative is computed. In most applications, it is
computed by the Fourier spectral method:
\be
u_{n+1}(x) = \F^{-1} \left[ \,\exp(i\beta k^2 \dt) \; \F[ \bu(x)]\;\right]\,.
\label{e_03}
\ee
Here $\F$ and $\F^{-1}$ are the discrete Fourier transform and its inverse,
$k$ is the discrete wavenumber:
\be
-\pi/\dx \le k \le \pi/\dx,
\label{e_04}
\ee
and $\dx$ is the mesh size in $x$. However, the spatial derivative in \eqref{e_02}
can also be computed by a finite-difference (as opposed to spectral) method
\cite{WH}--\cite{CPC_2013}.
For example, 
using the central-difference discretization of $u_{xx}$ and the Crank--Nicolson
method, the dispersion step yields:
\be
i \frac{u_{n+1}^m - \bu^m}{\dt} \,=\, \frac{\beta}2 \,
\left( \frac{u_{n+1}^{m+1}-2u_{n+1}^m+u_{n+1}^{m-1}}{\dx\,^2} + 
\frac{\bu^{m+1}-2\bu^m+\bu^{m-1}}{\dx\,^2} \right),
\label{e_05}
\ee
where $u_n^m \equiv u(x_m,n\dt)$, $x_m$ is a point in the discretized spatial
domain: \ $-L/2 < x_m < L/2$, and $L$ is the the length of the domain. 
Recently, solving the dispersive step of \eqref{e_02} by a finite-difference 
method has found an application in the electronic post-processing of the optical
signal in fiber telecommunications \cite{AOP_2009}. 
Also, the version of the NLS where the second derivative is 
replaced by its
finite-difference approximation, similarly to the right-hand side (r.h.s.) 
of \eqref{e_05}, describes a wide range phenomena from transport of
vibrational energy in molecular chains to light propagation in 
waveguide arrays (see, e.g., \cite{Kenkre_1986, Aceves_2014}).

In what follows we assume periodic boundary conditions:

\be
u(-L/2,t)=u(L/2,t);
\label{e_06}
\ee
the case of other types of boundary conditions is considered in Appendix A.
We will refer to the SSM with spectral \eqref{e_03} and finite-difference 
\eqref{e_05} implementations of the dispersive step in \eqref{e_02} as
s-SSM and fd-SSM, respectively. 
Our focus in this paper will on the fd-SSM.

\begin{figure}[h]
\vspace{-1.6cm}
\mbox{ 
\begin{minipage}{7cm}
\rotatebox{0}{\resizebox{7cm}{9cm}{\includegraphics[0in,0.5in]
 [8in,10.5in]{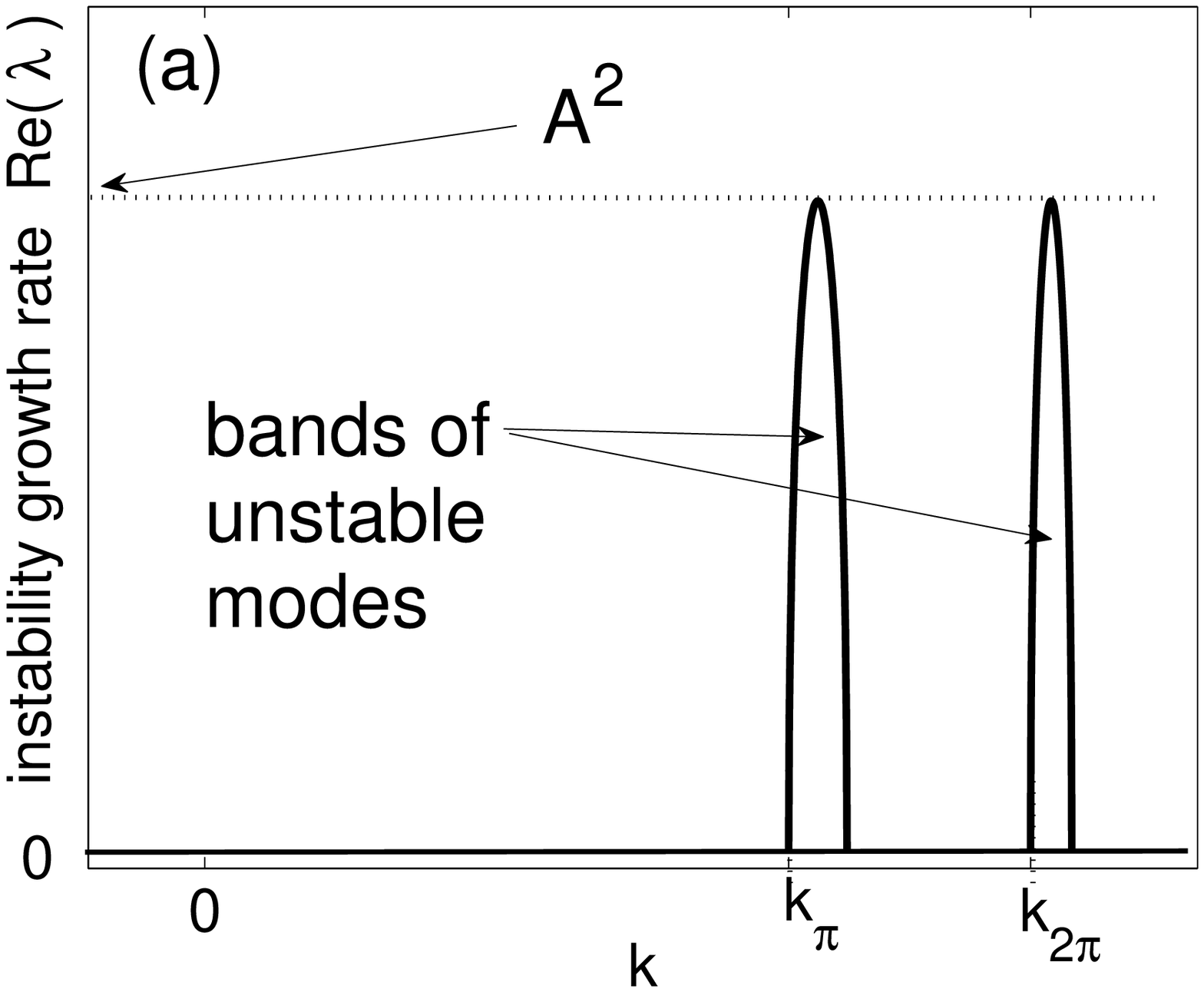}}}
\end{minipage}
\hspace{0.1cm}
\begin{minipage}{7cm}
\rotatebox{0}{\resizebox{7cm}{9cm}{\includegraphics[0in,0.5in]
 [8in,10.5in]{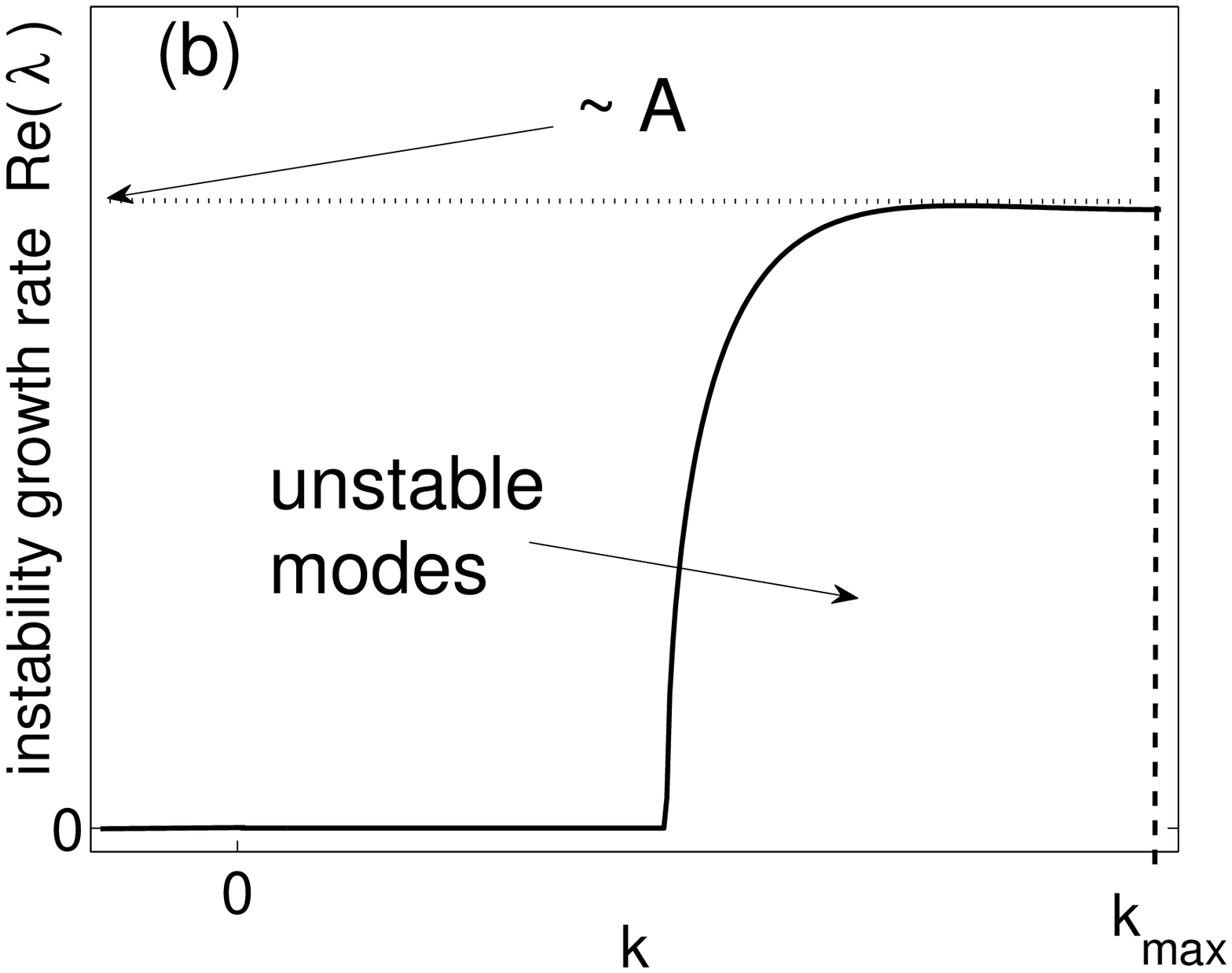}}}
\end{minipage}
 }
\vspace{-1.6cm}
\caption{Growth rate of numerical instability of the s-SSM (a) and
fd-SSM (b) on the plane-wave background. 
The dotted horizontal line indicates how the maximum growth rate depends on the
wave's amplitude. In (a), $k_{m\pi}$, $m=1,2,\ldots$ are the wavenumbers where the
$m$th resonance condition holds (see \cite{ja}): $|\beta|k_{m\pi}^2\dt = m\pi$. 
}
\label{fig_1}
\end{figure}

Weideman and Herbst \cite{WH} used the von Neumann analysis to show that 
both versions, s- and fd-, of the SSM can become unstable when the background
solution of the NLS is a plane wave:
\be
\upw = (A/\sqrt{\gamma}) \; e^{i\omega_{\rm pw}t}, \qquad A=\const, \quad
\omega_{\rm pw} = |A|^2.
\label{e_07}
\ee
Specifically, they linearized the SSM equations on the background of \eqref{e_07}:
\be
u_n=\upw + \tu_n, \qquad |\tu_n| \ll |u_n|
\label{e_08}
\ee
and sought the numerical error in the form
\be 
\tu_n = \tilde{A}\,e^{\lambda t_n - ikx}, \qquad \tilde{A}=\const.
\label{e_09}
\ee
The SSM is said to be unstable when for a certain wavenumber $k$ one has: \ 
(i) \ Re$(\lambda) > 0$ in \eqref{e_09}, but \ (ii) \ the corresponding Fourier
mode in the original equation \eqref{e_01} is linearly stable. Weideman and Herbst
found that the s- and fd-SSMs on the background \eqref{e_07} become unstable when
the step size $\dt$ exceeds:
\be
\dt_{\rm thr,\,s} \approx \dx\,^2/(\pi |\beta|), \qquad 
\mbox{for s-SSM \eqref{e_02} \& \eqref{e_03}}
\label{e_10}
\ee
and
\be
\dt_{\rm thr,fd}=\dx/\sqrt{2|\beta|\gamma \,|A|^2} \quad 
\underline{\mbox{only for $\beta>0$}}, \qquad 
\mbox{for fd-SSM \eqref{e_02} \& \eqref{e_05}}
\label{e_11}
\ee
respectively. 
Note that for $\beta<0$, the fd-SSM simulating a solution close to the plane wave 
\eqref{e_07} is unconditionally stable. 
Typical dependences of the instability growth rate, Re$(\lambda)>0$,
on the wavenumber is shown in Fig.~\ref{fig_1}. 
Let us emphasize that the SSM is unstable for $\dt>\dt_{\rm thr}$ even
though both its constituent steps, \eqref{e_02} and either \eqref{e_03} or
\eqref{e_05}, are numerically stable for any $\dt$.

Solutions of the NLS (and of other evolution equations) that are of
practical interest are considerably more complicated than a constant-amplitude
solution \eqref{e_07}. To analyze stability of a numerical method that is
being used to simulate a {\em spatially varying} solution, one often
employs the so-called ``principle of frozen coefficients" \cite{vonNeumannRichtmeyer}
(see also, e.g., \cite{Trefethen_book,KGE_2012}). According to that principle,
one assumes some constant value for the solution $u$ and then linearizes the
equations of the numerical method to determine the evolution of the numerical error
(see \eqref{e_08} and \eqref{e_09}). However, as we show below, this principle
applied to the SSM fails to predict, even qualitatively, important features
of the numerical instability (NI).

In this regard we stress --- and will subsequently illustrate --- that a NI
of a particular method applied to a {\em nonlinear} equation depends,
in general, not only on the method and the equation, {\em but also on the solution
which is being simulated}. This is similar to the situation with
linear stability analysis of particular solutions of a nonlinear equation: 
some of those solutions may be stable while others may be not. For example, the
plane wave \eqref{e_07} of the NLS with $\beta<0$ is unstable (\cite{Agrawal_book}, Sec.~5.1),
while the soliton, given by Eq.~\eqref{e_12} below, is stable with respect to small
perturbations of their respective initial profiles.

As a step towards understanding  NI on the
background of a spatially varying solution, we analyzed \cite{ja} the instability
of the s-SSM on the background of a soliton of the NLS:
\bsube
\be
\usol(x,t) = \Usol(x-St)\; 
  \exp\left[ i\omega_{\rm sol}t + \Ksol(x-St) \right]; \qquad (\beta < 0)
\label{e_12a}
\ee
\be
 \Usol(x) = A \sqrt{2/\gamma} \; \sech( Ax/\sqrt{-\beta} ); 
 \qquad \omega_{\rm sol}=A^2 + |\beta|\Ksol^2, \qquad \Ksol = S/(2|\beta|).
\label{e_12b}
\ee
\label{e_12}
\esube
The parameter $S$, describing the soliton's speed, was set to 0 in \cite{ja}. 
First, we demonstrated numerically that the instability growth rate in this case
is very sensitive to the time step $\dt$ and the length $L$ of the spatial domain;
also, its dependence on the wavenumber is quite different from that shown in
Fig.~\ref{fig_1}(a). Moreover, the instability on the background of, say, two 
well-separated (and hence non-interacting) solitons can be completely different from 
that on the background of one of these solitons. To our knowledge, such features
of the NI had not been reported for other numerical methods. In particular,
they could not be predicted based on the principle of frozen coefficients. We then
demonstrated that all those features could be explained by analyzing an equation
satisfied by the numerical error of the s-SSM {\em with large wavenumber $k$}:
\be
i\tv_t - \omsol \tv -\beta(\tv_{xx}+k^2_{\pi}\tv) + \gamma |\usol|^2 
(2\tv + \tv^*) = 0,
\label{e_13}
\ee
where $\tv(x,t)$ is proportional to the continuous counterpart of 
$\tu_n(x)\equiv \tu(x,n\dt)$ defined similarly to \eqref{e_08}, and $k_{\pi}$ is defined
in the caption to Fig.~\ref{fig_1}. Note that \eqref{e_13} is similar, but not 
equivalent, to the NLS linearized about the soliton:
\be
i\tu_t - \omsol \tu -\beta\tu_{xx} + \gamma |\usol|^2 
(2\tu + \tu^*) = 0.
\label{e_14}
\ee
The extra $k^2_{\pi}$-term in \eqref{e_13} indicates that the potentially 
unstable numerical error of the s-SSM has a wavenumber close to $k_{\pi}$.

In this paper we theoretically analyze the NI of the
fd- (as opposed to s-) SSM on the soliton background.
We will also comment on generic features of NI on more general backgrounds.
The NI of the fd-SSM has a number of different features both from the NI
of the s-SSM and from the textbook examples of NI of linear equations.
Specific features of the NI depend, as we have noted earlier, on both the
equation (i.e., \eqref{e_00} or \eqref{e_01}) and the background solution.
They will be listed in respective sections in the text, and most of them will
be explained, quantitatively or qualitatively.

Our analysis is based
on an equation for the large-$k$ numerical error which, as \eqref{e_13}, is a modified 
form of the linearized NLS. However, both that equation and its analysis are
substantially different from those \cite{ja} for the s-SSM.
For example, when the background solution is a standing soliton 
($S=0$ in \eqref{e_12}), 
the modes that render the s- and fd-SSMs unstable are qualitatively
different. Namely, for the s-SSM, the numerically unstable modes contain just a few
Fourier harmonics and hence are not spatially localized; they resemble plane waves. 
On the contrary, 
the modes making the fd-SSM unstable are localized and are supported by the
sides (i.e., ``tails") of the soliton. To our knowledge, such ``tail-supported"
localized modes, as opposed to those supported by the soliton's core, 
have not been reported before.


The main part of this manuscript is organized as follows. 
We begin by studying the instability of the {\em standing} soliton, where in 
\eqref{e_12} $S=0$. 
In Sec.~2 we present
simulation results showing the development of NI of the fd-SSM
applied to such a soliton. In Sec.~3 we derive an equation (a counterpart of 
\eqref{e_13}) governing the evolution of the numerical error, and in Sec.~4
obtain its localized solutions that grow exponentially in time. 
In essence, instead of following a numerical analyst's approach where one focuses
on obtaining a bound for the time step that would guarantee numerical stability,
as, e.g., in \cite{Faou_2011,Gauckler_2010},
we employ the procedure used to study (in)stability of nonlinear waves
and focus on finding the modes that cause the numerical instability.
In doing so, we also find an estimate of the instability threshold
as well as the growth rate of those unstable modes.
The latter may be useful because, as we will show, in many cases the NI
 is so weak that it does not affect the simulated solution
for a long time.
Thus, numerical simulations will produce valid results even if the integration
time step exceeds the NI threshold. Using this observation would reduce the 
computational time.

In Sec.~5 we consider differences in the NI behavior for (a subclass of)
the generalized NLS \eqref{e_00} compared to that for the pure NLS \eqref{e_01}. 
In Sec.~6 we turn to the NI on the background of a 
{\em moving} soliton: $S\neq 0$ in \eqref{e_12}. (We assume $S=O(1)$.) 
As we have pointed out
above, one should expect that the NI behavior should depend on the background
solution. Indeed, we find that the NIs for the standing and moving
solitons are substantially different, both in the required analytical tools
and in features. In Sec.~7 we return to the case when the solution of the 
(generalized) NLS is not moving along $x$.
However, unlike in Secs.~2--5, it is 
oscillating in time. We demonstrate via simulations that in this case,
the NI behavior is similar to that for a certain subclass of the generalized NLS,
with the background being a {\em stationary}, rather than oscillating, soliton.
Conclusions of our work, as well as open issues, are summarized in Sec.~8. 
Appendices A--C pertain to the case of the 
standing soliton of the pure NLS \eqref{e_01}.
In Appendix A we show how our analysis
can be modified for boundary conditions other than periodic. In Appendix B
we describe the numerical method used to obtain the localized solutions in Sec.~4.
In Appendix C we discuss how the instability analyzed in Secs.~2--4 sets in.
Finally, in Appendix D we speculate why the NI reported in Sec.~7 for an oscillating
pulse may be similar to that for a standing soliton in a certain potential,
reported in Sec.~5.3.


\section{Numerics of fd-SSM with standing soliton background}
\setcounter{equation}{0}

We numerically simulated Eq. \eqref{e_01} with $\beta = -1$, $\gamma=2$, and the periodic
boundary conditions \eqref{e_06} via the fd-SSM algorith \eqref{e_02} \& \eqref{e_05}.
The initial condition was the soliton \eqref{e_12} with $A=1$ and $S=0$:
\be 
u_0(x) = \sech(x) + \xi(x);
\label{e_15}
\ee
the noise component $\xi(x)$ with zero mean and the standard deviation $10^{-10}$
was added in order to reveal the unstable Fourier components sooner than if they 
had developed from a round-off error. 

Below we report results for two values of the spatial mesh size $\dx=L/N$, where
$N$ is the number of grid points: \ $\dx=40/2^9$ and $\dx=40/2^{10}$. We verified that,
for a fixed $\dx$, the results are insensitive to the domain's length $L$
(unlike they are for the s-SSM \cite{ja}) as long as $L$ is sufficiently large. 
Also, at least within the range of $\dx$ values considered, the results depend
on $\dx$ monotonically (again, unlike for the s-SSM).

\begin{figure}[h]
\vspace{-1.6cm}
\mbox{ 
\begin{minipage}{7cm}
\rotatebox{0}{\resizebox{7cm}{9cm}{\includegraphics[0in,0.5in]
 [8in,10.5in]{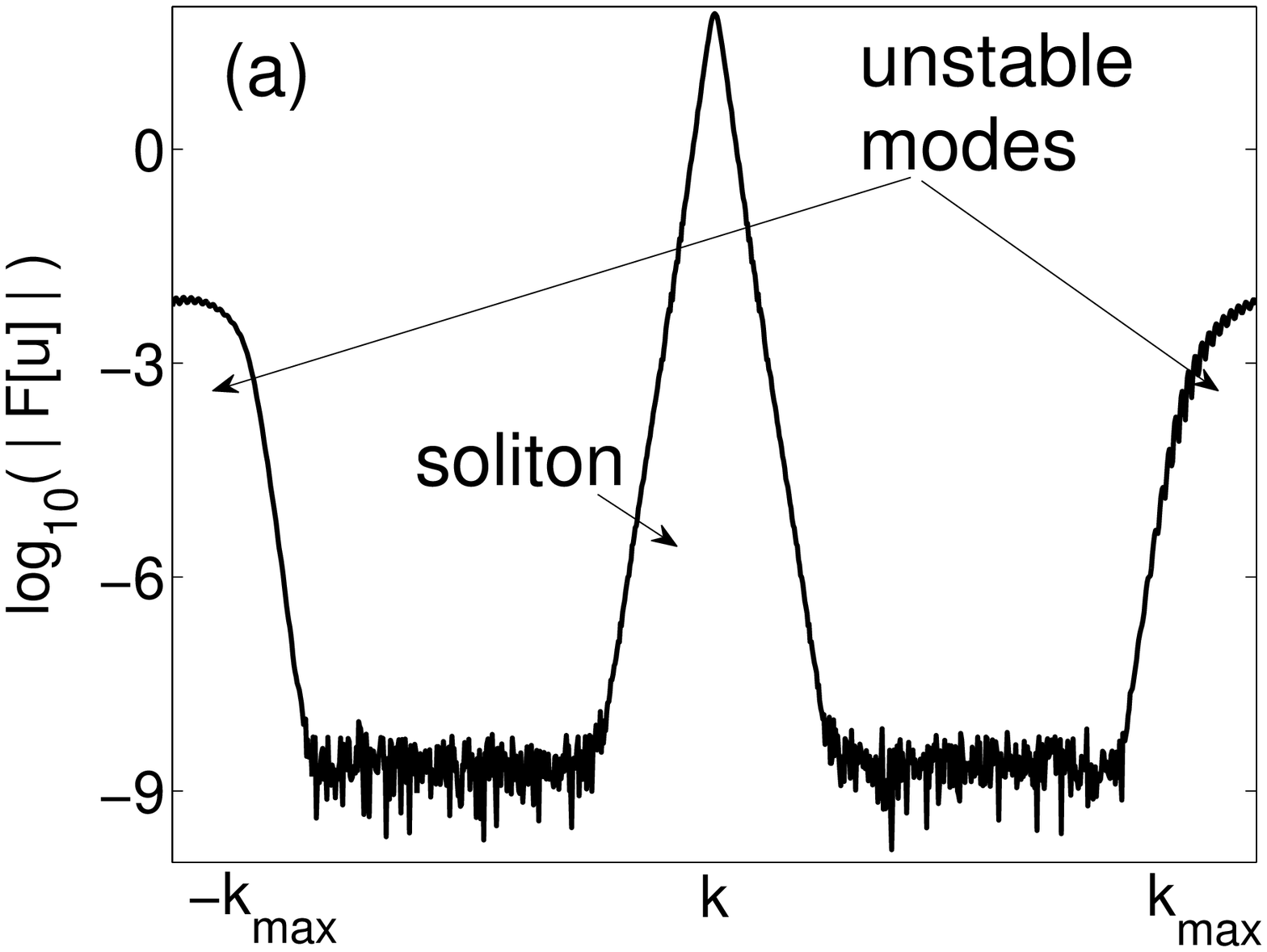}}}
\end{minipage}
\hspace{0.1cm}
\begin{minipage}{7cm}
\rotatebox{0}{\resizebox{7cm}{9cm}{\includegraphics[0in,0.5in]
 [8in,10.5in]{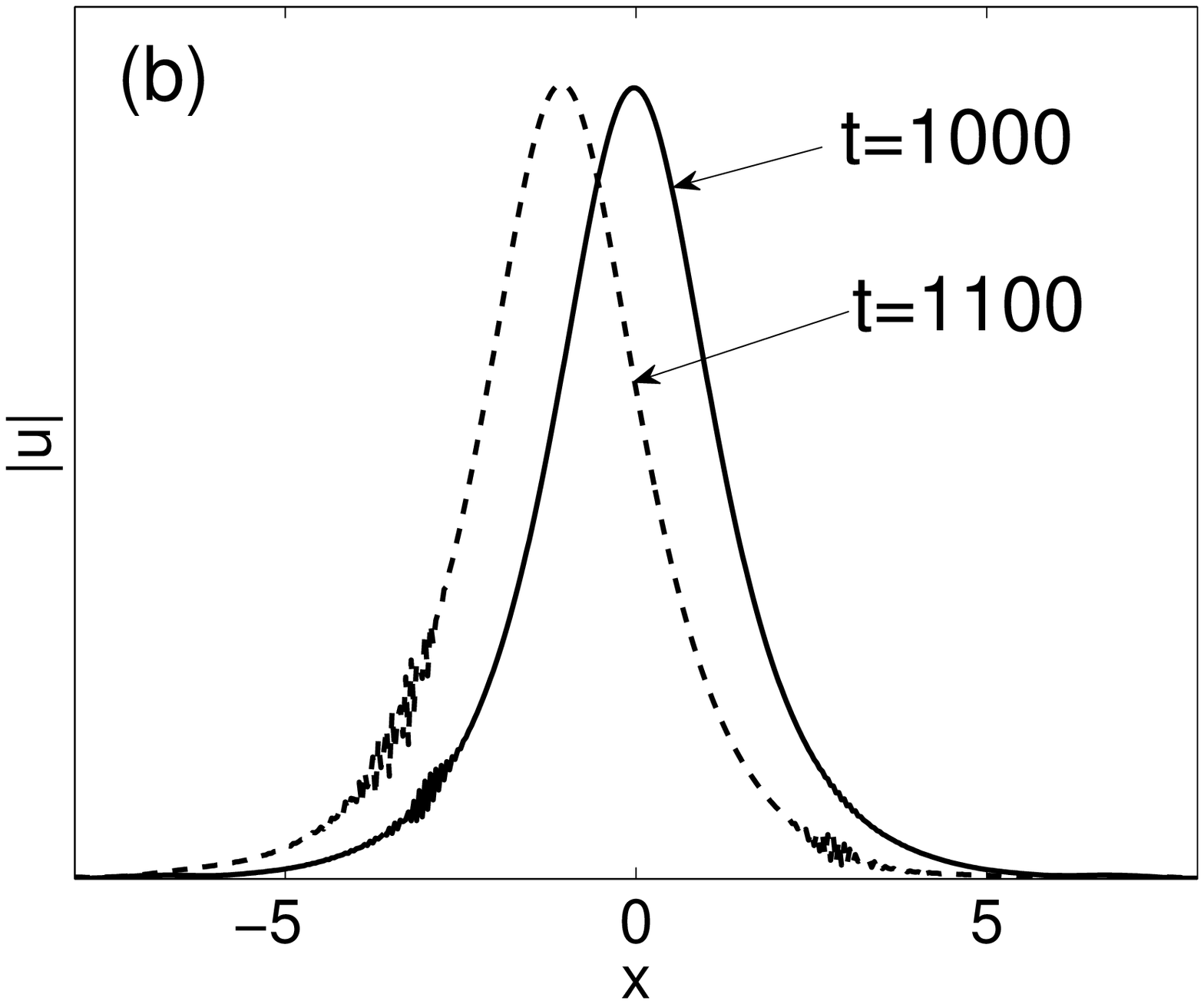}}}
\end{minipage}
 }
\vspace{-1.6cm}
\caption{ (a) \ Fourier spectrum of the unstable numerical solution of 
\eqref{e_01} with initial condition \eqref{e_15}. \ 
(b) \ Effect of numerically unstable modes on the soliton. \ 
Details are presented in the text.
}
\label{fig_2}
\end{figure}

First, let us remind the reader that the analysis of \cite{WH} on a constant-amplitude
background \eqref{e_07} for $\beta <0$ predicted that the fd-SSM should be stable for 
any $\dt$.\footnote{
Note that the stability of instability of the numerical method is in no way related
to that of the actual solution. In fact, the plane wave \eqref{e_07} is well-known
to be modulationally unstable for $\beta<0$, while it is modulationally stable 
for $\beta > 0$.}
For the soliton initial condition \eqref{e_15} and the parameters stated above
(with $\dx=40/2^{10}$), our simulations showed that the 
numerical solution becomes unstable for $\dt> \dx$. For future use we introduce a notation:
\be
C = (\dt/\dx)^2\,.
\label{e_16}
\ee
In Fig.~\ref{fig_2}(a) we show the Fourier spectrum of the numerical solution of 
\eqref{e_01}, \eqref{e_15} obtained by the fd-SSM with $C=1.05$ 
(i.e., slightly above the instability threshold)  at $t=800$.
The numerically unstable modes are seen near the edges of the spectral axis.
At $t=1000$, these modes are still small enough so as not to cause 
visible damage to the soliton: see the solid curve in Fig.~\ref{fig_2}(b).
However, at a later time, the soliton begins to drift: see the dashed line in
Fig.~\ref{fig_2}(b), that shows the numerical solution at $t=1100$. Such a drift 
may persist over a long time: e.g., for $C=1.05$, the soliton 
still keeps on moving at $t\sim 4000$. However, eventually it gets overcome by noise and 
loses its identity. 

We observed the same scenario for several different values of $\dx$, $L$, and $C$
(for $C>1$). The direction of the soliton's drift appears to be determined by the
initial noise; this direction is {\em not} affected by the placement of the initial
soliton closer to either boundary of
the spatial domain. The time when the drift's onset becomes visible 
decreases, and the drift's velocity increases, as $C$ increases.

\begin{figure}[h]
\vspace{-1.6cm}
\centerline{ 
\begin{minipage}{7cm}
\rotatebox{0}{\resizebox{7cm}{9cm}{\includegraphics[0in,0.5in]
 [8in,10.5in]{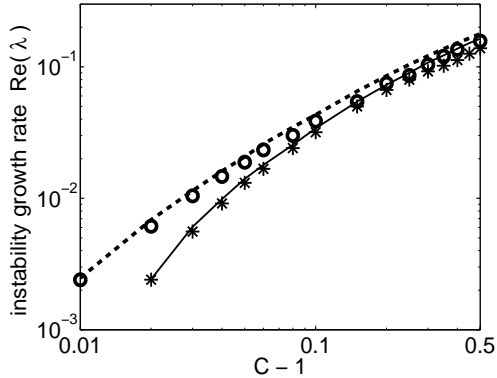}}}
\end{minipage}
 }
\vspace{-1.6cm}
\caption{Growth rate of the NI for $\dx=40/2^9$ (solid line --- 
analysis of Sec.~4, stars --- computed by \eqref{e_17}) and for 
$\dx=40/2^{10}$ (dashed line --- 
analysis of Sec.~4, circles --- computed by \eqref{e_17}).
}
\label{fig_3}
\end{figure}

The soliton's drift is a nonlinear stage of the development of the numerical 
instability and will be explained in Sec.~4.2. In the linear stage, the numerically
unstable modes are still small enough so that they do not visibly affect the soliton
or one another. To describe this stage, we computed a numeric approximation to the
instability growth rate \ Re$(\lambda)$ defined in \eqref{e_09}:
\be
{\rm Re}\,(\lambda)|_{\rm computed} = 
\frac{ \ln\left( \ba{c} \max \left| \F[u](k)\right| \;\; \mbox{for} \\ 
                        \mbox{$k \sim k_{\max}$ at time$=t$} \ea \right) -
       \ln \left( \ba{c} \mbox{noise floor} \\ \mbox{at time$=0$} \ea \right) }{t},
\label{e_17}
\ee
where $k_{\max}=\pi/\dx$ (see \eqref{e_04}).
The so computed
values of the instability growth rate are shown in Fig.~\ref{fig_3} along with the results
of a semi-analytical calculation presented in Sec.~4.1.

The above numerical results motivate the following three questions: \ (i) explain the observed
instability threshold $\dt$ (see the sentence before \eqref{e_16}); \ (ii) identify the modes
responsible for the NI; and \ (iii) calculate the instability growth rate 
(see Fig.~\ref{fig_3}). 
In Sec.~4 we will give an approximate analytical answer to question (i).
However, answers to questions (ii) and (iii) will be obtained only semi-analytically, i.e., via
numerical solution of a certain eigenvalue problem.


\section{Derivation of equation for numerical error for fd-SSM}
\setcounter{equation}{0}

\begin{figure}[h]
\vspace{-1.6cm}
\centerline{ 
\begin{minipage}{7cm}
\rotatebox{0}{\resizebox{7cm}{9cm}{\includegraphics[0in,0.5in]
 [8in,10.5in]{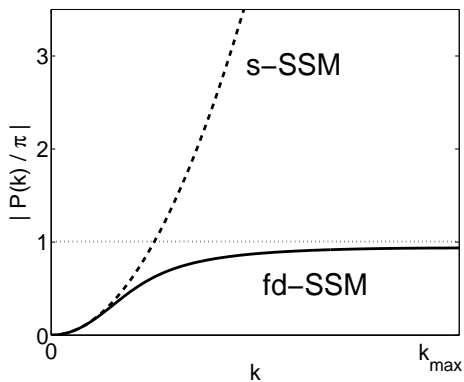}}}
\end{minipage}
 }
\vspace{-1.6cm}
\caption{ Normalized phase: \ $|\beta|k^2\dt$ for the s-SSM (dashed) and as given by \eqref{e_19} 
for the fd-SSM (solid). In both cases, $r=5$. The horizontal line indicates the condition
of the first resonance: $|P(k)|=\pi$.
}
\label{fig_4}
\end{figure}

Here we will derive a modified linearized NLS --- Eq.~\eqref{e_32} below ---
for a small numerical error 
with a high wavenumber, 
when the fd-SSM simulates an initial condition close to a standing soliton,
\eqref{e_15} with $S=0$, of the NLS \eqref{e_01}.
This modified equation will be a counterpart of \eqref{e_13},
which was obtained for the s-SSM in \cite{ja}. 
The key difference between \eqref{e_13} and \eqref{e_32} occurs due
to the following. 
In view of periodic boundary conditions
\eqref{e_06}, the finite-difference implementation \eqref{e_05} of the dispersive step 
in \eqref{e_02} can be written as
\be
u_{n+1}(x) = \F^{-1} \left[ e^{iP(k)}\,\F\left[ \bu(x) \right] \; \right],
\label{e_18}
\ee
\be
e^{iP(k)} \equiv \frac{1+2i\beta r \sin^2(k\dx/2) }{1-2i\beta r \sin^2(k\dx/2) }
 = \exp\left[ 2i\,{\rm arctan}\big( 2\beta r \sin^2(k\dx/2)\,\big) \right], 
 \qquad r = \frac{\dt}{\dx\,^2},
\label{e_19}
\ee
where $\F$, $\F^{-1}$ were defined after \eqref{e_03}. For $|k\dx|\ll 1$, the 
exponent in \eqref{e_19} equals that in \eqref{e_03}. However, for $|k\dx|>1$, they
differ substantially: see Fig.~\ref{fig_4}.
It is this difference that leads to the instabilities of the s- and fd-SSMs being
qualitatively different.

Using Eqs. \eqref{e_02} and \eqref{e_18}, one can write, similarly to Eq. (3.1) in
\cite{ja}, a linear equation satisfied by a small numerical error $\tu_n$ of the 
fd-SSM with an {\em arbitrary} $k$:
\be
\F \left[ \tu_{n+1} \right] = e^{iP(k)} \F
 \left[ e^{i\gamma |\ub|^2\dt} \big( \tu_n + 
        i\gamma \dt ( \ub^2 \tu_n^* + |\ub|^2 \tu_n )\, \big) \; \right]\,.
 \label{e_20}
 \ee
 Here $\tu_n$ is defined similarly to \eqref{e_08}, with $\ub$ being either
 $\upw$ or $\usol$, depending on the background solution. The exponential growth
 of $\tu_n$ can occur only if there is sufficiently strong coupling between
 $\tu_n$ and $\tu_n^*$ in \eqref{e_20}. This coupling is the strongest when
 the temporal rate of change of the relative phase between those two terms is
 minimized. In \cite{ja} we showed that this rate can be small only for those $k$
 where the exponent $P(k)$ is close to a multiple of $\pi$. Using \eqref{e_19}
 (see also Fig.~\ref{fig_4}),
 we see that this can occur only for sufficiently high $k$ where 
 $\sin^2(k\dx/2)=O(1)$ rather than $O(\dx\,^2)$. Then:
 \bea
 -P(k) & = & \pi - \frac1{|\beta| r \sin^2(k\dx/2)} + O\left( \frac1{r^3}\right)
  \nonumber  \\
  & = & \pi - \frac1{|\beta| r} - \frac{(k-k_{\max})^2 \dx\,^2}{4|\beta| r} + 
        O\left( \frac1{r^3} + \frac{\big( (k-k_{\max}) \dx\,\big)^4}{r} \right),
 \label{e_21}
 \eea
 where $k_{\max}=\pi/\dx$; also recall that $\beta<0$. We have also used that 
 \be
 r = \dt/\dx\,^2 = C/\dt \gg 1,
 \label{e_22}
 \ee
 given that the NI was observed in Sec.~2 for $C=O(1)$.

 We will now discuss which terms in \eqref{e_21} should be retained.
 First of all, in order to neglect the entire $O$-term, one needs to require that
 \bsube
 \be
 (k-k_{\max})^2\dx\,^2  < O(1),
 \label{e_23a}
 \ee
 where we have also used \eqref{e_22} to neglect the $O(1/r^3)$-term. 
 Next, if we keep the third term on the 
 right-hand side (r.h.s.) of \eqref{e_21}, it should be greater (in the order of
 magnitude sense) than the discarded $O$-term, whence
 \be
 (k-k_{\max})^2\dx\,^2  > O(1/r^2) = O(\dx^2).
 \label{e_23b}
 \ee
 It is not particularly important where in the range defined by \eqref{e_23a} and
 \eqref{e_23b} the value of \ $(k-k_{\max})^2\dx\,^2$ \ should be. For example, if we
 take it in the middle of that range:
 \be
 |k-k_{\max}| = O(1/\sqrt{\dx}).
 \label{e_23c}
 \ee
 \label{e_23}
 \esube
 then the three terms on the r.h.s. of \eqref{e_21} have orders of magnitude $O(1)$,
 $O(\dx)$, and $O(\dx^2)$. 
  What {\em is} important is that we have chosen to keep the third term in \eqref{e_21}
  and hence required \eqref{e_23b}. We stress that this choice has followed not from our derivation
  but rather from our numerical results, as illustrated by Fig.~\ref{fig_2}(a). 
  Indeed, one sees from that figure that the width of the bands of unstable modes,
  i.e. $|k-k_{\max}|$, 
  is significantly greater than the spectral width of the soliton, which is of order one.
  In Sec.~6
  we will encounter a situation where, in contrast to the above,
  the third term on the r.h.s. of \eqref{e_21}
  will {\em not} need to be kept.

 Substituting the first three terms on the r.h.s. of \eqref{e_21} into \eqref{e_20}, using 
 \eqref{e_22}, and introducing a new variable 
 \be
 \tv_n = \left( e^{-i\pi} \right)^n \tu_n = (-1)^n \tu_n,
 \label{e_24}
 \ee
 one obtains:
 \bea
 \F [\tv_{n+1}] & = & \exp\left( -\frac{i\dt}{C \beta} 
                       \left\{1 + \frac{ (k-k_{\max})^2\dx\,^2}4 \right\} \right) \,\times
  \nonumber \\
   &  & \F \left[ e^{i\gamma |\ub|^2\dt}\, 
          \left\{ \tv_n + i\gamma \dt ( \ub^2\tv_n^* + |\ub|^2 \tv_n ) \right\} \,
           \right].
 \label{e_25}
 \eea
 Note that \eqref{e_25} describes a {\em small} change of $\tv_n$ occurring over the step
 $\dt$, because for $\dt\To 0$, the r.h.s. of that equation reduces to 
 $\F[\tv_n]$. Therefore we can approximate the {\em difference} equation \eqref{e_25}
 by a {\em differential} equation, as we will now explain.

 First, recall from \eqref{e_23a}
 that the wavenumbers of $\tv_n$ are on the order of $k_{\max}$; hence we seek\footnote{
 Strictly speaking, since the spectrum of the numerical error is symmetric relative
  to $k=0$, as seen from Fig.~\ref{fig_2}(a), one should have assumed
  \ $\tv_n(x) = \exp[i k_{\max} x]\, \tw^+_n(x) + \exp[-i k_{\max} x]\, \tw^-_n(x)$ \ 
  instead of \eqref{e_26}. However, both approaches
  can be shown to lead to the same conclusions and hence here we will use the simpler 
 one based on \eqref{e_26}. In Sec.~6 it will be more natural to use the other approach.}
 \be
 \tv_n(x) = e^{i k_{\max} x}\, \tw_n(x).
 \label{e_26}
 \ee
 The effective wavenumber of $\tw_n$ is then $(k-k_{\max})$, and according to
 \eqref{e_23a} $\tw_n$ varies slowly over the scale $O(\dx)$. 
 (One may recognize \eqref{e_26} as the standard slowly varying envelope approximation.)
Introducing the scaled
 variables by
 \be
 \chi= x/\epsilon, \quad k_{\rm sc} = (k-k_{\max})\epsilon, \qquad \epsilon = \dx/2,
 \label{e_27}
 \ee
 one rewrites \eqref{e_25} as:
 \be
 \F_{\rm sc} [\tw_{n+1}] = \exp\left( -\frac{i\dt}{C \beta} 
                        \{1 + k_{\rm sc}^2 \} \right) \;
 \F_{\rm sc} \left[ e^{i\gamma |\ub|^2\dt}\, 
           \left\{ \tw_n + i\gamma \dt ( \ub^2\tw_n^* + |\ub|^2 \tw_n ) \right\} \,
            \right],
  \label{e_28}
  \ee
 where now $\F_{\rm sc}$ is the Fourier transform with respect to the scaled variables
 \eqref{e_27}. In handling the $\tv_n^*$ term in \eqref{e_25}, we have used the fact that
 on the spatial grid $x_m=m\dx$, one has:
 $$
 \tv_n^*(x_m) = e^{-i k_{\max}x_m} \tw_n^*(x_m) = e^{-i\pi m} \tw_n^*(x_m) 
 = e^{i\pi m} \tw_n^*(x_m) = e^{ik_{\max}x_m} \tw_n^*(x_m).
 $$

Second, note that the s-SSM \eqref{e_02}, \eqref{e_03} can be written as 
\be
\F [u_{n+1}] = e^{i\beta k^2 \dt}\; \F
 \left[ e^{i\gamma |u|^2 \dt}\, u \right].
\label{e_29}
\ee
When $|\beta|k^2\dt \ll 1$ and $\gamma |u|^2\dt \ll 1$, 
this is equivalent to the NLS \eqref{e_01}
{\em plus} a term proportional to 
\be
\dt\; \left[ \beta \partial^2_x, \; \gamma |u|^2 \right]_{-} u + O(\dt\,^2),
\label{e_30}
\ee
where $[\ldots,\; \ldots]_{-}$ denotes a commutator (see, e.g., Sec. 2.4 in 
\cite{Agrawal_book}). Equation \eqref{e_28} has the form of a linearized Eq. 
\eqref{e_29} with a different coefficient in the dispersion term and 
with an extra
phase. Therefore, \eqref{e_28} must be equivalent to a modified linearized NLS,
with the modification affecting only the corresponding terms:
\be
i\tw_t + (\tw_{\chi\chi}- \tw)/(C\beta) + \gamma (\ub^2 \tw^* + 2|\ub|^2 \tw) =0,
\label{e_31}
\ee
{\em plus} a term proportional to the linearized form of the commutator \eqref{e_30}.
Neglecting that latter term as small (of order $O(\dt)$) compared to the rest of
the expression and denoting \ $\psi=\tw\;\exp(-i\omega_{\rm b}t)$, we rewrite 
\eqref{e_31} as:
\be
i\psi_t + \delta\psi +\psi_{\chi\chi}/(C\beta) + \gamma \Ub^2(\epsilon\chi)\;
 (2\psi + \psi^*) =0,
\label{e_32}
\ee
where
\be
\delta = -\omega_{\rm b} - 1/(C\beta).
\label{e_33}
\ee
Here $\omega_{\rm b}$ is either $\ompw$ or $\omsol$, and $\Ub$ is either constant
or $\Usol$, depending on whether the background solution is a plane wave \eqref{e_07}
or a soliton \eqref{e_12}. The modified linearized NLS \eqref{e_32} for the fd-SSM
is the counterpart of Eq.~\eqref{e_13} that was derived for the s-SSM.

Our subsequent analysis of the instability of the first-order accurate
fd-SSM \eqref{e_02} \& \eqref{e_05}
will be based on Eq.~\eqref{e_32}. 
The instability of the second-order accurate version of this method,
where the order of the nonlinear and dispersive steps is alternated
in any two consecutive full time steps \cite{Strang}, is the same as 
that of the first-order version. The instability of higher-order 
versions (e.g., $O(\dt\,^4)$-accurate) can be studied similarly to how
that was done in Ref.~\cite{ja} for the s-SSM.

The boundary conditions satisfied by $\psi$ are still periodic:
\be
\psi(-L/(2\epsilon),\,t) = \psi(L/(2\epsilon),\,t).
\label{e_34}
\ee
This follows from the fact that $\tu_n(x)$ satisfies the periodic boundary conditions
\eqref{e_06} and from \eqref{e_26}, given that for $k_{\max}=\pi/\dx$ and 
$L/2=M\dx$ with some integer $M$,
$$
e^{-ik_{\max}L/2} = e^{-iM\pi} = e^{iM\pi} = e^{ik_{\max}L/2} .
$$

There are three differences between Eq.~\eqref{e_32} and the linearized NLS \eqref{e_14}.
Most importantly, \eqref{e_32} has the opposite sign of the dispersion term. This is
explained by the shape of the curve $P(k)$ 
for the fd-SSM in Fig.~\ref{fig_4} at high wavenumbers,
where the curvature is opposite to that at $k\approx 0$. 
Secondly, unlike the $(-\omsol)$-term in \eqref{e_14}, the $\delta$-term in \eqref{e_32}
with $\beta<0$ can be either positive or negative, depending on the value of $C$.
Thirdly, the ``potential" $\Ub^2(\epsilon\chi)$ (when $\Ub\equiv \Usol$) is a {\em slow}
function of the scaled variable $\chi$. That is, solutions of \eqref{e_32} that vary on 
the scale $\chi=O(1)$ ``see" the soliton as being very wide. This should also be
contrasted with the situation for the s-SSM, where the modes described by Eq.~\eqref{e_13}
``see" the soliton as being very narrow \cite{ja}.

Before proceeding to find unstable modes of Eq.~\eqref{e_32} with $\Ub\equiv \Usol$, let
us note that \eqref{e_32} with $\Ub=\const$ confirms the result of Ref.~\cite{WH}
regarding the instability of the fd-SSM on the plane-wave background. Namely, for
$\beta<0$, Eq.~\eqref{e_32} with $\Ub=\const$ describes the evolution of a small 
perturbation to the plane wave in the {\em modulationally  stable} case (see, e.g.,
Sec.~5.1 in \cite{Agrawal_book}). That is, for $\beta<0$, there is no NI,
in agreement with \cite{WH}. On the other hand, for $\beta>0$, Eq.~\eqref{e_32}
describes the evolution of a small perturbation in the {\em modulationally  unstable}
case, and hence the plane wave of the NLS \eqref{e_01} can become numerically unstable.
The corresponding instability growth rate found from \eqref{e_32} and Eq. (5.1.8)
of \cite{Agrawal_book} can be shown to agree with the one that can be obtained from
Eq.~(37) and the next two unnumbered relations in \cite{WH}. An example of this growth
rate is shown in Fig.~\ref{fig_1}(b). Also, using our \eqref{e_32} 
and Eq.~(5.1.8) of Ref.~\cite{Agrawal_book}, 
the threshold value of $\dt$ can be shown to be given by \eqref{e_11},
in agreement with \cite{WH}.


\section{Analysis of numerical instability of standing soliton of NLS}
\setcounter{equation}{0}

\subsection{Unstable modes of modified linearized NLS \eqref{e_32}}

In this section we focus on the case where the background solution is a soliton
with zero velocity ($S=0$ in \eqref{e_12});
hence $\beta < 0$ and $\Ub\equiv \Usol(x)$. Substituting into
\eqref{e_32} and its complex conjugate the standard ansatz \cite{Kaup90}
\ $(\psi(\chi,t),\,\psi^*(\chi,t))=(\phi_1(\chi),\,\phi_2(\chi))\,e^{\lambda t}$ \ 
and using yet another rescaling:
\be
\ba{c}
\dst
X=\frac{A}{\sqrt{-\beta}}\chi \equiv \frac{2A}{\sqrt{-\beta}}\frac{x}{\dx},
\qquad
D = -\frac{C\beta^2}{A^2}\delta \equiv \beta^2\left(\frac1{\beta A^2} + C\right),
\vspace{0.2cm} \\
\dst
\Lambda = \frac{C\beta^2}{A^2}\lambda,  \qquad
V(y) = 2C\beta^2 \sech^2(y),
\ea
\label{e_35}
\ee
one obtains:
\be
\left( \partial_X^2 + D - V(\epsilon X) 
 \left( \ba{cc} 2 & 1 \\ 1 & 2 \ea \right) \,\right)
\vec{\phi} \,=\, i\Lambda \sigma_3 \vec{\phi},
\label{e_36}
\ee
where $\sigma_3={\rm diag}(1,-1)$ is a Pauli matrix, $\vec{\phi}=(\phi_1,\,\phi_2)^T$,
and $T$ stands for a transpose. If $(\vec{\phi},\,\Lambda)$ is an eigenpair of \eqref{e_36},
then so are $(\sigma_1\vec{\phi},\,-\Lambda)$, \ $(\vec{\phi}^*,\,-\Lambda^*)$, and 
$(\sigma_1\vec{\phi}^*,\,\Lambda^*)$, where
$$
\sigma_1 = \left( \ba{cc} 0 & 1 \\ 1 & 0 \ea \right)
$$
is another Pauli matrix. Note also that $\lambda$ is defined in the same way as in
\eqref{e_09}; hence Re$(\Lambda)\neq 0 $ indicates an instability.
Below we will use shorthand notations $\Lambda_R={\rm Re}(\Lambda)$ and 
$\Lambda_I={\rm Im}(\Lambda)$.

We begin analysis of \eqref{e_36} with two remarks. First, this equation is qualitatively
different from an analogous equation that arises in studies of stability of both bright
\cite{Kaup90} and dark \cite{Tran92} NLS solitons in that the relative sign of the first
and third terms of \eqref{e_36} is opposite of that in \cite{Kaup90,Tran92}. This fact
is the main reason why the unstable modes supported by \eqref{e_36} are qualitatively 
different from unstable modes of linearized NLS-type equations, as we will see below. 
While the latter modes are
supported by the soliton's core (see, e.g., Fig.~3 in \cite{Peli98}), the unstable
modes of \eqref{e_36} are supported by the soliton's ``tails". 

Second, from \eqref{e_36} and \eqref{e_35} one can easily establish the minimum value
of parameter $C$ where an instability (i.e., $\Lambda_R\neq 0 $) {\em can} occur.
The matrix operators on both sides of \eqref{e_36} are Hermitian; the operator 
$\sigma_3$ on the r.h.s. is not sign definite. Then the eigenvalues $\Lambda$
are guaranteed to be purely imaginary when the operator on the l.h.s.
is sign definite \cite{RefLA}; otherwise they may be complex. The third term on the
l.h.s. of \eqref{e_36} is negative definite, and so is the first term in view of 
\eqref{e_34}. The second term, $D$, is negative when 
\be
C \, <  \, 1 /(|\beta| A^2).
\label{e_37}
\ee
Thus, \eqref{e_16} and \eqref{e_37} yield the stability condition of the fd-SSM on
the background of a soliton. We will show later that an unstable mode indeed
first arises when $C$ just slightly exceeds the r.h.s. of \eqref{e_37}.

Since the potential term in \eqref{e_36} is a slow function of $X$, it may seem
natural to employ the Wentzel--Kramers--Brillouin (WKB) method to analyze it.
Below we show that, unfortunately, the WKB method fails to yield an analytic form
of unstable modes of \eqref{e_36}. 
Away from ``turning points" (see below) the WKB-type solution of \eqref{e_36} is:
\be
\vec{\phi} = \left( a_+ e^{\theta_+/\epsilon} + b_+ e^{-\theta_+/\epsilon}\right)
 \, \vec{\varphi}_+ + 
 \left( a_- e^{\theta_-/\epsilon} + b_- e^{-\theta_-/\epsilon}\right)
 \, \vec{\varphi}_- ,
\label{e_38}
\ee
where $a_{\pm},\,b_{\pm}$ are some constants, and 
\bsube
\be
( \theta'_{\pm} )^2 = -D + 2V \pm \sqrt{V^2-\Lambda^2}, \qquad 
V \equiv V(\epsilon X), \;\; \theta' \equiv d\theta/d(\epsilon X),
\label{e_39a}
\ee
\be
\vec{\varphi}_{\pm} = 
 \frac{1}{\left[ ( \theta'_{\pm} )^2 (V^2-\Lambda^2) \right]^{1/4} } \,
 \left( \ba{r} \sqrt{\Lambda \pm \sqrt{\Lambda^2-V^2}} \vspace{0.2cm} \\
     -i\sqrt{\Lambda \mp \sqrt{\Lambda^2-V^2}}  \ea \right).
\label{e_39b}
\ee
\label{e_39}
\esube
At a turning point, say, $X=X_0$, the solution \eqref{e_38}, \eqref{e_39} breaks
down, which can occur because the denominator in \eqref{e_39b} vanishes.
In such a case, one needs to obtain a solution of \eqref{e_36} in a transition region
around the turning point by expanding the potential: \
$V(\epsilon X)=V(\epsilon X_0) + \epsilon(X-X_0)\,V'(\epsilon X_0)+\ldots$,
and then solving the resulting approximate equation. For a 
single linear Schr\"odinger
equation, a well-known solution of this type is given by the Airy function. This
solution is used to ``connect" the so far arbitrary constants $a_{\pm},\,b_{\pm}$
in \eqref{e_38} on both sides of the turning point.

Now a turning point of \eqref{e_36} is where: either (i) $\theta_+'=0$ or 
$\theta_-'=0$, \ or (ii) \ $(V(\epsilon X))^2-\Lambda^2 = 0$. The former case can be
shown (see, e.g., \cite{WKB_system}) to reduce to the single Schr\"odinger equation
case, where the solution in the transition region is given by the Airy function.
However, at present, no such transitional solution is analytically available in
case (ii)\footnote{Note that in this case, $(V^2-\Lambda^2)^{1/4}\vec{\varphi}_+$
and $(V^2-\Lambda^2)^{1/4}\vec{\varphi}_-$ are linearly dependent.}
\cite{Fulling2,Skorupski}.
Therefore, the solutions \eqref{e_38} canot be ``connected" by an analytic formula
across such a turning point, and hence one cannot find the eigenpairs 
($\vec{\phi}$, $\Lambda$) analytically. 

However, the preceding analysis indicates {\em where} the unstable modes
$\vec{\phi}$ can exist.
Based on the past experience with unstable linear modes of nonlinear waves, 
it is reasonable to assume that unstable solutions of \eqref{e_36} must be 
localized. 
We now show that localized solutions \eqref{e_38} cannot exist around the
soliton's core and thus may only exist at the soliton's sides.
For simplicity, we assume that $\Lambda_R\neq 0$ and $\Lambda_I=0$
for such a solution, but a more detailed analysis upholds this conclusion for
the case $\Lambda_I \neq 0$. Note that just above the instability threshold,
$D$ is small (see the text before \eqref{e_37}), and so is $\Lambda$. On the
other hand, near the soliton's core, $V(\epsilon X)=O(1)$, and hence from \eqref{e_39a}
one sees that there $\theta_{\pm}^2>0$. 
Thus, both $\theta_{\pm}$ are real, and hence the corresponding \eqref{e_38}
would grow exponentially away from the soliton's core.
This, however, is not possible because on the scale of Eq.~\eqref{e_36},
the soliton is very wide, and then a mode growing away from its center
would become exponentially large before it reaches the turning point. 
Thus, the only possibility for a localized mode of \eqref{e_36} is to be
centered at some point at the soliton's side and decay in both directions away from
that point. A straightforward but tedious analysis shows that this is
indeed possible when $D>0$ and $\Lambda_R\neq 0$. 

\begin{figure}[h]
\vspace{-0.8cm}
\hspace*{-0.5cm}
\mbox{ 
\begin{minipage}{5.1cm}
\rotatebox{0}{\resizebox{5.1cm}{6.5cm}{\includegraphics[0in,0.5in]
 [8in,10.5in]{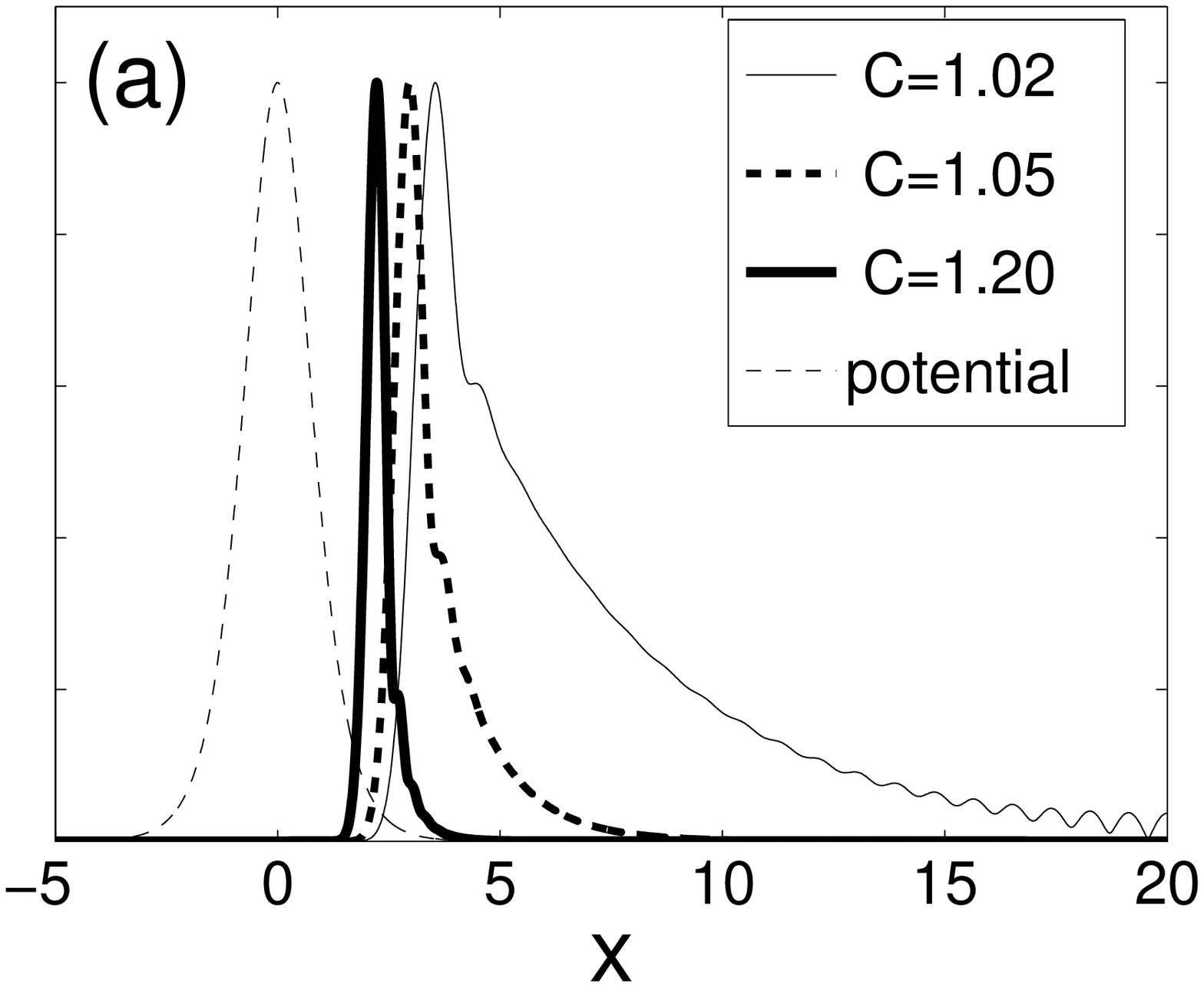}}}
\end{minipage}
\hspace{-0.4cm}
\begin{minipage}{5.1cm}
\rotatebox{0}{\resizebox{5.1cm}{6.5cm}{\includegraphics[0in,0.5in]
 [8in,10.5in]{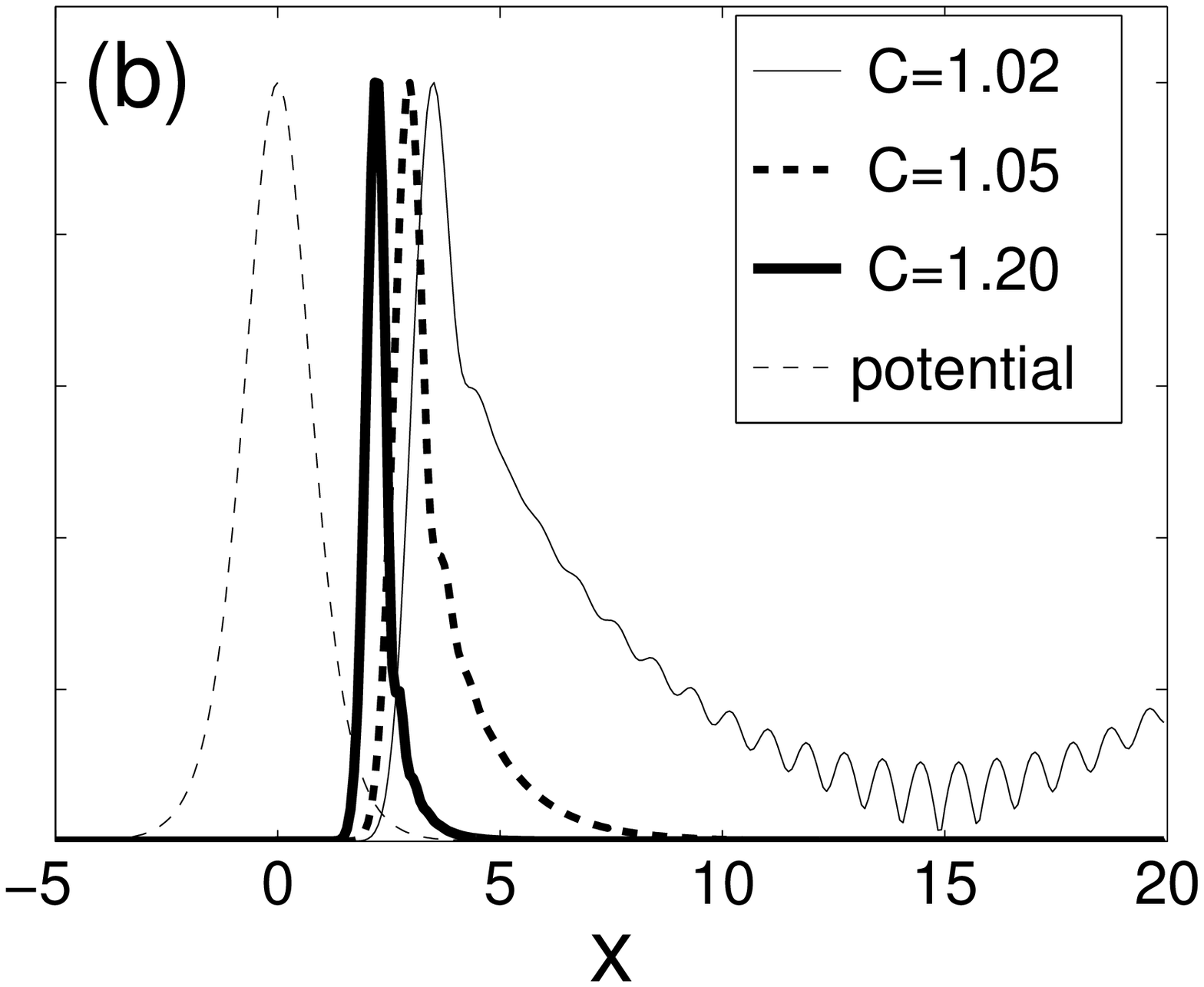}}}
\end{minipage}
\hspace{-0.4cm}
\begin{minipage}{5.1cm}
\rotatebox{0}{\resizebox{5.1cm}{6.5cm}{\includegraphics[0in,0.5in]
 [8in,10.5in]{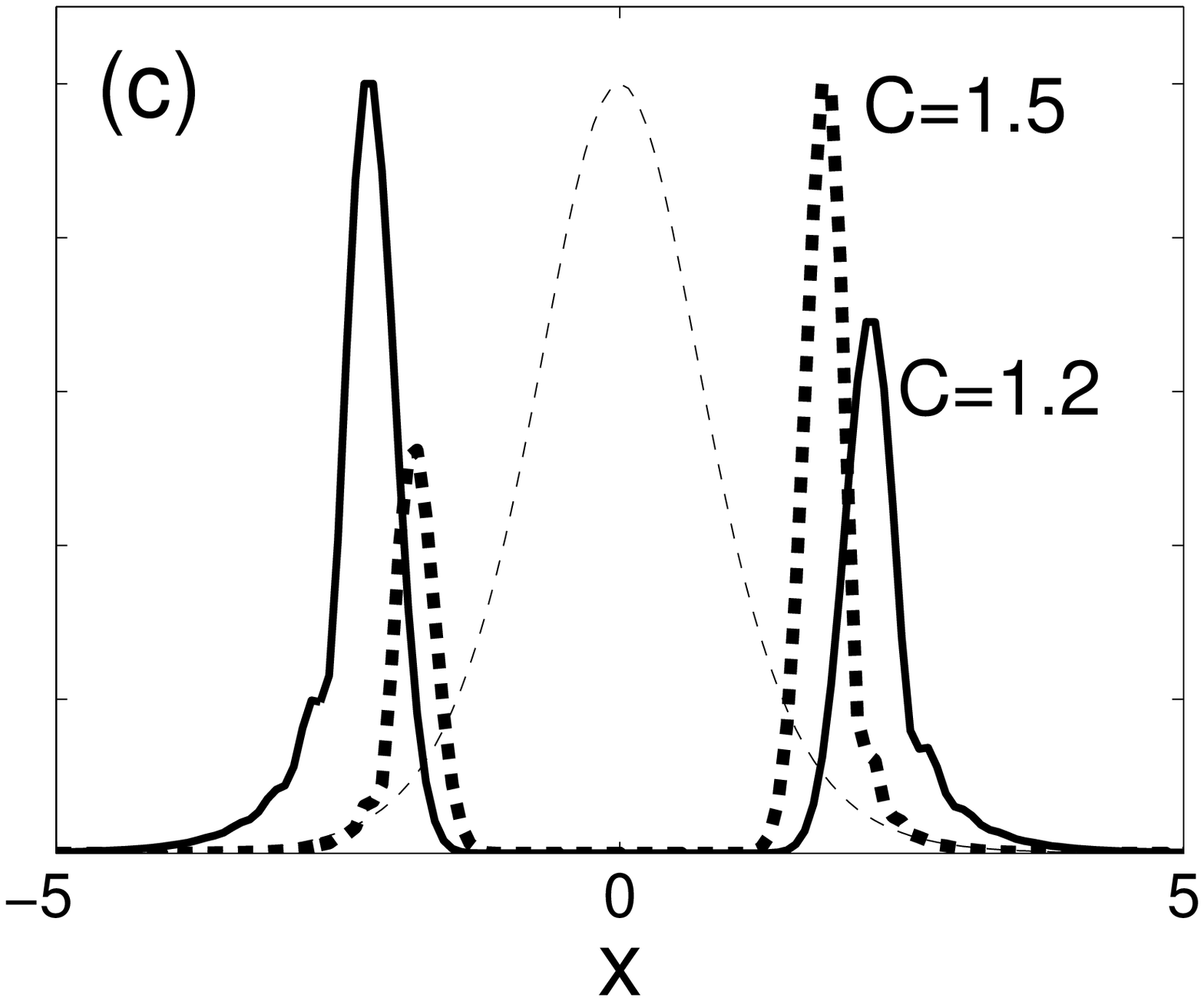}}}
\end{minipage}
 }
\vspace{-1cm}
\caption{(a) \ Profiles of the first localized mode on the right side of the soliton
for different values of $C$,
as found by the numerical method of Appendix B. \ (b) \ Same as in (a), but found 
from the numerical solution of \eqref{e_01}, as explained in the text. \ 
(c) \ The modes at {\em both} sides of the soliton found from the numerical solution
of \eqref{e_01}. Note that these modes do not ``see" each other because of the
barrier created by the soliton, and hence in general have different amplitudes as
they develop from independent noise seeds. In all panels, the potential is 
$\sech^2(\epsilon X)$ (see \eqref{e_35}) and the amplitude of the mode is normalized
to that of the potential.
}
\label{fig_5}
\end{figure}

\begin{figure}[h]
\vspace{-0.6cm}
\hspace*{-0.5cm}
\mbox{ 
\begin{minipage}{7cm}
\rotatebox{0}{\resizebox{7cm}{9cm}{\includegraphics[0in,0.5in]
 [8in,10.5in]{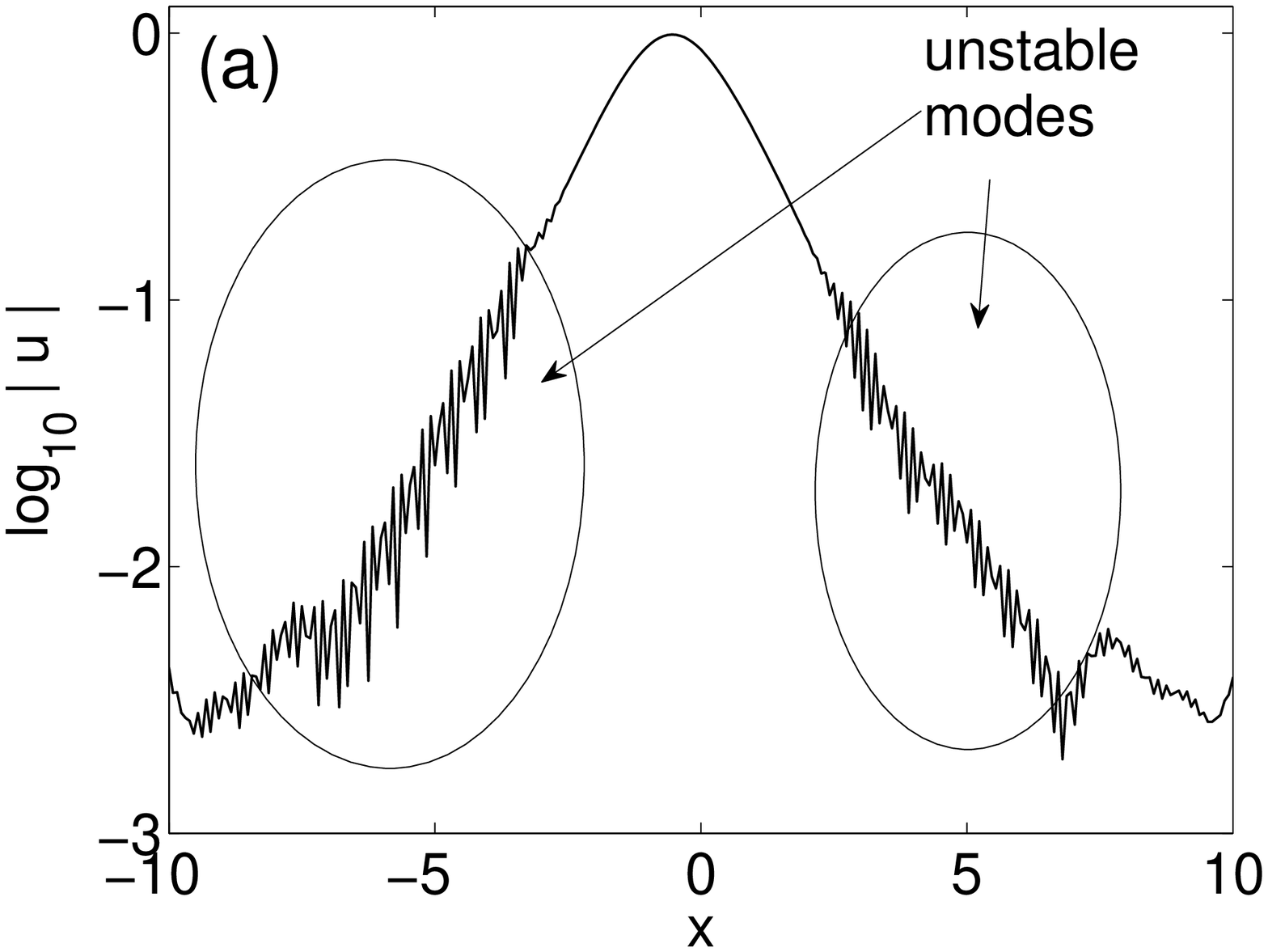}}}
\end{minipage}
\hspace{0.1cm}
\begin{minipage}{7cm}
\rotatebox{0}{\resizebox{7cm}{9cm}{\includegraphics[0in,0.5in]
 [8in,10.5in]{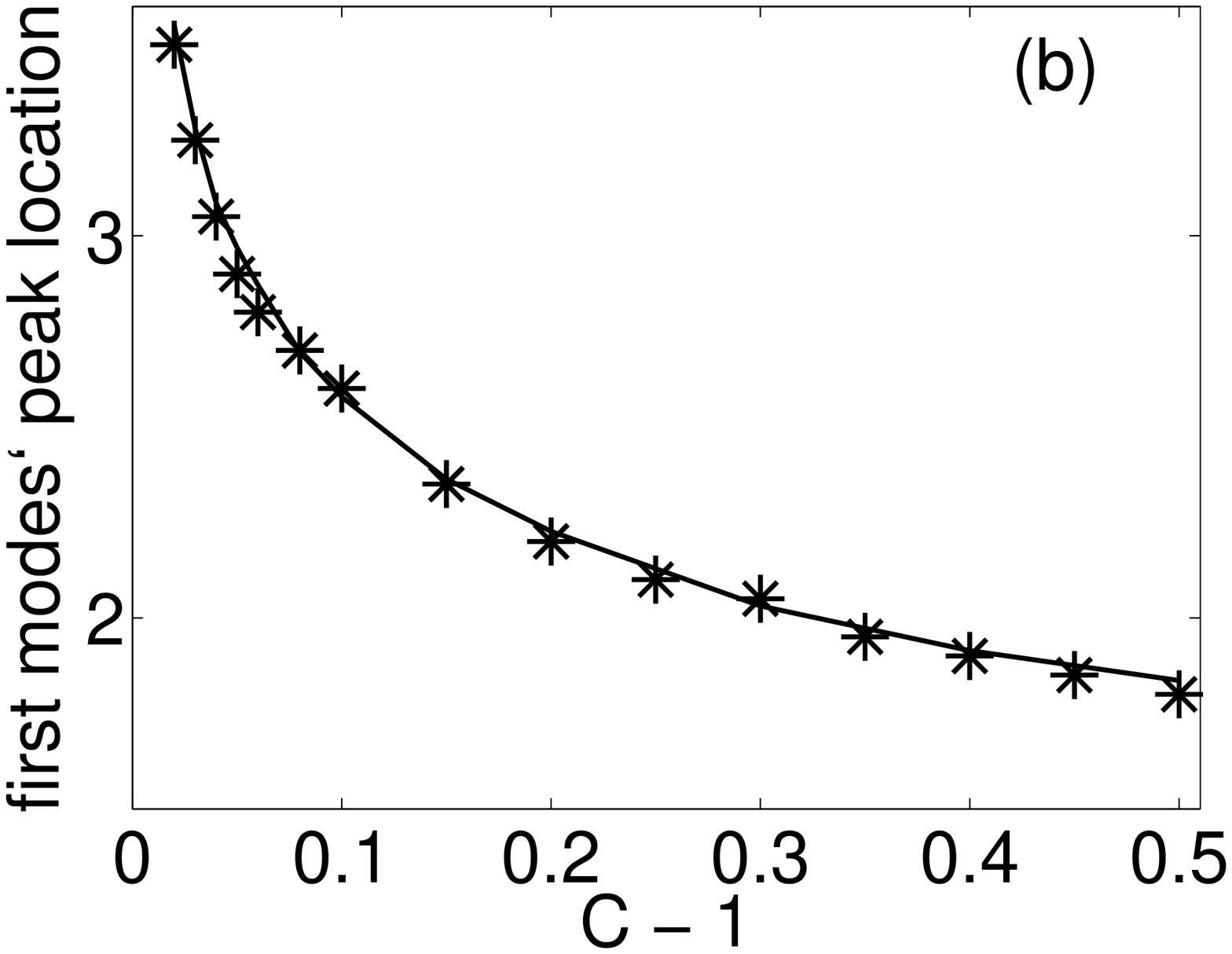}}}
\end{minipage}
 }
\vspace{-1.6cm}
\caption{(a) \ The numerical solution for $t=1500$ and the same parameters 
 as in Fig.~\ref{fig_5}, with $C=1.05$.  
 \ (b) \ Location of the peak of the first localized mode,
found by the method of Appendix B (solid line) and from the solution of \eqref{e_01}
(stars). 
Similar data for $L=40$ and $N=2^{10}$ are very close and hence
are not shown.
}
\label{fig_6add}
\end{figure}

In Fig.~\ref{fig_5} we show the first (i.e., corresponding to the greatest
$\Lambda_R$) such a mode for $L=40$, $N=2^9$ points (hence 
$\epsilon=\dx/2\approx 0.04$), $A=1$, $\beta=-1$, $\gamma=2$.
For these parameters, the threshold given by the r.h.s. of \eqref{e_37} is
$C=1$, and parameter $D$ in \eqref{e_36} is related to $C$ by:
\be
D=C-1.
\label{sec4_extra1}
\ee
The numerical method of solving \eqref{e_36} is described in Appendix B,
and the modes found by this method are shown in Fig.~\ref{fig_5}(a) for
different values of $C$. 
In Fig.~\ref{fig_5}(b) we show the same modes obtained from the numerical solution
of the NLS \eqref{e_01} by the fd-SSM. These
modes were extracted from the numerical solution by a high-pass filter, and then
the highest-frequency harmonic was factored out as per \eqref{e_26}. 
The agreement between Figs.~\ref{fig_5}(a) and \ref{fig_5}(b) is seen to be good.
Note that Fig.~\ref{fig_5} shows, essentially, the envelope of the unstable mode.
The mode {\em not extracted} from the numerical solution is shown
in Fig.~\ref{fig_6add}(a); it can also be seen at the ``tails" of the soliton
in Fig.~\ref{fig_2}(b).

In Fig.~\ref{fig_6add}(b) we show the location of the peak of the first unstable mode,
computed both from \eqref{e_36} and from the numerical solution of \eqref{e_01},
versus parameter $C$. The corresponding values of the instability growth rate 
$\lambda$ were shown earlier in Fig.~\ref{fig_3}. Let us stress that $\lambda$
for the localized modes of \eqref{e_36} was found to be purely real 
up to the computer's round-off error ($\sim10^{-15}$).
There also exist unstable modes with complex $\lambda$, but such modes 
were found to be not
localized and to have smaller growth rates than the localized modes. 

\bigskip

\bigskip

\begin{figure}[h]
\vspace{-0.6cm}
\hspace*{-0.5cm}
\mbox{ 
\begin{minipage}{5.1cm}
\rotatebox{0}{\resizebox{5.1cm}{6.5cm}{\includegraphics[0in,0.5in]
 [8in,10.5in]{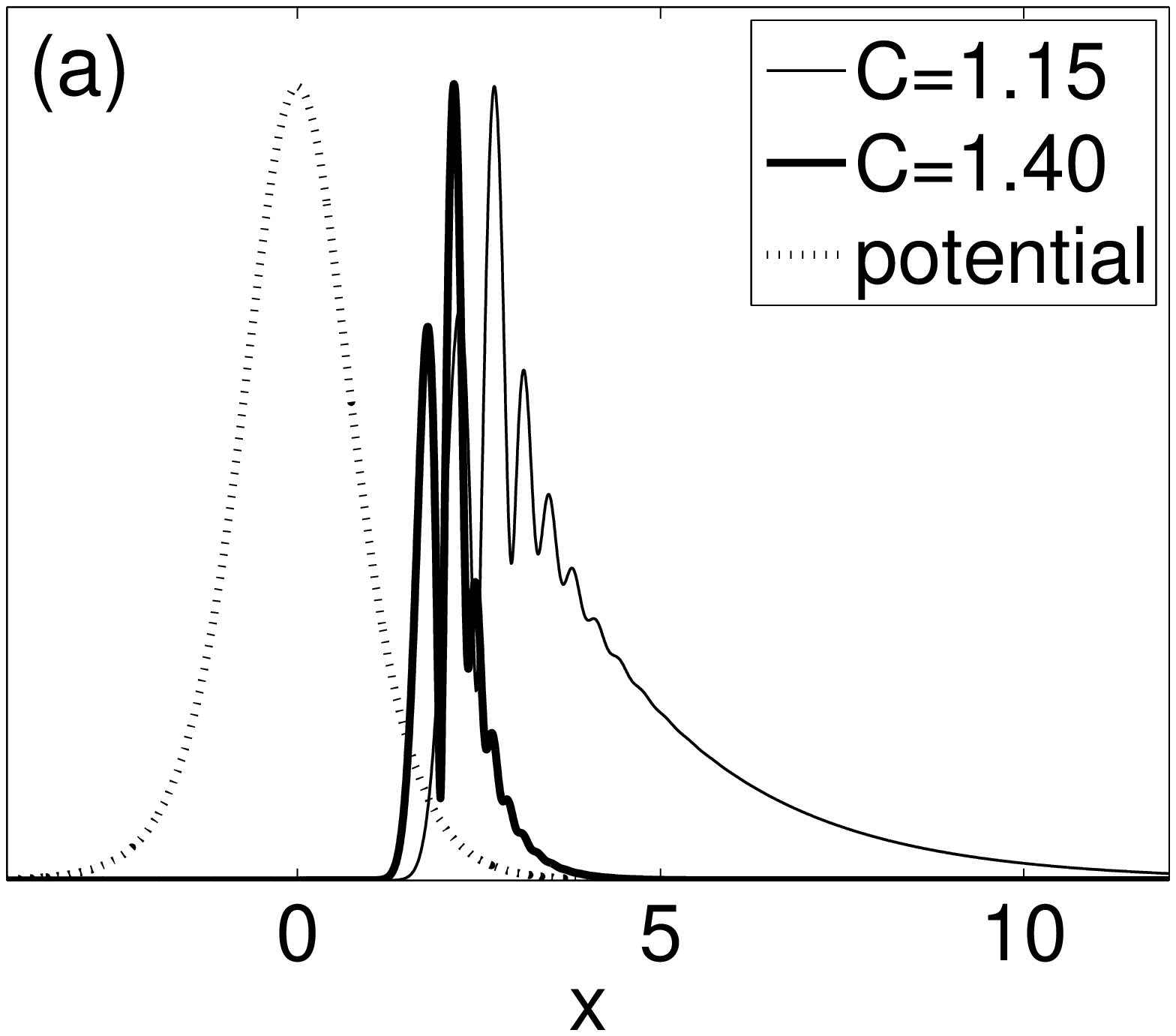}}}
\end{minipage}
\hspace{0.1cm}
\begin{minipage}{5.1cm}
\rotatebox{0}{\resizebox{5.1cm}{6.5cm}{\includegraphics[0in,0.5in]
 [8in,10.5in]{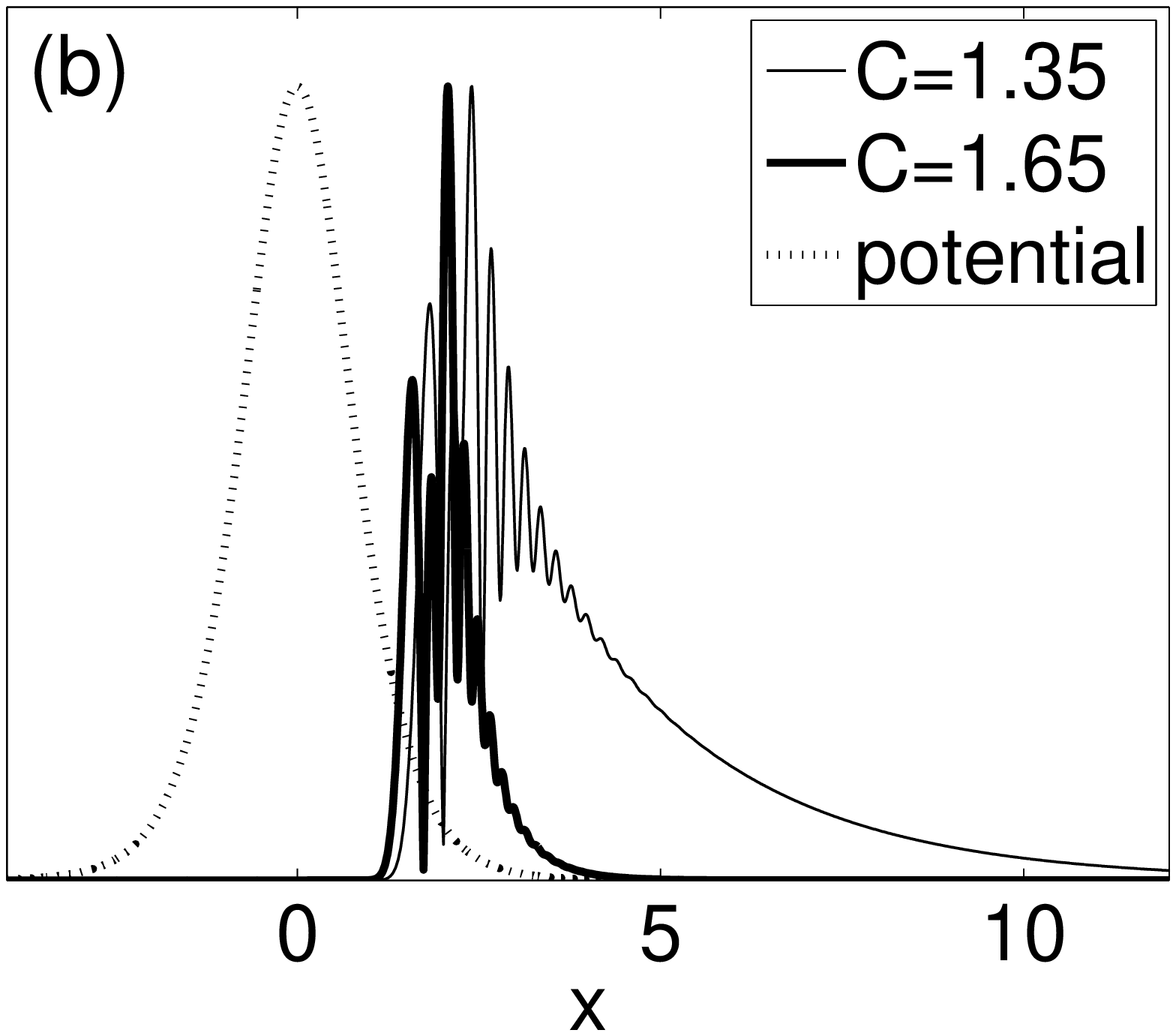}}}
\end{minipage}
\hspace{0.1cm}
\begin{minipage}{5.1cm}
\rotatebox{0}{\resizebox{5.1cm}{6.5cm}{\includegraphics[0in,0.5in]
 [8in,10.5in]{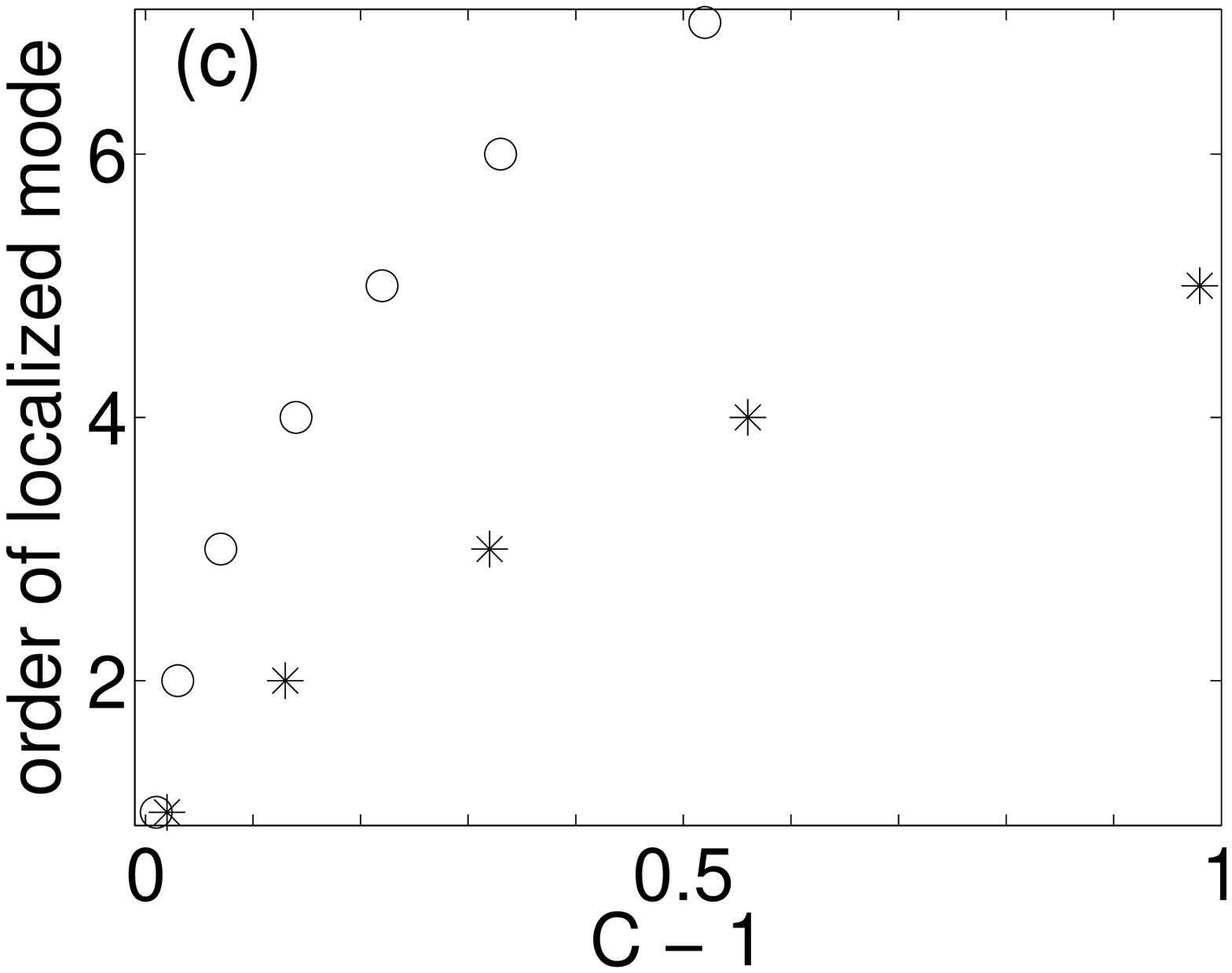}}}
\end{minipage}
 }
\vspace{-1.6cm}
\caption{Similar to Fig.~\ref{fig_5}(a), but for the second (a) and third (b) localized modes.
 \ (c): $C$ values where localized modes of increasing order appear. Stars --- for 
 $\epsilon = 40/1024$, circles --- for $\epsilon = 40/2048$. 
}
\label{fig_6}
\end{figure}

As $C$ increases from the critical value given by \eqref{e_37}, the localized unstable 
mode becomes narrower  and also moves toward the center of
the soliton. Moreover, higher-order localized modes of \eqref{e_36} arise. Typical
profiles of the second and third modes are shown in Fig.~\ref{fig_6}, along with the
parameter $C$ 
for which such modes first become localized within the spatial domain. 
In Appendix C we demonstrate that the process of ``birth" of an eigenmode that eventually
(i.e., with the increase of $C$) becomes localized,  is rather complicated. In particular,
it is difficult to pinpoint the exact value of parameter $C$ where such a mode appears.
Therefore, the $C$ values shown in Fig.~\ref{fig_6} are accurate only up to the second
decimal place.


\subsection{Effect of unstable modes on soliton}

Let us now show how our results can qualitatively explain the 
observed dynamics of the numerically
unstable soliton --- see the text after Eq.~\eqref{e_16} and Fig.~\ref{fig_2}(b). 
Let $\tilde{u}_{\rm unst}$ be the field of the unstable
modes at the soliton's sides. 
At an early stage of the instabilty, 
it is much less than the amplitude of the soliton: 
$|\tilde{u}_{\rm unst}|\ll A$. Also, its
characteristic wavenumbers are much greater than those of the soliton: see Fig.~\ref{fig_2}(a)
and \eqref{e_23}.
Then, to determine its effect on the soliton,
one substitutes $u=u_{\rm sol}+u_{\rm unst}$ into the NLS \eqref{e_01} and discards all the 
high-wavenumber terms to obtain:
\be
i(\usol)_t -\beta (\usol)_{xx}+\gamma \usol |\usol|^2 = - 2\gamma \usol |u_{\rm unst}|^2.
\label{conc_2}
\ee
This is the equation for a perturbed soliton with the perturbation being, in general, 
not symmetric about
the soliton's center (see Fig.~\ref{fig_5}(c)). Indeed, the modes on the left and right sides 
of the soliton do not ``see" each other 
through the wide barrier created by the soliton's core
and hence can have different amplitudes. Such an asymmetric perturbation is
known (see, e.g., \cite{Agrawal_book}, Sec.~5.4.1) to cause the soliton to move, 
which is precisely the 
effect reported in Fig.~\ref{fig_2}(b).


\section{Numerical instability of soliton in generalized NLS}
\setcounter{equation}{0}


The analysis of Secs.~3 and 4.1 easily extends to the case when 
the nonlinearity in \eqref{e_00} has a different form 
than in \eqref{e_01} (e.g., is saturable) or when
an external potential $\Pi(x)$ is included. Below we focus on the latter
situation, i.e. on the subclass of \eqref{e_00} described by
\bsube
\be
iu_t - \beta u_{xx} + \big( \gamma |u|^2 + \Pi(x) \big)u =0.
\label{e4_301a}
\ee
For brevity, and without loss of generality, we will assume 
\be
\beta < 0.
\label{e4_301b}
\ee
\label{e4_301}
\esube
The opposite choice, i.e. $\beta>0$ (with the corresponding adjustment of signs of both $\gamma$
and $\Pi$), will not affect either real or numerical instabilities of the solution, 
since Eq.~\eqref{e4_301a} is Hamiltonian.

The soliton solution of \eqref{e4_301} has the form similar to \eqref{e_12}:
\bsube
\be 
u(x,t) = \Usol(x) \,\exp[i\omsol t],
\label{e4_302a}
\ee
where now $\Usol(x)$ and $\omsol$ are found (usually numerically) 
from the nonlinear eigenvalue problem
\be
| \beta | \, (\Usol)_{xx} + \big( \gamma |\Usol|^2 + \Pi(x) \big) \Usol = \omsol \Usol.
\label{e4_302b}
\ee
\label{e4_302}
\esube
Note that in the presence of potential $\Pi(x)$, the soliton has zero velocity.
The evolution equation for the unstable mode, $\psi$, is similar to \eqref{e_32}:
\bsube
\be
i\psi_t + \delta\psi - \psi_{\chi\chi}/(C|\beta|) + \gamma \Usol^2(\epsilon\chi)\;
 (2\psi + \psi^*) + \Pi(\epsilon\chi)\,\psi =0,
\label{e4_303a}
\ee
where now
\be
\delta = -\omsol + 1/(C|\beta|),
\label{e4_303b}
\ee
\label{e4_303}
\esube
with $\Usol$ and $\omsol$ being defined by \eqref{e4_302}.

Below we will describe three scenarios in which the NI 
governed by \eqref{e4_303} may be qualitatively different from that
governed by \eqref{e_32} and described earlier in this paper. 
In the first two scenarios, the differences will be quite obvious,
while in the third, it will be less so. Nonetheless, it is this third
scenario of the onset of NI that will be shown in Sec.~7 to occur in
a yet wider range of situations.

Let us also mention that in all scenarios, 
the presence of an external potential 
affects the nonlinear stage of NI (see Sec.~4.2) in a predictable way:
due to the confinement by the potential, the soliton would not drift. 
Rather, it would disintegrate once the numerical noise becomes strong enough.


\subsection{External potential with multiple minima}

We will show that in this case, unstable modes can be localized only at
the absolute minima of the potential. This will result in ``stability windows",
i.e. intervals of $C$ values past the NI threshold where NI does not occur.

\begin{figure}[h]
\vspace{-1.6cm}
\centerline{ 
\begin{minipage}{7cm}
\rotatebox{0}{\resizebox{7cm}{9cm}{\includegraphics[0in,0.5in]
[8in,10.5in]{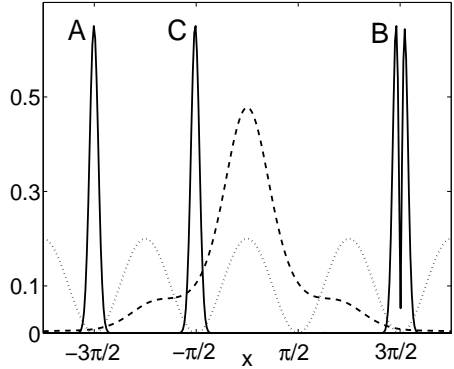}}}
\end{minipage}
 }
\vspace{-1.6cm}
\caption{Dashed line: ``Internal" potential 
$\gamma |\Usol|^2$ ($\gamma=2$), 
where $\Usol$ is the solution of \eqref{e4_301} and \eqref{e4_304}. 
The marked vertical scale pertains to this curve; all other curves are plotted with an
arbitrary vertical scale. \ Dotted line: External potential $\Pi(x)$. \ 
Solid lines: Absolute value of unstable modes for various values of $C$. 
(Only one side of each mode
is shown; in numerical simulations; one observes such a mode on both sides of the soliton;
see Fig.~\ref{fig_5}(c).) \ A: $C\in [1.04, \,1.05]$; \ B: $C\in[1.10, \,1.11]$ (this is 
the second-order mode, similar to that in Fig.~\ref{fig_6}(a)); \ 
C: $C > 1.15$. 
}
\label{fig_4_3_1}
\end{figure}

As an example, 
we considered Eq.~\eqref{e4_301} with 
\be
\beta=-1, \quad \gamma=2, \quad \Pi(x)=1.5 \cos^2 x, \quad \omsol = 1.
\label{e4_304}
\ee
The numerical parameters were $L=14\pi$ and $N=2^{10}$.
The corresponding soliton, found by the numerical method of \cite{gPetviashvili},
is shown in Fig.~\ref{fig_4_3_1}. In simulations, the initial condition was taken
as that soliton plus small noise, similarly to \eqref{e_15}.
Note that the estimate \eqref{e_37} of the threshold value of $C$ beyond which NI may 
occur is modified as follows (recall \eqref{e4_301b}):
\be
C_{\rm thresh} \, \approx  \, 
\frac1{|\beta| \left(\, \omsol - \min_x (\Pi + 1\cdot \gamma |\Usol|^2 )\,\right)}.
\label{e4_305}
\ee
The last term in the denominator estimates the ``internal'' potential, 
created by the soliton itself. 
The `$1\cdot$' in front of it indicates that this estimate 
has used the fact that the smaller eigenvalue of the matrix on the
l.h.s. of \eqref{e_36} equals 1;
see Eq.~\eqref{e_C1a} in Appendix C.
 Later on we will explain why the expression
on the r.h.s. is bound to (slightly) underestimate the threshold value for $C$.

We have observed no NI until $C=1.04$, at which point the unstable mode appeared as
curve A in Fig.~\ref{fig_4_3_1}. Note that the mode is localized near a minimum 
of $\Pi(x)$. Given that the ``internal"' potential at this 
$x$ is $2|\Usol(x)|^2\approx 0.01$,
estimate \eqref{e4_305} yields a smaller NI threshold: $C\approx 1.01$.
The discrepancy (i.e., $C=1.01$ versus $C=1.04$)
occurs due to neglecting the contribution of $\psi_{\chi\chi}$ in the derivation of
estimate \eqref{e4_305} as explained before \eqref{e_37}. Since operator $\partial_{\chi\chi}$
is non-positive definite, then accounting for its contribution
would decrease the denominator of  \eqref{e4_305} and hence increase $C_{\rm thresh}$.
This effect is more conspicuous in the case of the soliton of \eqref{e4_301}, \eqref{e4_304}
than it was for the NLS soliton in Sec.~4.1 because in the former case, the unstable mode
is more localized (compare Fig.~\ref{fig_4_3_1} and the thin solid curve in 
Fig.~\ref{fig_5}(b)), leading to a more negative contribution from $\partial_{\chi\chi}$.
From the above discussion and Eq.~\eqref{e4_305}, 
the contribution of operator $\partial_{\chi\chi}$ to the
threshold value of $C$ can be roughly estimated as:
\be
``\partial_{\chi\chi}" \approx \frac1{1.04} - \frac1{1.01} \approx -0.03.
\label{e4_306}
\ee
Let us note, in passing, that the small absolute value of $\partial_{\chi\chi}$
agrees with the fact that while the mode is seen as narrow in $x$-space, it is still
very wide in $\chi$-space (recall that $\chi=x/\epsilon$).

In an interval $C\in [1.06,\, 1.09]$,  NI disappears. This occurs due to the following.
As $C$ increases, the unstable mode ``wants" to move towards the soliton's center, similarly
to the situation shown in Figs.~\ref{fig_6add}(b).\footnote{ 
   One cannot {\em explain} this behavior without an analytical solution
	 of the eigenvalue problem \eqref{e4_309} below, which is an extension of
	the eigenvalue problem \eqref{e_36} in the presence of the external	 potential.
	 For the reason explained in Sec.~4.1, such an analytical solution does not 
	appear to be possible at this time. 
   However, the tendency of the localized mode to shift towards the center of the
	soliton with the increase of $C$ has been consistently verified by our numerical solution
	 of \eqref{e_36} and \eqref{e4_309}.}
As it moves, the value of $\Pi(x)$ increases and this, according to \eqref{e4_305},
increases the NI threshold, leading to NI's disappearance.

As $C$ continues to increase, 
the second-order unstable mode moves from outside the soliton
to the location $x\approx 3\pi/2$, where
$\Pi(x)\approx 0$, and then NI reappears, being now caused by that second-order mode;
see curve B in Fig.~\ref{fig_4_3_1}. With further increase of $C$, that mode moves
towards the soliton's center and
away from the minimum of $\Pi(x)$, and NI disappears again.

It reappears when the first-order unstable mode moves into the minimum of $\Pi(x)$ closest
to the soliton's center; see curve C in Fig.~\ref{fig_4_3_1}. In our numerical
simulations this was observed starting at $C\approx 1.15$. 
On the other hand, estimate \eqref{e4_305}
yields $C\approx 1.11$, where we have used that at $x\approx \pi/2$ one has $2|\Usol|^2\approx 0.1$.
However, if we add the contribution of $\partial_{\chi\chi}$ to the denominator of
\eqref{e4_305} and use \eqref{e4_306}, we obtain: \ $C\approx 1/(1-0.1-0.3)\approx 1.15$,
which is in excellent agreement with the numerical result.


\subsection{Bell-shaped potential; $\gamma<0$}

We will show that in this case, the unstable mode can appear either at the center or
at the ``tail" of the soliton.

As an example, 
we considered Eq.~\eqref{e4_301} with 
\be
\beta=-1, \quad \gamma=-1, \quad \Pi(x)=6 \sech^2 x,
\label{e4_307}
\ee
and used $L=40$ and $N=2^{10}$. 
Let us note that equations with $\gamma<0$ are not too uncommon; for instance, the
generalized NLS with saturable nonlinearity \cite{Segev_1994}
provides an example of a realistic physical system with negative effective nonlinearity.  

We will first describe how the 
soliton of \eqref{e4_301}, \eqref{e4_307} depends on $\omsol$,
as this will explain different behaviors of NI observed in this case.
By comparing the equation in question with the linear Schr\"odinger
equation with a $\sech^2 x$ potential, one can see that
its soliton exists for $\omsol \in(0,\,4)$. At $\omsol=4-0$,
it becomes vanishingly small and has the shape of $\sech^2 x$. As
$\omsol$ decreases, the soliton becomes wider and its amplitude grows,
 so that at $\omsol=1$, one has $\Usol=2\sech x$. 
As $\omsol = +0$, the soliton becomes very wide and its amplitude approaches
$\sqrt{6}$. Amplitudes of the soliton at three values of $\omsol$ are shown
in Table \ref{table_1}.

\begin{table}
\bc 
\begin{tabular}{|c|c|c|c|c|c|}  \hline 
$\omsol$   &  $\max{|\Usol|}$ & $\min\big( \Pi - 3\gamma|\Usol|^2\big)$ &
mode's & $C_{\rm thresh,\;\eqref{e4_308}}$  & $C_{\rm thresh,\;numer}$  \\
   &   &   &  location &   &   
  \\ \hline \hline
 1  &  2       &  $-6$     & center  & $0.144$  & $0.145$  \\ \hline 
 2  & $1.657$  &  $-2.237$ & center  & $0.237$  & $0.241$  \\ \hline 
 3  & $1.187$  &  0        & ``tail" & $0.334$  & see Sec.~5.3 \\ \hline 
\end{tabular}
\ec
\caption{NI of the soliton of Eq.~\eqref{e4_301}, \eqref{e4_307}; see text for details.
The last two columns list theoretical and numerically observed values for the NI
threshold.
 }
\label{table_1}
\end{table}

The estimate of the the threshold beyond which NI {\em can} appear
is almost the same as \eqref{e4_305}:
\be
C_{\rm thresh} \, \approx  \, 
\frac1{|\beta| \left(\, \omsol - \min_x (\Pi - 3\cdot |\gamma| |\Usol|^2 )\,\right)}.
\label{e4_308}
\ee
Here the `$3\cdot$' in front of the last term occurs
because to minimize the expression in parentheses, one needs to 
use the larger eigenvalue of the matrix on the l.h.s. of \eqref{e_36},
since now $\gamma<0$. The validity of this estimate is supported
by the first two lines of Table \ref{table_1}.
We would like to stress three aspects of these results.

First, since for $\omsol=1$ and 2, \ $\min_x (\Pi - 3\cdot |\gamma| |\Usol|^2 )$ \ 
occurs at $x=0$, the unstable mode appears at the soliton's center rather than at
its ``tails", as was the case in Sec.~4. This mode looks like modes A and C
in Fig.~\ref{fig_4_3_1} except that it is located at $x=0$.

Second, when the unstable mode occurs at the soliton's center, NI develops very
rapidly with respect to parameter $C$. That is, lowering $C$ by 0.001 
compared to the value listed in the Table will suppress the NI entirely.
On the contrary, at the indicated $C_{\rm thresh}$, magnitude of unstable modes
reaches $O(1)$ within $t\sim 100$, which is more than an order of magnitude faster
than the unstable modes in Secs.~2 and 4 would do within 1\% past $C_{\rm thresh}$.

Third, the case $\omsol=3$ is different from that of $\omsol=1$ or 2 in that
the unstable mode is predicted by \eqref{e4_308} to be at the ``tails" of the soliton.
In that respect, it is similar to the mode discussed in Sec.~4. 
However, we have also observed substantial differences from the 
unstable mode of the pure NLS \eqref{e_01}. These new features of the NI
are {\em not} specific to having $\gamma<0$, and therefore we report them
in a separate subsection, which follows next.


\subsection{``Sluggish" numerical instability}

\begin{figure}[h]
\vspace{-1.6cm}
\begin{minipage}{7cm}
\rotatebox{0}{\resizebox{7cm}{9cm}{\includegraphics[0in,0.5in]
[8in,10.5in]{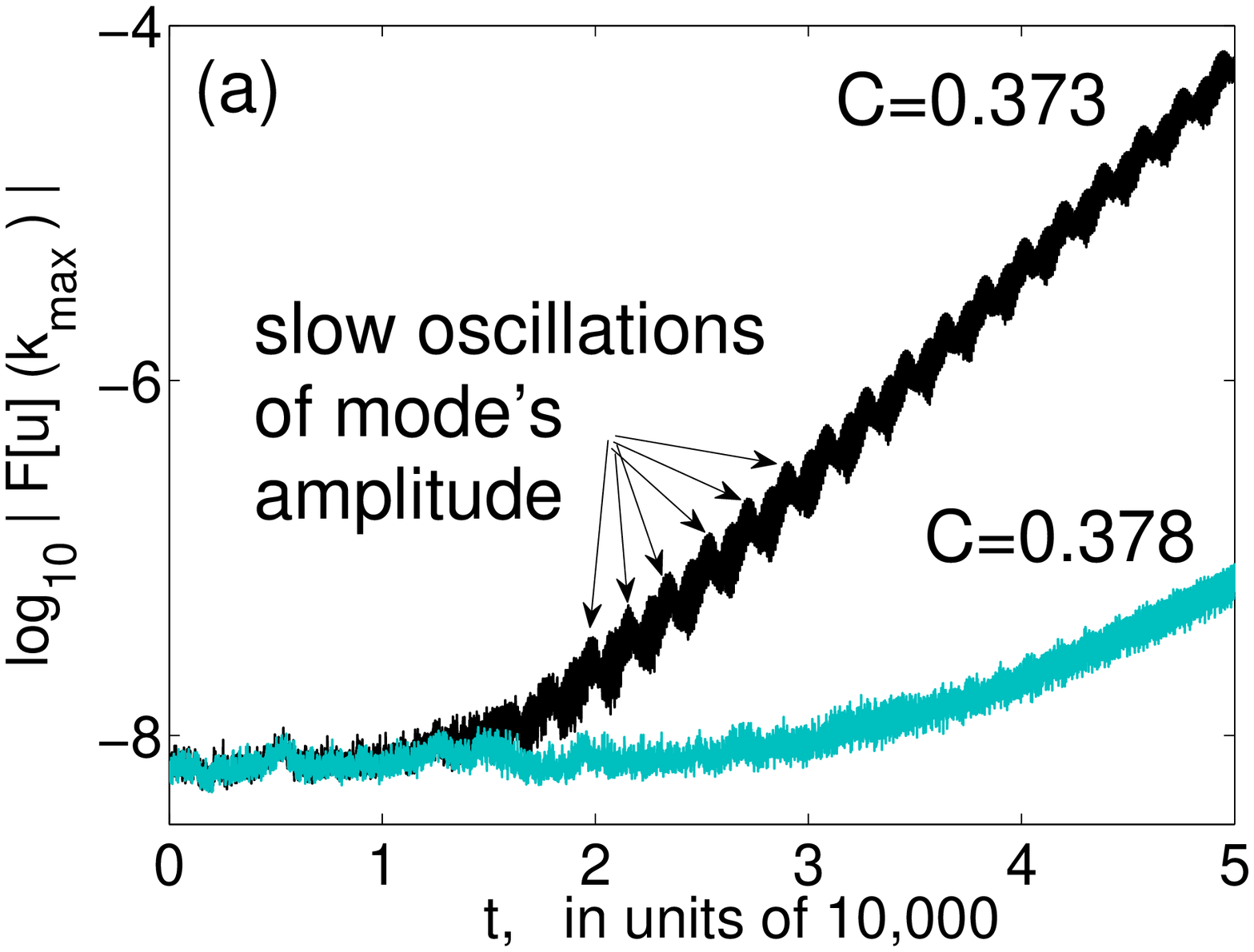}}}
\end{minipage}
\hspace{0.1cm}
\begin{minipage}{7cm}
\rotatebox{0}{\resizebox{7cm}{9cm}{\includegraphics[0in,0.5in]
[8in,10.5in]{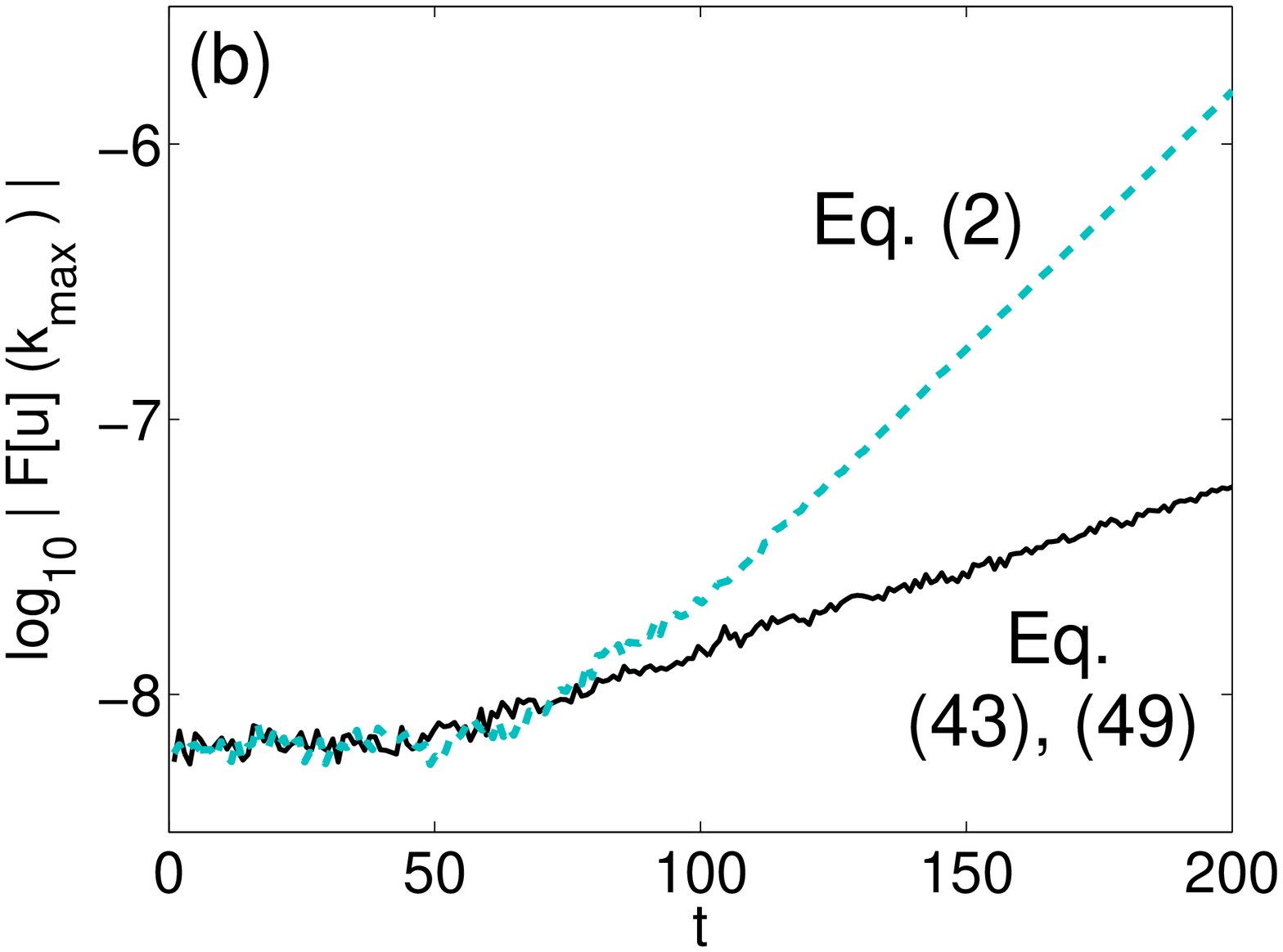}}}
\end{minipage}
\vspace{-1.6cm}
\caption{Evolution of the highest Fourier harmonic (see text). \
 (a) \ For Eq.~\eqref{e4_301}, \eqref{e4_307} with $\omsol=3$. The lines appear thick because
of oscillations on the scale of $t\sim 50$, which is not resolved in this figure.
Note also slower oscillations with the period of $t\sim 2000$. \ 
(b) \ For Eq.~\eqref{e_01}, $C=1.1$; \ for Eq.~\eqref{e4_301}, \eqref{e4_307}, $C=0.45$.
}
\label{fig_4_3_2}
\end{figure}

We begin by reporting our results for the model \eqref{e4_301}, \eqref{e4_307}
with $\omsol=3$.
For several values of $C$ near the theoretical threshold $C_{\rm thresh}=1/3$, we 
ran simulations up to $t=50,000$. Recall that other numerical parameters are
$L=40$ and $N=2^{10}$.
At $C=0.345$, which is over 3\% above the threshold,
we have not observed any sign of NI. At $C=0.350$,
we have
observed an order-of-magnitude growth (from $10^{-8}$ to $10^{-7}$) of
high-$k$ harmonics in the Fourier spectrum. 
In comparison, for the same relative increase above the threshold,
$(C-C_{\rm thresh})/ C_{\rm thresh} \approx 5$\%, the NI {\em growth rate}
of the pure NLS
soliton is about two orders of magnitude greater: see Fig.~\ref{fig_3}.
As we continued to increase
$C$, the NI has gradually become stronger; however, this was {\em not monotonic}.
For example, the evolution of $|\F[u](k_{\rm max})|$ at two values of $C$
is shown in Fig.~\ref{fig_4_3_2}(a), where a stronger NI corresponds to the 
smaller $C$. It is only past $C\approx 0.45$, i.e. 35\% above the threshold
predicted by \eqref{e4_308}, that the increase of NI's growth rate with $C$
becomes monotonic.

We have called this NI ``sluggish" due to its very slow, compared to the pure NLS case,
development with the increase of $C$. We have found that it occurs when
the external potential $\Pi(x)$ is either wider or significantly taller 
(or both, as in the case reported above)
than the ``internal" potential $\gamma |\Usol(x)|^2$. It is {\em not} 
specific to the particular sign of $\gamma$; for example, it also occurs 
when in \eqref{e4_308} one takes $\gamma=+1$ (and, e.g., $\omsol=5$), as well
as for Eq.~\eqref{e4_310} below.
We will now list features of this ``sluggish" NI and then will provide
some insight into them. In Sec.~8 we will speculate on a reason behind
the occurrence of ``sluggish" NI.

\subsubsection{Features of ``sluggish" NI}

(i) \ The unstable mode could remain ``hidden" for some time. This is most
conspicuous when $C$ is close to the threshold value predicted by \eqref{e4_308} or,
more generally, when the NI is weak. For example, in the cases shown in
Fig.~\ref{fig_4_3_2}(a), NI becomes visible only after $t\sim 15,000$ for $C=0.373$
and $t\sim 25,000$ for $C=0.378$. \ Motivated by this observation, 
we revisited our earlier simulations for the soliton of the pure NLS.
We have found the same ``delayed" NI there as well, except that its starting time
was considerably less; see Fig.~\ref{fig_4_3_2}(b). 

(ii) \ The increase of NI with $C$ is not monotonic; that is,
as one increases $C$, NI may sometimes get substantially weaker
than it was for a smaller value of $C$. This was illustrated by Fig.~\ref{fig_4_3_2}(a), 
but has also been observed in many other cases.

(iii) \ Growth of unstable modes {\em with time} is not monotonic, either. 
A mild example of
it is also shown in Fig.~\ref{fig_4_3_2}(a); in some cases, we even
observed oscillations of mode's amplitude of almost on order of magnitude. 

\begin{figure}[h]
\vspace{-1.6cm}
\begin{minipage}{7cm}
\rotatebox{0}{\resizebox{7cm}{9cm}{\includegraphics[0in,0.5in]
[8in,10.5in]{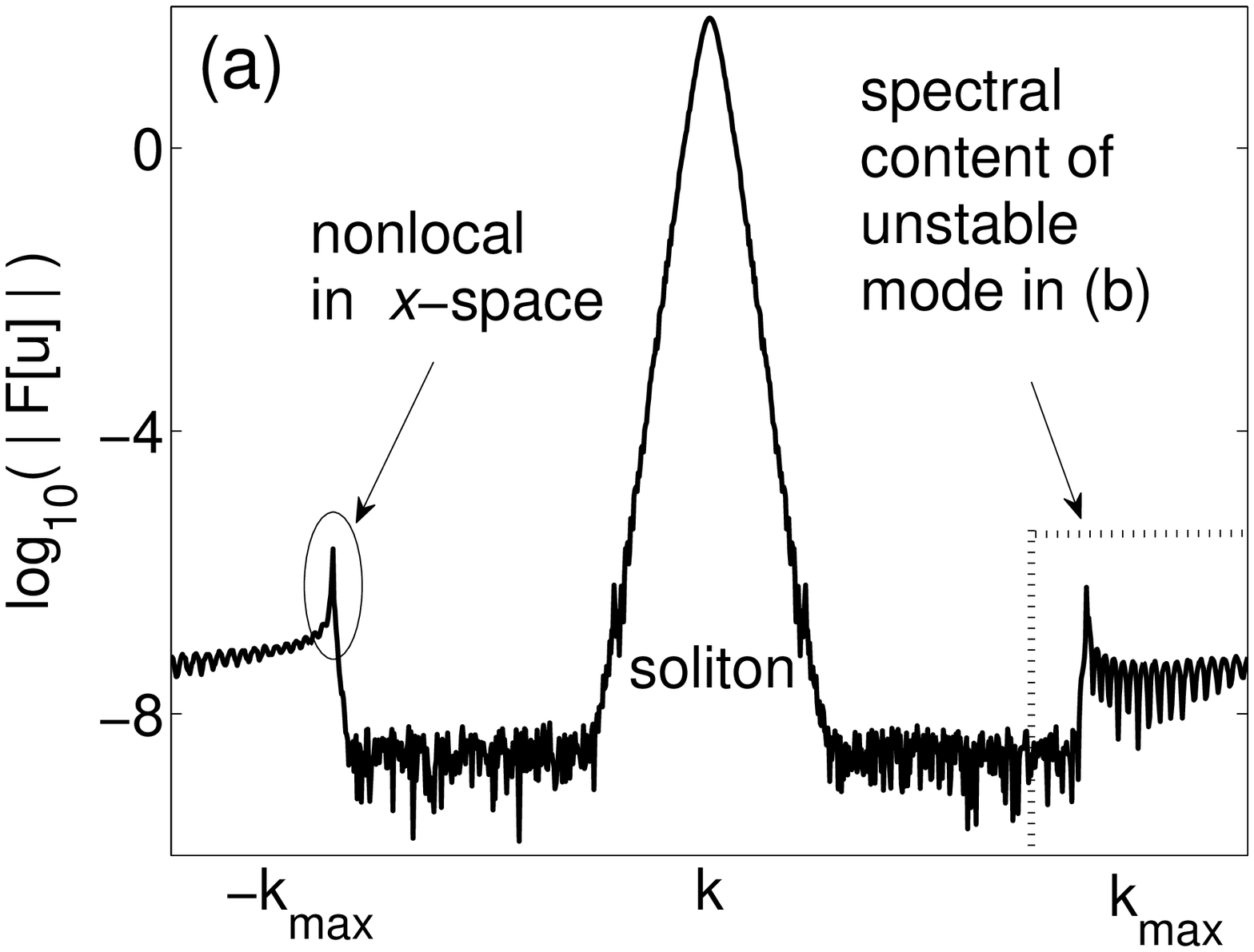}}}
\end{minipage}
\hspace{0.1cm}
\begin{minipage}{7cm}
\rotatebox{0}{\resizebox{7cm}{9cm}{\includegraphics[0in,0.5in]
[8in,10.5in]{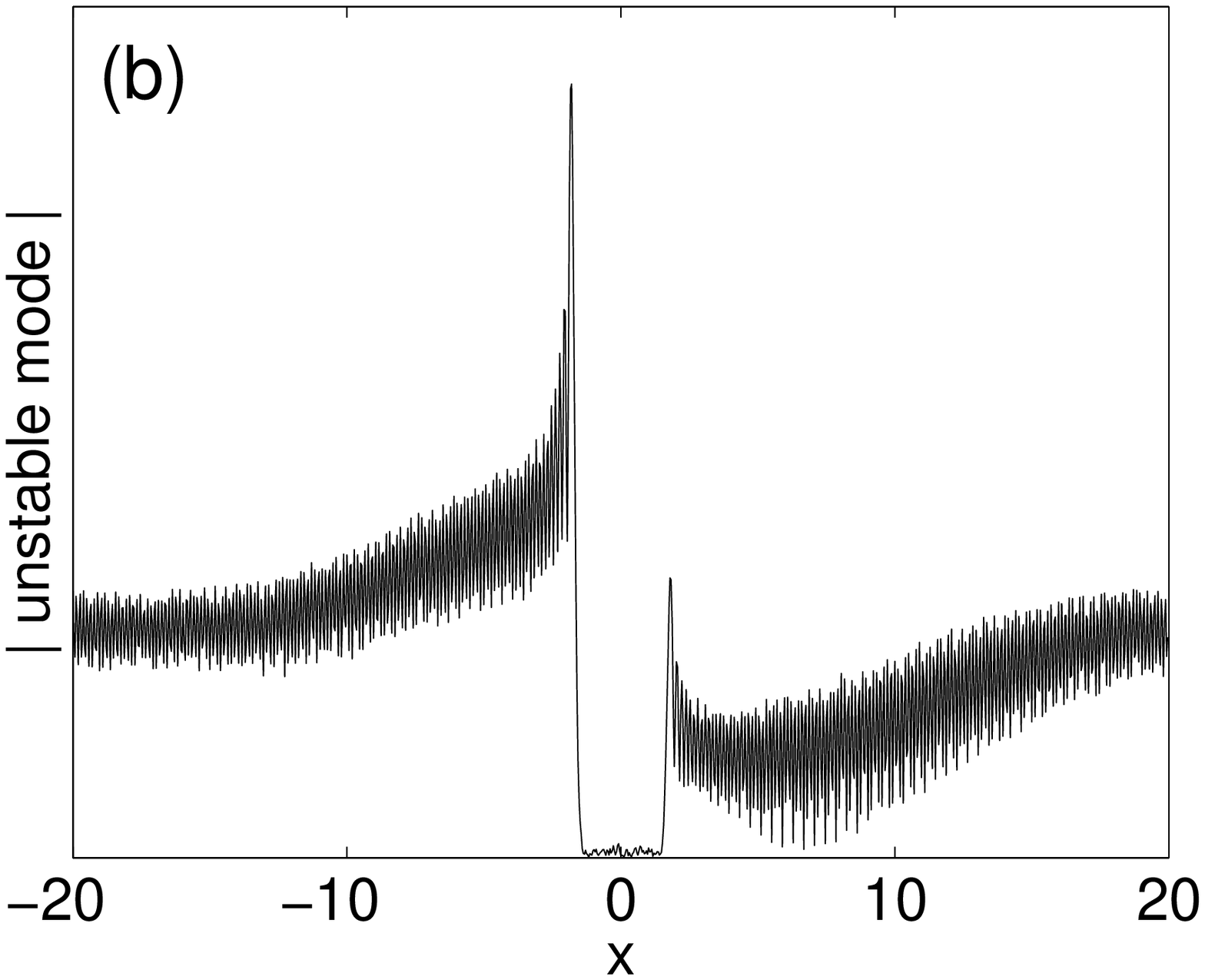}}}
\end{minipage}
\vspace{-1.6cm}
\caption{(a) Fourier spectrum of ``sluggish" NI; compare to Fig.~\ref{fig_2}(a). 
\ (b) Absolute value in $x$-space of the unstable mode whose spectral
content is shown in the box on the right of panel (a); compare to Fig.~\ref{fig_5}.
}
\label{fig_4_3_3}
\end{figure}

(iv) \ The unstable modes of this ``sluggish" NI look different from 
the unstable modes described in Sec.~4. A typical example is shown in
Fig.~\ref{fig_4_3_3}. The difference in $x$-space is that while the mode
is still almost zero within the soliton (and the external potential), 
it is {\em not localized} outside the soliton. In $k$-space, the latter
circumstance is reflected by a peak marked in Fig.~\ref{fig_4_3_3}(a),
while the steep decay of the mode towards the soliton's center is reflected
in a broad ``plateau", similarly to what occurred for the unstable mode of the 
pure-NLS soliton. \ Let us note that these characteristics of a ``sluggish"
unstable mode are {\em generic}. Eventually, as $C$ becomes large enough,
the shape of the unstable mode becomes qualitatively similar to that 
described in Sec.~4.

(v) \ The fact that the unstable mode may be non-localized in $x$ implies that
the growth rate of ``sluggish" NI can be affected by the length $L$ of the
computational domain, and this was indeed observed in our numerics.

(vi) \ Finally, as one decreases $\dx$, the relative range \ 
$\D C_{\rm rel,\,sluggish} \equiv 
 (C-C_{\rm thresh,\,\eqref{e4_308}})/C_{\rm thresh,\,\eqref{e4_308}}$ \ 
where the ``sluggish" NI is observed,\footnote{
We have delineated between ``sluggish" and ``non-sluggish" NIs by
whether the unstable mode is localized (has width of $O(1)$) in $x$-space.
The values of $C$ where NI becomes ``non-sluggish" approximately 
coincide with those values where the NI's growth rate begins to increase
significantly.}
decreases. 
For example, if in
the simulations reported at the beginning of this subsection
 one takes $N=2^{11}$ or $N=2^{12}$ (i.e. decreases $\dx$ two- and four-fold), 
then ``sluggish" NI turns
into ``non-sluggish" one around $C=0.42$ and $0.38$, respectively.
These values correspond to the 
$\D C_{\rm rel,\,sluggish} < 30$\% and $\D C_{\rm rel,\,sluggish}\approx 15$\%,
which should be contrasted with $C\approx 0.47$ for $N=2^{10}$,
where $\D C_{\rm rel,\,sluggish} > 40$\%.)

\subsubsection{Explanation of features of ``sluggish" NI}

To provide some insight into these features, we have computed 
eigenvalues and eigenfunctions of the problem
\be
\left( \frac1{C|\beta|} \partial_{\chi\chi} - \delta -\Pi(\epsilon\chi) - 
       \gamma|\Usol(\epsilon\chi)|^2 \left( \ba{cc} 2 & 1 \\ 1 & 2 \ea \right)
\right) \vec{\phi} = i\lambda \sigma_3 \vec{\phi},
\label{e4_309}
\ee
obtained from \eqref{e4_303}.
Both the notations and the method of numerical solution of this eigenproblem
are the same as for \eqref{e_36}. Below we report results
for the following specific values of parameters:
\be
\beta=-1, \quad \gamma=2, \quad \Pi(x)=e^{-0.3x^2}, \quad   \omsol=1.
\label{e4_310}
\ee
We have chosen a different $\Pi(x)$
than in \eqref{e4_307} to emphasize that to bring about a ``sluggish" NI
it may be sufficient to have the external potential wider, but
not necessarily much taller, than the internal one. (Both these potentials
corresponding to \eqref{e4_310} are shown in Fig.~\ref{fig_4_3_5} below.)
However, we have also solved \eqref{e4_309} for parameters \eqref{e4_307}
and found qualitatively similar results.

\begin{figure}[h]
\vspace{-1.6cm}
\begin{minipage}{7cm}
\rotatebox{0}{\resizebox{7cm}{9cm}{\includegraphics[0in,0.5in]
[8in,10.5in]{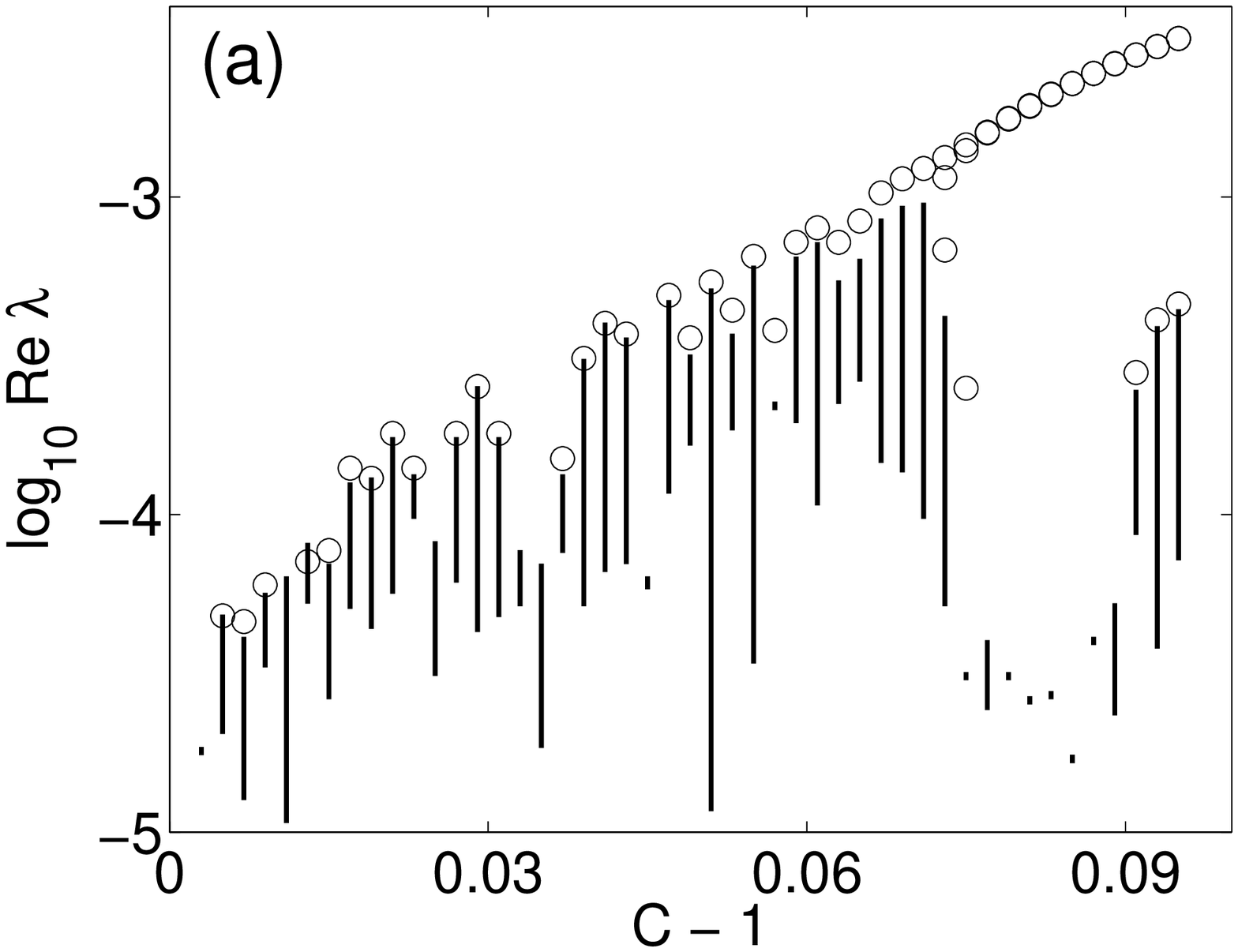}}}
\end{minipage}
\hspace{0.1cm}
\begin{minipage}{7cm}
\rotatebox{0}{\resizebox{7cm}{9cm}{\includegraphics[0in,0.5in]
[8in,10.5in]{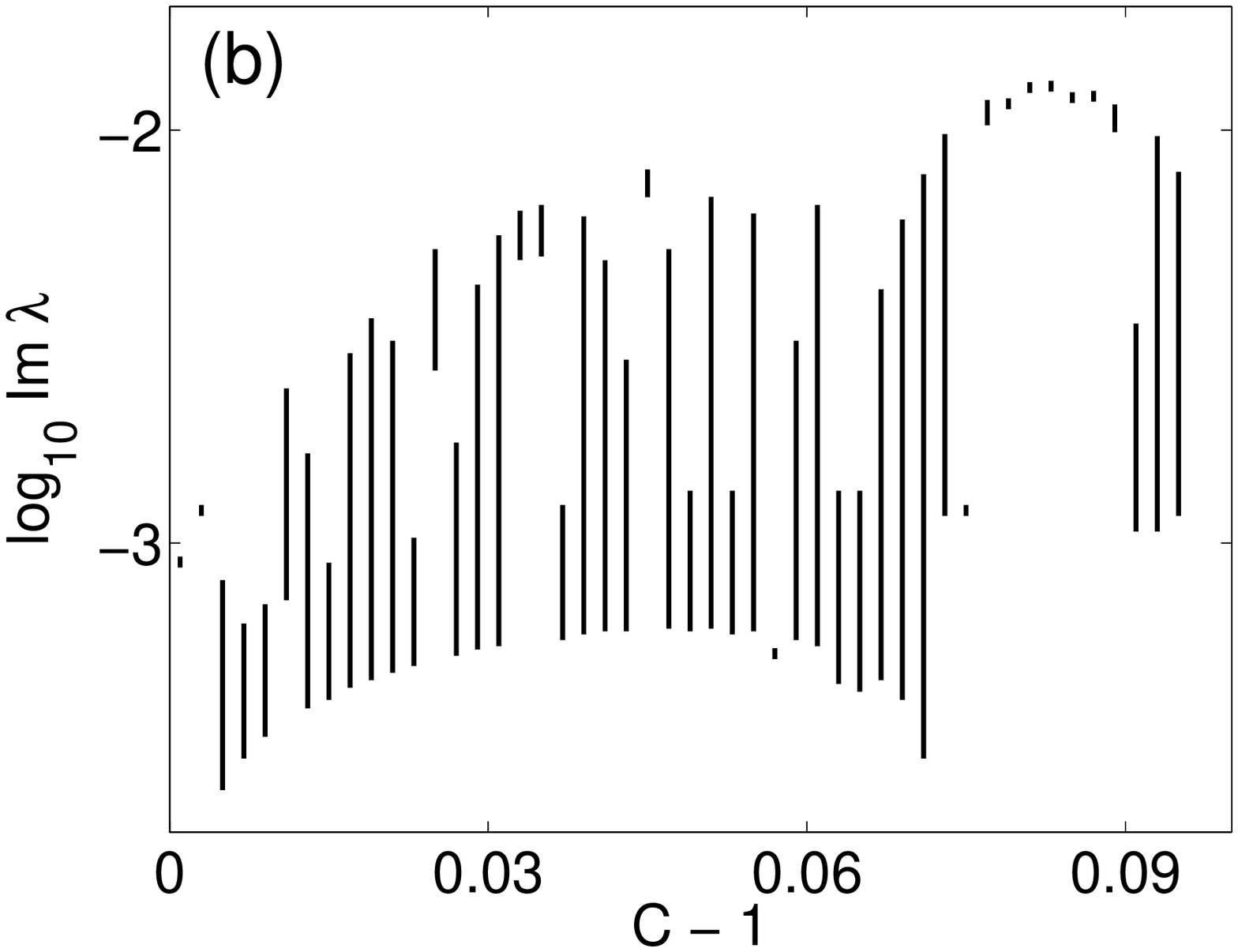}}}
\end{minipage}
\vspace{-1.6cm}
\caption{Eigenvalues of \eqref{e4_309}, \eqref{e4_310} with Re$\,\lambda>0$. \ 
Note that $C_{\rm thresh,\,\eqref{e4_308}}=1$. \ 
(a) \ Circles: Purely real eigenvalues. \ (a) and (b) \ Vertical segments show the 
intervals of Re$\,\lambda$ and Im$\,\lambda$ corresponding to the smallest
$|\lambda|$. (For some $C$, there are also other intervals with larger $|\lambda|$.) 
It can be seen that around $C=1.077$, two purely real eigenvalues merge into
a double eigenvalue, as has observed earlier for the pure NLS case (see Appendix C).
}
\label{fig_4_3_4}
\end{figure}

The seemingly mysterious feature (i), i.e. the ``delayed" NI, has a simple explanation.
The noise $\xi(x)$ in the initial condition (see \eqref{e_15})
consists of Fourier harmonics with random phases and random, but
similar, amplitudes. The part of each Fourier harmonic that overlaps
with the most unstable mode grows, while the rest of the harmonics
oscillates or grows at a lower rate. 
To clarify why this leads to an effective delay in the growth
of the Fourier harmonic, we will focus on
the case when the unstable mode is nonlocalized, as, e.g., in
 Fig.~\ref{fig_4_3_3}.
In this case, the explanation is most transparent, and it is also then
that the ``delayed" NI is most conspicuous. 
The overlap factor between a Fourier harmonic and the most unstable mode,
\be
{\rm OF}(k)=\left| \int_{-L/2}^{L/2} e^{ikx} u_{\rm mode} dx\right| \,\Big/\,
\sqrt{L \int_{-L/2}^{L/2} | u_{\rm mode}|^2 dx },
\label{e4_311}
\ee
is proportional to the spectrum of the unstable mode;
 see Fig.~\ref{fig_4_3_3}(a). A point to note is that
${\rm OF}(k)$ has a peak (circled in the figure), i.e. most of the content of the
unstable mode is in {\em one} Fourier wavenumber, $k_{\rm peak}$.
Also, we have verified that
\be
\left| {\rm OF}(k_{\rm peak}) \right| \sim 0.5.
\label{e4_312}
\ee
As will become clear shortly,
the significance of \eqref{e4_312} is that this number is substantially less than one.
The evolution of the corresponding Fourier harmonic is:
\bsube
\be
\F[u](k_{\rm peak},t) = \F[u](k_{\rm peak},0) \left( {\rm OF}(k_{\rm peak}) \,
                        e^{\lambda_{\rm most}t} + \sum_{j}
												   {\rm OF}(k_j) e^{\lambda_j t} \right),
\label{e4_313a}
\ee
where $\lambda_{\rm most}$ and $\lambda_j$ are the eigenvalues of the most unstable
mode and all other modes, respectively; they are shown in Fig.~\ref{fig_4_3_4}.
Over long time, the second term on the r.h.s.
of \eqref{e4_313a} is negligible compared to the first one, not only because 
Re$\lambda_{\rm most}>$Re$\lambda_j$ but also because of partial cancellation of
the summands due to Im$\lambda_j$ all being different. Therefore, asymptotically,
\be
\F[u](k_{\rm peak},t) \approx {\rm OF}(k_{\rm peak}) \,\F[u](k_{\rm peak},0)  
                        e^{\lambda_{\rm most}t} .
\label{e4_313b}
\ee
\label{e4_313}
\esube
Thus, the $k$-peak of the most unstable mode will become visible above
the noise floor when \ $ \left|{\rm OF}(k_{\rm peak}) \,e^{\lambda_{\rm most}t} \right| >1$,
i.e. for
\be
t_{\rm delay} > O(1)/{\rm Re} \lambda_{\rm most},
\label{e4_314}
\ee
where we have used \eqref{e4_312}. In other words, the weaker the NI,
the longer it takes the NI to become observable.

Feature (ii)  is immediately explained by Fig.~\ref{fig_4_3_4}(a),
which shows that the increase of $\max \,{\rm Re}\,\lambda$ is not monotonic with $C$.
This is most notably seen near $C=1.035$, where the NI growth rate drops by almost
an order of magnitude.

Feature (iii) is explained by noticing that below $C\approx 1.07$,
there are multiple eigenvalues with very similar Re$\,\lambda$. 
Their eigenmodes grow at very similar rates and interfere, thus causing
non-monotonic growth of the numerical error with time.

Feature (iv) is supported by Fig.~\ref{fig_4_3_5}(a). It shows the most
unstable mode at $C=1.051$, which is essentially nonlocalized and thus
looks qualitatively similar to the mode shown in Fig.~\ref{fig_4_3_3}(b).
This should be compared to the localized mode at $C=1.05$ for the pure NLS;
see Fig.~\ref{fig_5}. Even at $C=1.077$, where the two pairs of purely real
eigenvalues have almost merged and far exceed real parts of other eigenvalues, 
the most unstable mode is still not quite localized (Fig.~\ref{fig_4_3_5}(b)).

\begin{figure}[h]
\vspace{-0.6cm}
\hspace*{-0.5cm}
\mbox{ 
\begin{minipage}{5.1cm}
\rotatebox{0}{\resizebox{5.1cm}{6.5cm}{\includegraphics[0in,0.5in]
 [8in,10.5in]{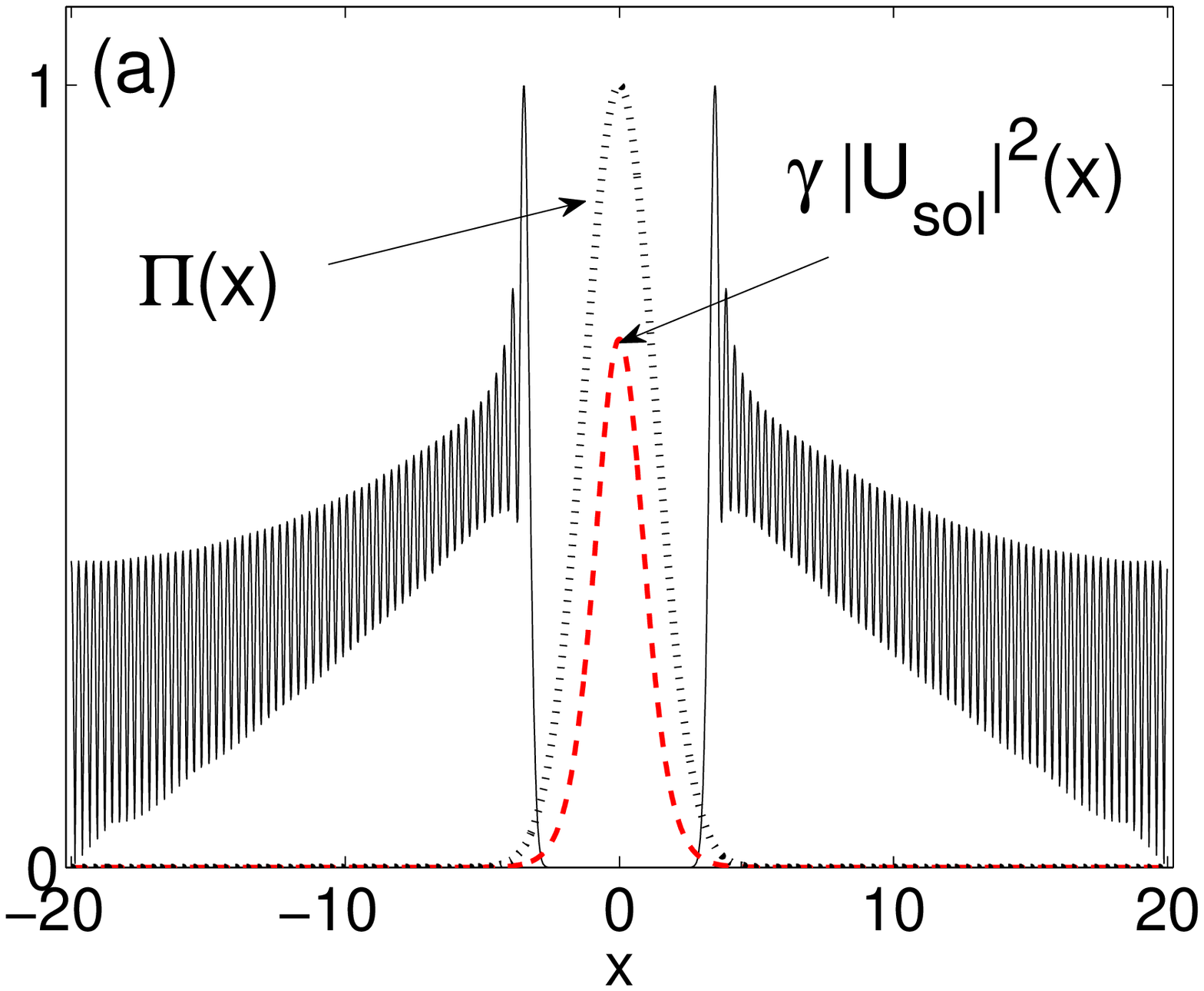}}}
\end{minipage}
\hspace{0.1cm}
\begin{minipage}{5.1cm}
\rotatebox{0}{\resizebox{5.1cm}{6.5cm}{\includegraphics[0in,0.5in]
 [8in,10.5in]{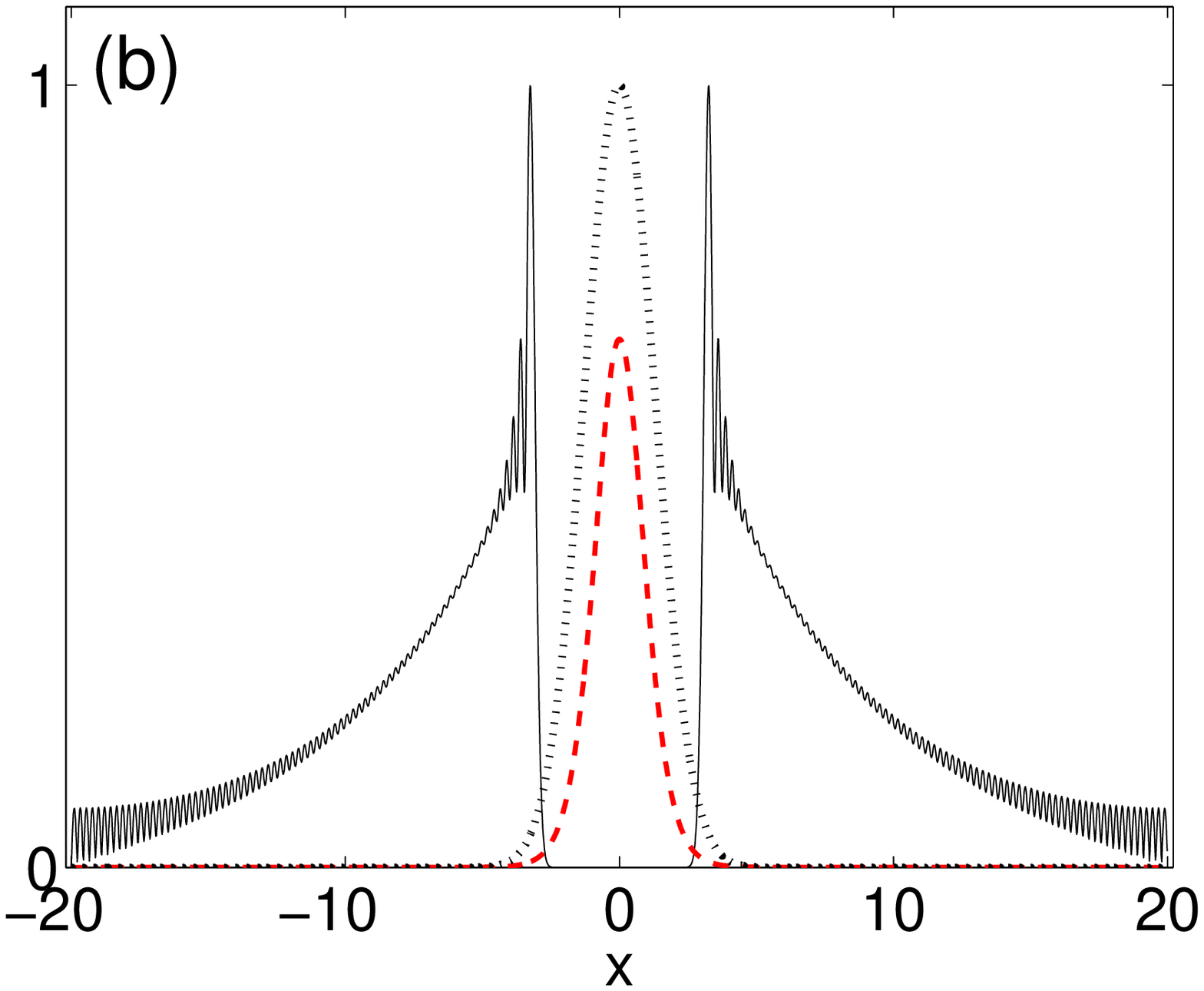}}}
\end{minipage}
\hspace{0.1cm}
\begin{minipage}{5.1cm}
\rotatebox{0}{\resizebox{5.1cm}{6.5cm}{\includegraphics[0in,0.5in]
 [8in,10.5in]{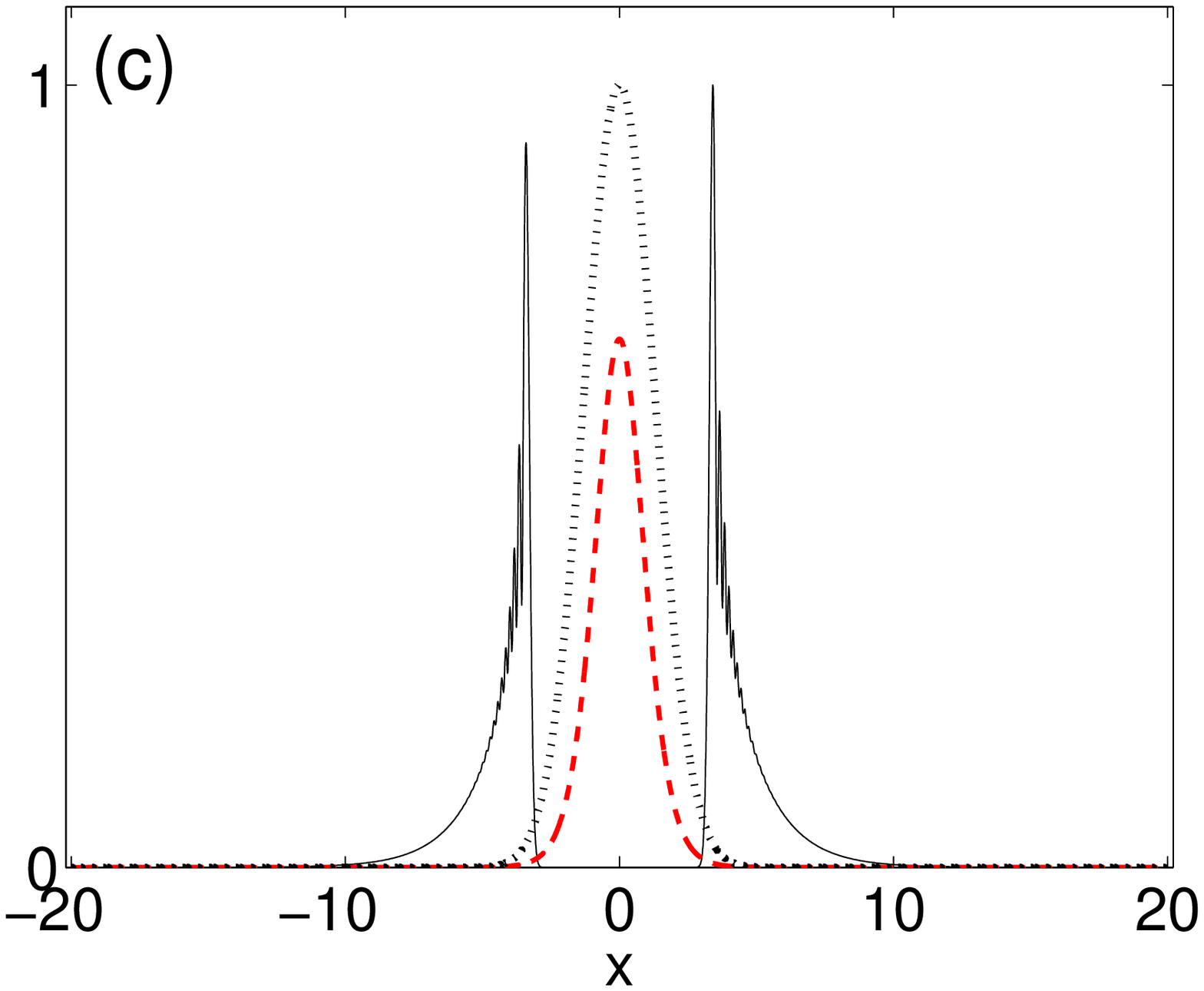}}}
\end{minipage}
 }
\vspace{-1.6cm}
\caption{Most unstable mode (solid) and external (dotted) and ``internal" (dashed)
 potentials. The vertical scale pertains to both potentials, while the amplitude
of the mode is scaled to one. \ (a) \ $N=2^{10}$, $C=1.051$; \ 
(b) \ $N=2^{10}$, $C=1.077$; \ (c) \ $N=2^{11}$, $C=1.051$.
}
\label{fig_4_3_5}
\end{figure}

Feature (v) is self-explanatory, as has been mentioned earlier.
In our numerics we have observed that in its ``sluggish" stage, 
where the most unstable mode is nonlocalized, NI gets, on average,
weaker as $L$ increases. However, this dependence is not monotonic.
As the most unstable mode becomes essentially localized, the NI's
growth rate, naturally, ceases to depend on $L$.

Feature (vi) is supported by Fig.~\ref{fig_4_3_5}(c), where we show
that the most unstable mode becomes localized, and the NI ceases to 
become ``sluggish", earlier on for smaller $\dx$ (or, equivalently, 
smaller $\epsilon$).

We will encounter ``sluggish" NI again
in Sec.~7. For now we leave the 
case of a standing soliton and turn to the moving soliton of the pure NLS \eqref{e_01}.


\section{Numerical instability of moving soliton}
\setcounter{equation}{0}

The study presented in this section has been motivated by the numerical results of
U. Ascher \cite{Ascher}, who, to our knowledge, was the first to report the development
of NI in the fd-SSM for a moving soliton. More specifically,
he considered a collision of two solitons and observed generation of a high-wavenumber
ripple for a certain relation between $\dt$ and $\dx$. However, since a collision is a 
short-term event, it could not cause NI (which was demonstrated to develop over a very long
time: $t\sim 1000$), and hence it is the stationary propagation of an individual soliton
that must have lead to the aforementioned NI. Note that due to the periodic boundary 
conditions, the soliton
remained in the computational domain at all times, which justifies the use of the word
`stationary' above.

When we learned of Ascher's
results, we have already completed the analysis of NI for a standing
soliton and hence initially thought that for a moving soliton, the instability should
develop similarly, because the shape of the moving and standing soliton is the
same and the only difference is a phase factor: see \eqref{e_12}.
However, Ascher's results suggested a qualitatively different scenario
of NI. In retrospect, this could be expected given the statement emphasized
in the Introduction: NI depends not only on the equation and numerical
scheme, but also on the particular solution being simulated. (Let us mention in passing
that the NI analysis for a moving plane wave of \eqref{e_01} also exhibits some
differences from that for a standing plane wave \cite{jaJOSAB}.)

We will begin by reporting our own numerical results which demonstrate
the same NI as observed by Ascher but for the parameters closer to those
used in Secs.~2 and 4. After that we will present an
approximate theory of this NI. It will begin as in Sec.~3 but will lead to a different
equation to the numerical error than \eqref{e_31}, which will, therefore, require a
{\em qualitatively} different analysis. To carry out that analysis, we will have to
approximate the soliton by a rectangular box. Such a crude approximation cannot
lead to quantitatively accurate predictions about the NI's threshold, spectral
location, and increment. However, it still qualitatively explains a number of 
observed features of this NI.

\subsection{Numerics for the moving soliton, and key observation about spectrum of numerical error}

\begin{figure}[h]
\vspace{-1.6cm}
\mbox{ 
\begin{minipage}{7cm}
\rotatebox{0}{\resizebox{7cm}{9cm}{\includegraphics[0in,0.5in]
 [8in,10.5in]{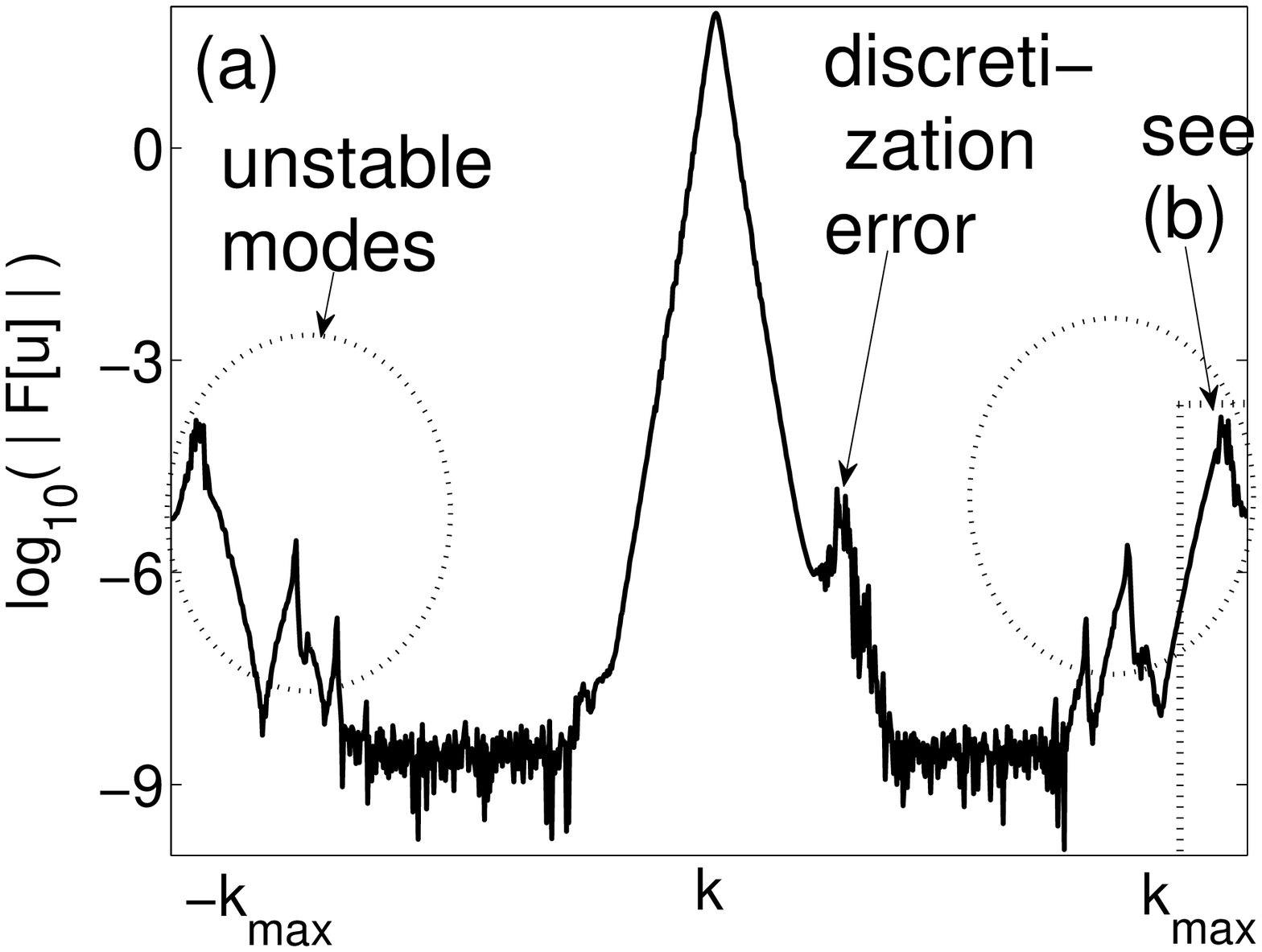}}}
\end{minipage}
\hspace{0.1cm}
\begin{minipage}{7cm}
\rotatebox{0}{\resizebox{7cm}{9cm}{\includegraphics[0in,0.5in]
 [8in,10.5in]{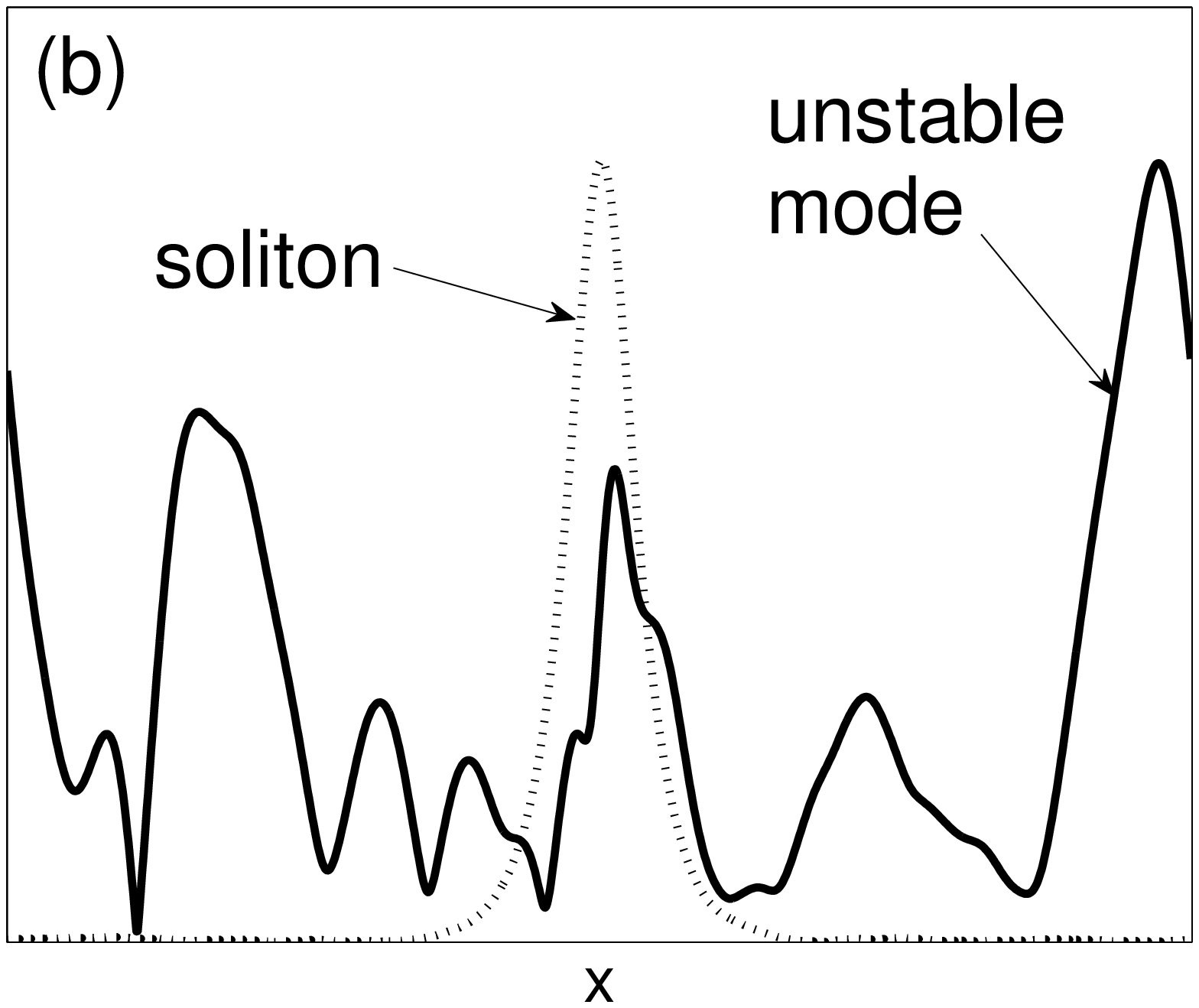}}}
\end{minipage}
 }
\vspace{-1.6cm}
\caption{(a) Logarithm of the 
Fourier spectrum of the numerical solution described in the text.
Unstable modes are circled at both ends of the $k$-domain.
The discretization error seen on the right of the soliton is due to the
approximation of the $u_{xx}$-term in \eqref{e_01} by the finite-difference
method \eqref{e_05}. 
\ (b) The unstable mode filtered out by a band-pass filter shown near the
right edge of (a). The soliton is shown by the dotted line. Both the soliton
and the unstable mode are normalized to have the same amplitude. 
}
\label{fig_5_1}
\end{figure}

The initial condition in our numerical simulations was taken similarly to \eqref{e_15}:
\be
u_0(x) = \sech(x)\,e^{i\Ksol x} + \xi,
\label{esec5_0}
\ee
where $\xi$ is Gaussian noise with amplitude of order $10^{-10}$ and $\Ksol$ is
related to the soliton's speed as $S=2\Ksol |\beta|$; see \eqref{e_12}.
The other parameters: $A=1$, $\beta=-1$, and $\gamma=2$, were as 
in the previous sections.
The Fourier spectrum of a typical numerical solution --- for $L=40$, $N=2^{10}$, 
$C=0.9$, and $t=1500$ --- is shown
in Fig.~\ref{fig_5_1} for $S=1.89$. 
(This is an approximation to $S=2$, used so that the exponential factor in 
\eqref{esec5_0} be exactly
periodic in the computational domain.) As the unstable modes continue 
to grow and become visible on the linear (versus logarithmic) scale 
on the background of the soliton, 
in the $x$-space they are observed as high-frequency ripple: see Fig.~7(b) in
\cite{Ascher}.

Figure \ref{fig_5_1} illustrates two main differences of the
NI for the moving and standing solitons. First, NI for the moving soliton
is observed even for $C\,(=(\dt/\dx)^2\,)$ that is 
less than the threshold value $C=1$  for the standing case
(see the text before \eqref{sec4_extra1}).
In fact, we have observed (a weak) NI of the moving soliton with the same
parameters even for $C=0.5$; we will comment on it later.
Second, the spectrum of each of the numerically unstable modes in Fig.~\ref{fig_5_1}(a)
(see also Fig.~\ref{fig_5_6}(b) below)
is considerably narrower than that in Fig.~\ref{fig_2}(a). We will explain 
in what follows that this leads to a different equation for the numerical error
than \eqref{e_31}.

Before deriving that equation, let us mention that our
derivation will be valid for $S=O(1)$. In the range $0 < S < O(1)$, which is
intermediate between the case of a standing soliton and the case of a moving soliton
considered below, the analysis must be more involved.
Indeed, such an analysis should be able to explain a transition between
those two cases. However, in the former case, the unstable modes are localized 
at the sides of the (standing) soliton (see Sec.~4), and such a situation
cannot be described even qualitatively using the box approximation for the pulse,
which we will (have to) use below for the moving soliton. Thus, the analysis of NI in
the intermediate range $0 < S < O(1)$ will {\em not} be attempted here
and remains an open problem.

On the other hand, we note that $\Ksol=S/(2|\beta|)$ is to be less than
approximately $1/\sqrt{|\beta|\dt}$ in order for the fd-SSM to yield an 
accurate solution of the NLS \eqref{e_01} \cite{jaJOSAB}.


\subsection{Modified equation for numerical error on background of moving soliton, and its analysis}

The numerical error satisfies Eq.~\eqref{e_20} for {\em any} background solution.
From Fig.~\ref{fig_5_1} one can see that the spectrum of unstable modes is
approximately symmetric relative to some value $k=O(1)$. It is, therefore, convenient 
to seek 
\be
\tu_n = e^{i\omsol t_n + i\Ksol (x - St_n)} \left(
    \tilde{p}_n(x) e^{i K_0 x} + \tilde{q}^*_n(x) e^{-iK_0 x} \right),
\label{esec5_01}
\ee
where $(\pm K_0 + \Ksol)$, with $K_0=O(k_{\max})\gg 1$,
 are the {\em approximate} locations of the unstable peaks
and $\tilde{p}_n(x)$, $\tilde{q}_n(x)$ 
may vary with $x$ on scale $O(1)$. Two notes are in order
about the latter assumption. First, it follows solely from numerical results
(Fig.~\ref{fig_5_1} and \ref{fig_5_6}(b)), 
where one sees that the unstable peaks have width of order one
in the Fourier space, which implies the above statement about 
$\tilde{p}_n(x)$ and $\tilde{q}_n(x)$. 
Second, the locations of the unstable peaks may differ by an amount 
of order one from $(\pm K_0 + \Ksol)$; our analysis will yield  approximate
expressions both for $K_0$ and for those modified locations.

When \eqref{esec5_01} is substituted into \eqref{e_20}, 
the next step is to expand the phase $P(k)$. The first step of that
expansion is given by the first line of \eqref{e_21}, but the subsequent
expansion is different. Indeed, as discussed in the previous paragraph,
the values of $k$ are located within a ``distance" of order one of $(\pm K_0 + \Ksol)$,
and therefore also
of $\pm K_0$ (recall that we have assumed that $S$ and hence $\Ksol$
are $O(1)$). 
Therefore, the expansion is:
\bea
-P(k) & = & \pi - \frac1{|\beta|r\sin^2(k\dx/2)} + O\left(\frac1{r^3}\right) 
 \nonumber \\
  & = & \pi - \frac1{|\beta|r\sin^2(K_0\dx/2)} + O\left(\frac1{r^2}\right)\,,
\label{esec5_02}
\eea
where we have also used $\dx = O(1/r)$, as in Sec.~3. Then, using Eqs.~\eqref{e_20},
\eqref{esec5_01}, \eqref{esec5_02}, a transformation \ 
$\{\tilde{p}_n,\,\tilde{q}_n\} = (-1)^n \{ p_n,\, q_n\}$ \ (as in \eqref{e_24}),
the reasoning outlined between Eqs.~\eqref{e_29} and \eqref{e_31}, and, finally, the
change of variables $(x,t) \longrightarrow (z=x-St,t)$,   we obtain:
\bsube
\be
p_t-Sp_{z} = i\mu p + i\gamma \Usol^2(z) \, (2p+q),
\label{esec5_03a}
\ee
\be
q_t-Sq_{z} = -i\mu q - i\gamma \Usol^2(z) \, (p+2q),
\label{esec5_03b}
\ee
\label{esec5_03}
\esube
where $\{p(x,t),\,q(x,t)\}$ are time-continuous counterparts of 
$\{p_n(x),\,q_n(x)\}$ and 
\be
\mu = \frac1{C|\beta|\sin^2(K_0\dx/2)} - A^2 + |\beta|\Ksol^2.
\label{esec5_04}
\ee

The boundary conditions that go with Eqs.~\eqref{esec5_03} are 
periodic, as in Sec.~3. To be more precise, in light of \eqref{esec5_01}
it is \ $p(x,t)\exp[i(\Ksol+K_0)x]$ and \ $q(x,t)\exp[i(\Ksol-K_0)x]$
that are to be spatially periodic. However, in all our numerical
simulations we have used the initial condition where $\Ksol$ was on
the spectral grid, whence $\exp[i\Ksol x]$ is periodic. As for the
yet unknown $K_0$, when later on we determine a range for its values,
we will select from that range only the values on the spectral grid;
hence $\exp[\pm i K_0 x]$ will be periodic. Thus, without loss of generality,
we require
\be
p(-L/2,t)=p(L/2,t), \qquad q(-L/2,t)=q(L/2,t).
\label{esec5_05}
\ee
Since these conditions hold at all times $t$, the first argument
of $p$ and $q$ in \eqref{esec5_05} may equally be interpreted as either $x$ or $z$.

Before we use Eqs.~\eqref{esec5_03}--\eqref{esec5_05} 
to study the NI of a moving soliton, let us note that they,
along with \eqref{esec5_01}, 
are the counterparts of the modified linearized NLS \eqref{e_32}--\eqref{e_34}.
These two sets of equations are different from one another in two
aspects, in addition to the obvious difference of having $S\neq 0$ for the
former set. First, Eqs.~\eqref{esec5_03} unlike Eq.~\eqref{e_32} do not have
a second-order spatial derivative. This is a direct consequence of the
numerically observed width of the unstable peaks being of order one for
the case of moving soliton (Fig.~\ref{fig_5_1}), whereas such peaks are 
considerably wider in the Fourier space
 for the standing soliton (Fig.~\ref{fig_2}(a)). This
was discussed before Eq.~\eqref{esec5_02} and near Eqs.~\eqref{e_23},
respectively. Second, while coefficient $\delta$ in \eqref{e_32} is fixed
(for given values of simulated parameters), coefficient $\mu$ in \eqref{esec5_03}
depends on a yet to be determined value $K_0$, related to the spectral location
of unstable peaks.

Let us now explain why this latter circumstance renders the finding 
of unstable modes for
Eqs.~\eqref{esec5_03}, \eqref{esec5_05} more difficult than 
for \eqref{e_32}, \eqref{e_34}.
Seeking the solution of \eqref{esec5_03} in the standard form
\ $(p(z,t),\,q(z,t))^T= \vec{\rho}(z) e^{\lambda t}$, one obtains
the following counterpart of  \eqref{e_36}:
\be
\left( iS\sigma_3\partial_z - \mu - \gamma \Usol^2(z) 
       \left( \ba{cc} 2 & 1 \\ 1 & 2 \ea \right) \, \right) \vec{\rho} = 
       i\lambda\sigma_3 \vec{\rho},
 \label{esec5_06}
 \ee
with $\vec{\rho}$ satisfying periodic boundary conditions following
from \eqref{esec5_05}.
Now, when in Sec.~4 we solved the eigenvalue problem \eqref{e_36},
we selected a value of
$C=(\dt/dx)^2$ and for it found all the unstable modes and their eigenvalues.
However, for \eqref{esec5_06} the task is more complicated because
$\mu$ depends not only on $C$ but also on $K_0$, which not yet known. Then,
for a given $C$, one would have to
scan through values of $K_0$ to determine those special values of $K_0$
where one has an unstable mode. Not only would this make the numerical
solution in this case considerably more time-consuming, but it would also 
not provide any insight into {\em why} the instability occurs
only for some special values of $K_0$ but not for all $K_0$. Such an insight
could only come from an analytical solution of \eqref{esec5_06}, but
we have been unable to find it for that system of differential
equations with a $z$-dependent coefficient $\Usol^2(z)$ given by \eqref{e_12b}.

\begin{figure}[h]
\vspace{-1.6cm}
\centerline{ 
\begin{minipage}{7cm}
\rotatebox{0}{\resizebox{7cm}{9cm}{\includegraphics[0in,0.5in]
 [8in,10.5in]{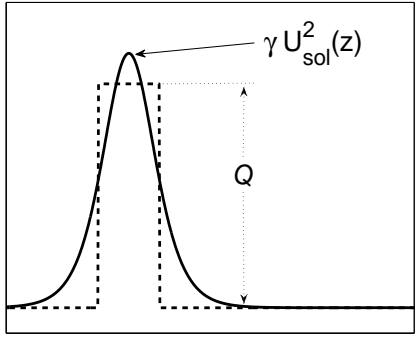}}}
\end{minipage}
 }
\vspace{-1.6cm}
\caption{Schematics of a box profile approximating $\gamma \Usol^2$ in 
\eqref{esec5_03}.
}
\label{fig_5_2}
\end{figure}

Therefore, we have resorted to a widely used approximation and
replaced $\gamma\Usol^2(z)$ with a box profile of width $\ell$ and height $Q$, 
as illustrated in Fig.~\ref{fig_5_2}. While such a crude approximation cannot
be expected to yield a quantitatively accurate description of NI, it still 
allows us to understand the nature of unstable modes as well as the 
dependence of NI's features on such parameters as $\Ksol$ (or, equivalently, $S$),
the length of the spatial domain $L$, and the mesh size $\dx$.

Without loss generality the left-hand edge of the box can be put at $z=0$.
Then the solutions of \eqref{esec5_06} with $\Usol^2$ replaced by the box
profile are given by the following expressions inside and outside the box:
\bsube
\be
\ba{l} 
\displaystyle 
0\le z \le \ell:  \vspace{0.2cm} \\
\displaystyle   \vec{\rho} = 
 a^-_{\rm in} \left( \ba{c} \mu+\eta+2Q \\ -Q \ea \right) e^{i\kappa^-_{\rm in}\,z} + 
 a^+_{\rm in} \left( \ba{c} -Q \\ \mu+\eta+2Q \ea \right) e^{i\kappa^+_{\rm in}\,z},
 \ea
\label{esec5_07a}  
\ee
\be
\hspace*{-3cm}
\ba{l} 
\displaystyle 
\ell \le z \le L:  \vspace{0.2cm} \\
 \displaystyle  \vec{\rho} = 
 a^-_{\rm out} \left( \ba{c} 1 \\ 0 \ea \right) e^{i\kappa^-_{\rm out}\,z} + 
 a^+_{\rm out} \left( \ba{c} 0 \\ 1 \ea \right) e^{i\kappa^+_{\rm out}\,z},
 \ea
\label{esec5_07b}  
\ee
\label{esec5_07}
\esube
where
\be
\eta = \sqrt{(\mu+2Q)^2-Q^2}, \qquad \kappa^{\pm}_{\rm in}=(-i\lambda\pm\eta)/S,
\qquad \kappa^{\pm}_{\rm out}=(-i\lambda\pm\mu)/S.
\label{esec5_08}
\ee
The constants $a^{\pm}_{\rm in,\, out}$ are found using the continuity of this solution
at $z=\ell-0$ and $z=\ell+0$ and at $z=L$ and $z=0$, 
with the latter condition being equivalent to
the periodic boundary condition. The existence of nontrivial solutions of the resulting
linear system determines the eigenvalue $\lambda$:
\bsube
\be
e^{\lambda L/S} = R \pm \sqrt{R^2-1},
\label{esec5_09a}
\ee
\be
R = \cos\Phi_+  +  
\frac{Q^2}{(\mu+\eta+2Q)^2-Q^2} 
     \left( \cos\Phi_+ - \cos\Phi_- \right).
     \vspace{0.2cm}
\label{esec5_09b}
\ee
\be
\Phi_{\pm} = (\mu(L-\ell) \pm \eta\ell)/S
\label{esec5_09c}
\ee
\label{esec5_09}
\esube
Eigenvalues with \ Re$\lambda>0$ exist for 
\be
|R|>1.
\label{esec5_10}
\ee

Thus, \eqref{esec5_10} along with (\ref{esec5_09}b,c) is the condition of NI
of a moving soliton.


\subsection{Features of numerical instability of moving soliton and their explanation}

As we announced in the Introduction, the focus of our study
is to understand what modes cause NI and, if possible,
estimate their growth rate and a threshold for their appearance.  
Numerical results of Sec.~6.1 presented evidence, and the analysis of Sec.~6.2 confirmed,
that these unstable modes are delocalized, plane-wave-like packets.
The NI is caused by a pair of these waves, denoted as $p$ and $q$ in Sec.~6.2,
repeatedly (due to the periodic boundary conditions) 
passing through the soliton and interacting with each other.
This situation should be contrasted with the unstable modes of standing solitons
considered in Secs.~2--5.2, which are localized and ``pinned" at the ``tails" of its
host pulse. 

Below we will show how to use Eqs.~\eqref{esec5_09}, \eqref{esec5_10} to explain 
qualitatively, and sometimes even quantitatively, a number of 
features (including the growth rate) of the NI of a moving soliton,
observed in numerical simulations.

\begin{itemize}
\item[(i)]
The  height and spectral width of unstable peaks decrease as their wavenumber $|k|$ decreases;
\item[(ii)]
The wavenumbers of the unstable peaks vary in inverse proportion to $\dx$; 
\item[(iii)]
The wavenumbers of the peaks are {\em not} symmetric about $\Ksol$,
as one could have concluded from \eqref{esec5_01};
\item[(iv)]
The instability growth rate, Re$\lambda$, varies in inverse proportion to the length
$L$ of the computational domain; 
\item[(v)]
The instability decreases as $C$ is decreased or $\Ksol$ is increased.
\end{itemize}
Feature (i) is illustrated by Fig.~\ref{fig_5_2}(a), while details on
the other features will be given as we proceed.

\begin{figure}[h]
\vspace{-1.6cm}
\mbox{ 
\begin{minipage}{7cm}
\rotatebox{0}{\resizebox{7cm}{9cm}{\includegraphics[0in,0.5in]
 [8in,10.5in]{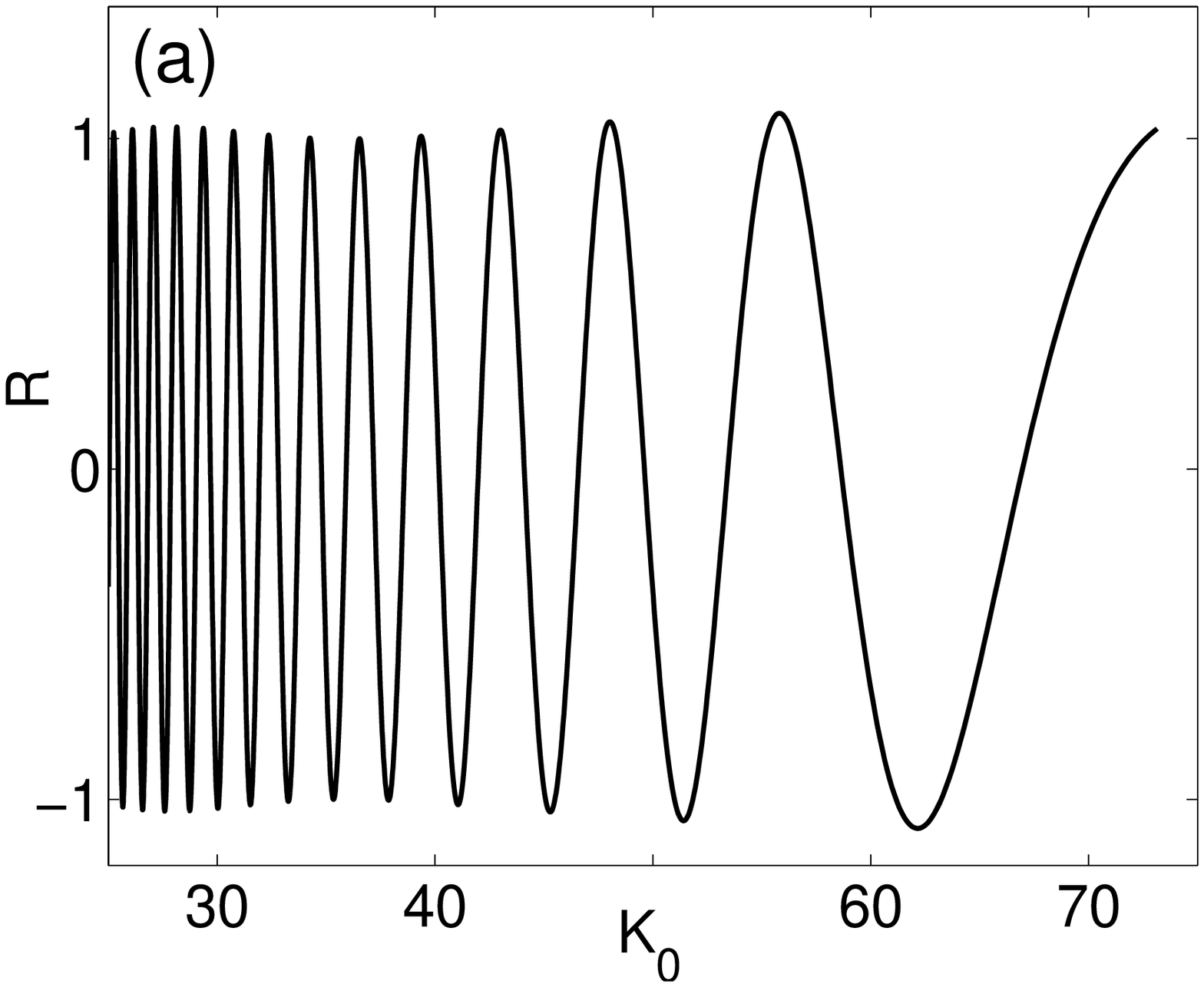}}}
\end{minipage}
\hspace{0.1cm}
\begin{minipage}{7cm}
\rotatebox{0}{\resizebox{7cm}{9cm}{\includegraphics[0in,0.5in]
 [8in,10.5in]{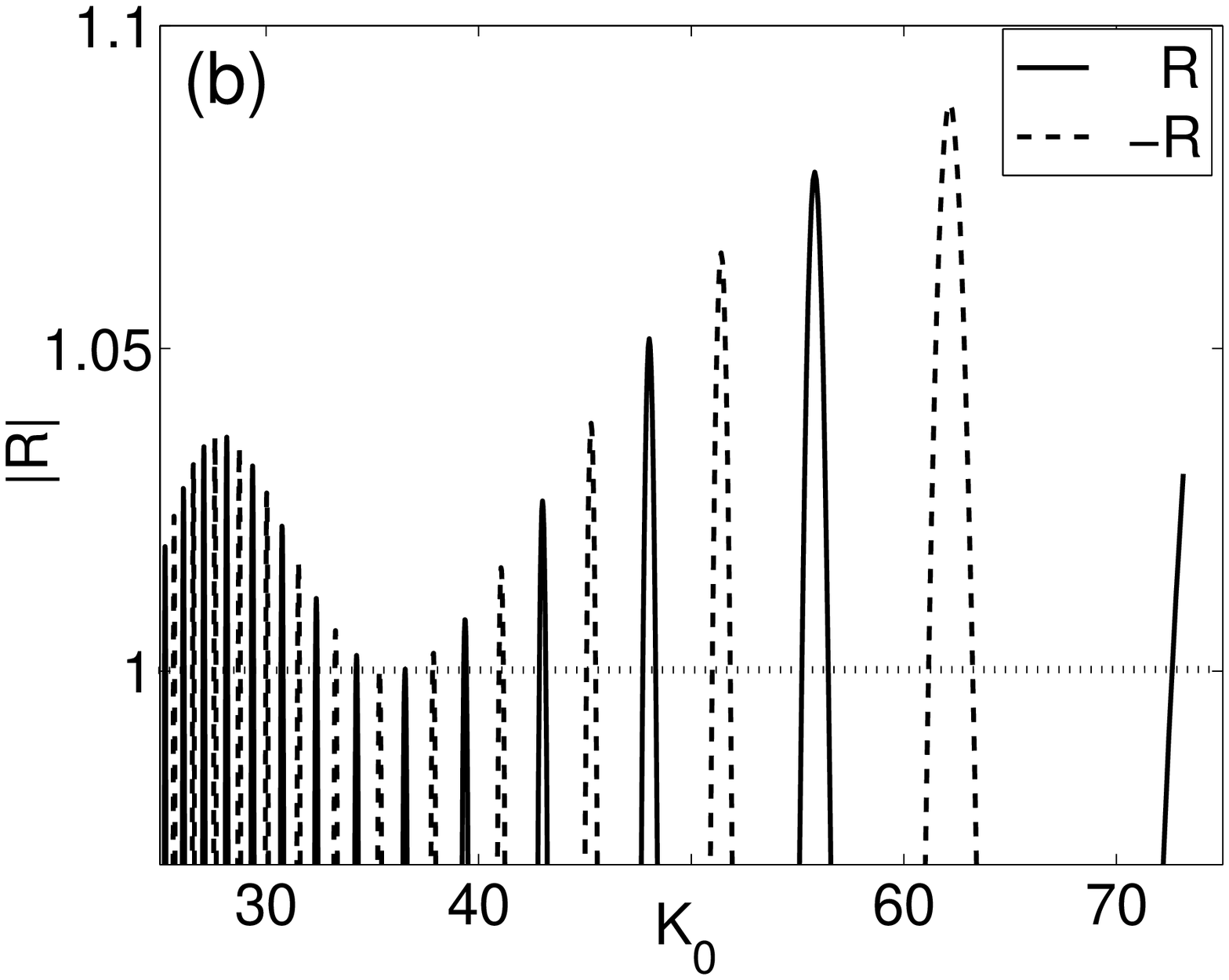}}}
\end{minipage}
 }
\vspace{-1.6cm}
\caption{(a) $R(K_0)$ \ and (b) a detailed view of (a) near $\pm R(K_0)=1$. 
}
\label{fig_5_3}
\vspace{1cm}
\end{figure}

A convenient way to analyze the NI condition is to consider a parametric 
representation $R=R(\mu)$ and $K_0=K_0(\mu)$, where for the latter one  
inverts \eqref{esec5_04}:
 \be
 K_0 = \frac2{\dx} {\rm arcsin} 
 \sqrt{\frac1{C|\beta|(\mu - |\beta|\Ksol^2+A^2)} }.
 \label{esec5_11}
 \ee
The resulting plot of $|R|$ versus $K_0$ is shown in Fig.~\ref{fig_5_3}(a)
for the same parameters as used for Fig.~\ref{fig_5_1}.
For other parameters,
the plot $R=R(K_0)$ looks qualitatively similar. 
We have also used the values
\be
\ell=1.76, \qquad Q=4/\ell,
\label{esec5_12}
\ee
where the first is the full width at half maximum of the $\sech^2$ profile
and the second follows from 
$Q\ell = \int_{-\infty}^{\infty} \gamma\Usol^2(z) dz = 4$ for $A=|\beta|=1$. 
In Fig.~\ref{fig_5_3}(b)
we show a detailed view of (a) that demonstrates that the NI condition 
\eqref{esec5_10} is satisfied only in narrow bands of wavenumbers $k$. 
As we noted before \eqref{esec5_05}, values of $K_0$ must be on the spectral
grid, and hence the increasingly narrow bands where $|R|>1$ occurring towards the
decreasing values of $K_0$ may simply miss points on the spectral grid. 
This, along with the fact that the ``tips" of $|R|$ that exceed 1 
become smaller as $K_0$ decreases explains feature (i) 
stated above.

Feature (ii) is illustrated by Table \ref{table_2}.
The simulation parameters are the same as those used for Fig.~\ref{fig_5_1},
except that we have varied $C$ and also compared the cases of $N=2^{10}$
and $N=2^{11}$ grid points, so that the corresponding $\dx$ differ
by a factor of 2. The locations of the respective peaks of unstable modes 
is seen to differ by an approximately reciprocal factor. 
An explanation for this follows directly from 
\eqref{esec5_11}.

\begin{table}
\bc 
\begin{tabular}{|c|rrr|rrr|}  \hline 
 $C$   &  \multicolumn{3}{c|}{$k^{\pm}_{\rm peak},\;N=2^{10}$} 
  &  \multicolumn{3}{c|}{$k^{\pm}_{\rm peak},\;N=2^{11}$} 
  \\ \hline 
 0.9  &  \quad $77.0$  \quad &  \quad $62.5$ \quad  & \quad  $56.2$ \quad 
      &  \quad $154.4$ \quad  & \quad  $125.0$ \quad  & \quad  $112.2$  \quad  \\ 
      & $-76.2$  & $-61.8$  & $-55.6$  & $-153.3$ & $-124.4$ &  $-111.7$  \\ \hline
 1.0  &  $63.9$  &  $56.4$  &  $51.5$  &  $128.0$  &  $112.6$  &  $102.9$   \\ 
      & $-63.1$  & $-55.7$  & $-50.9$  & $-127.2$ & $-112.0$ &  $-102.4$  \\ \hline 
 1.25 &  $71.8$  &  $58.7$  &  $52.0$  &  $144.5$  &  $117.2$  &  $103.8$   \\ 
      &  $-70.7$  & $-57.8$  & $-51.2$  & $-143.4$  & $-116.2$ &  $-103.0$  \\ \hline  
 1.50 &  $74.0$  &  $57.6$  &  $50.5$  &  $149.5$  &  $115.3$  &  $100.7$   \\ 
      &  $-72.8$  & $-56.6$  & $-49.5$  & $-147.7$  & $-114.2$ &  $-99.7$  \\ \hline 
\end{tabular}
\ec
\caption{Wavenumbers of the three most unstable peaks; $k^+>0,\,k^-<0$. 
For $C\ge 1$ the
outer peaks (those with larger  $|k_{\rm peak}^{\pm}|$) 
contain several grid points; only the
wavenumber of the maximum $|{\mathcal F}[u](k)|$ is listed in those cases.
 }
\label{table_2}
\end{table}

\begin{figure}[h]
\vspace{-1.6cm}
\mbox{ 
\begin{minipage}{7cm}
\rotatebox{0}{\resizebox{7cm}{9cm}{\includegraphics[0in,0.5in]
 [8in,10.5in]{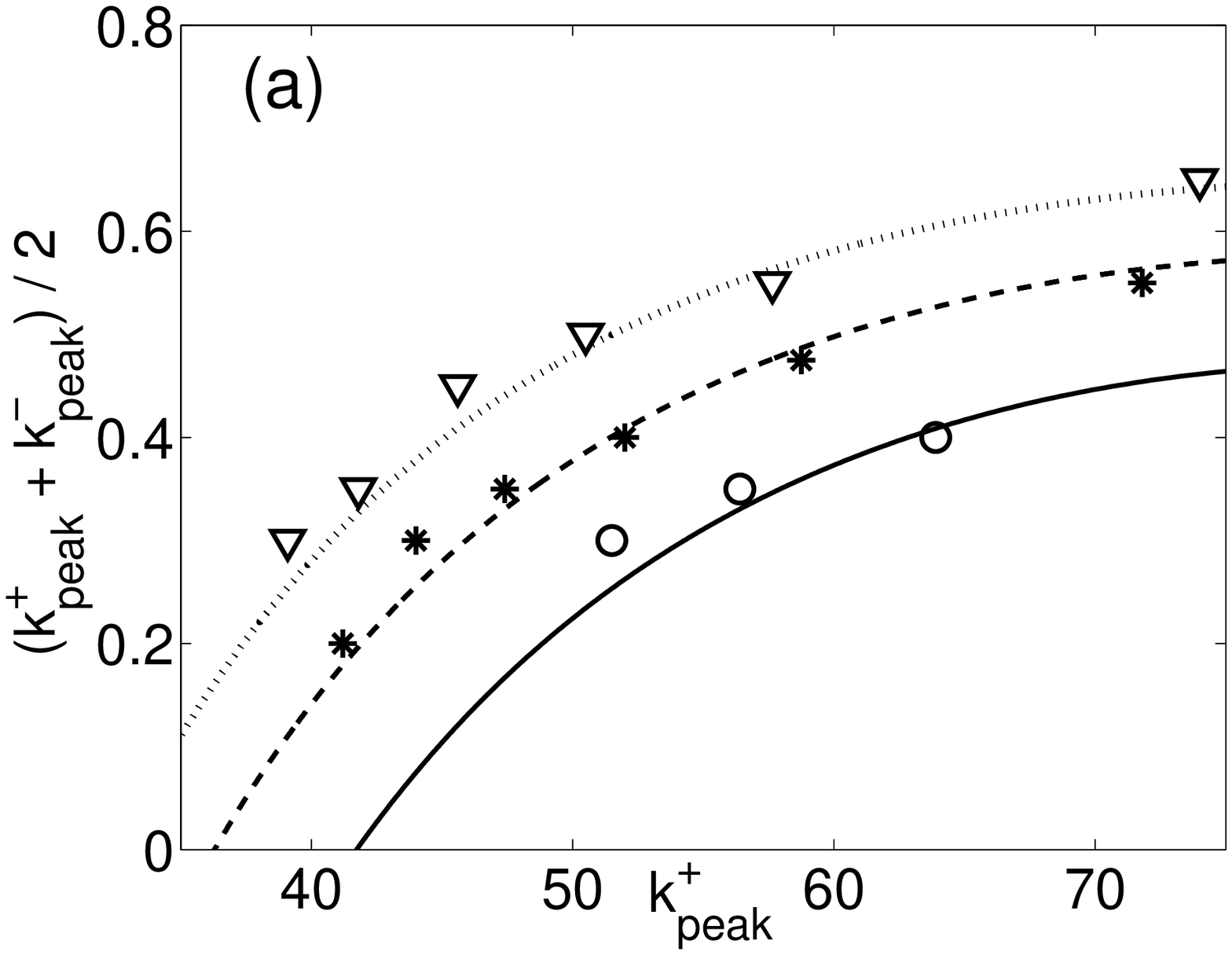}}}
\end{minipage}
\hspace{0.1cm}
\begin{minipage}{7cm}
\rotatebox{0}{\resizebox{7cm}{9cm}{\includegraphics[0in,0.5in]
 [8in,10.5in]{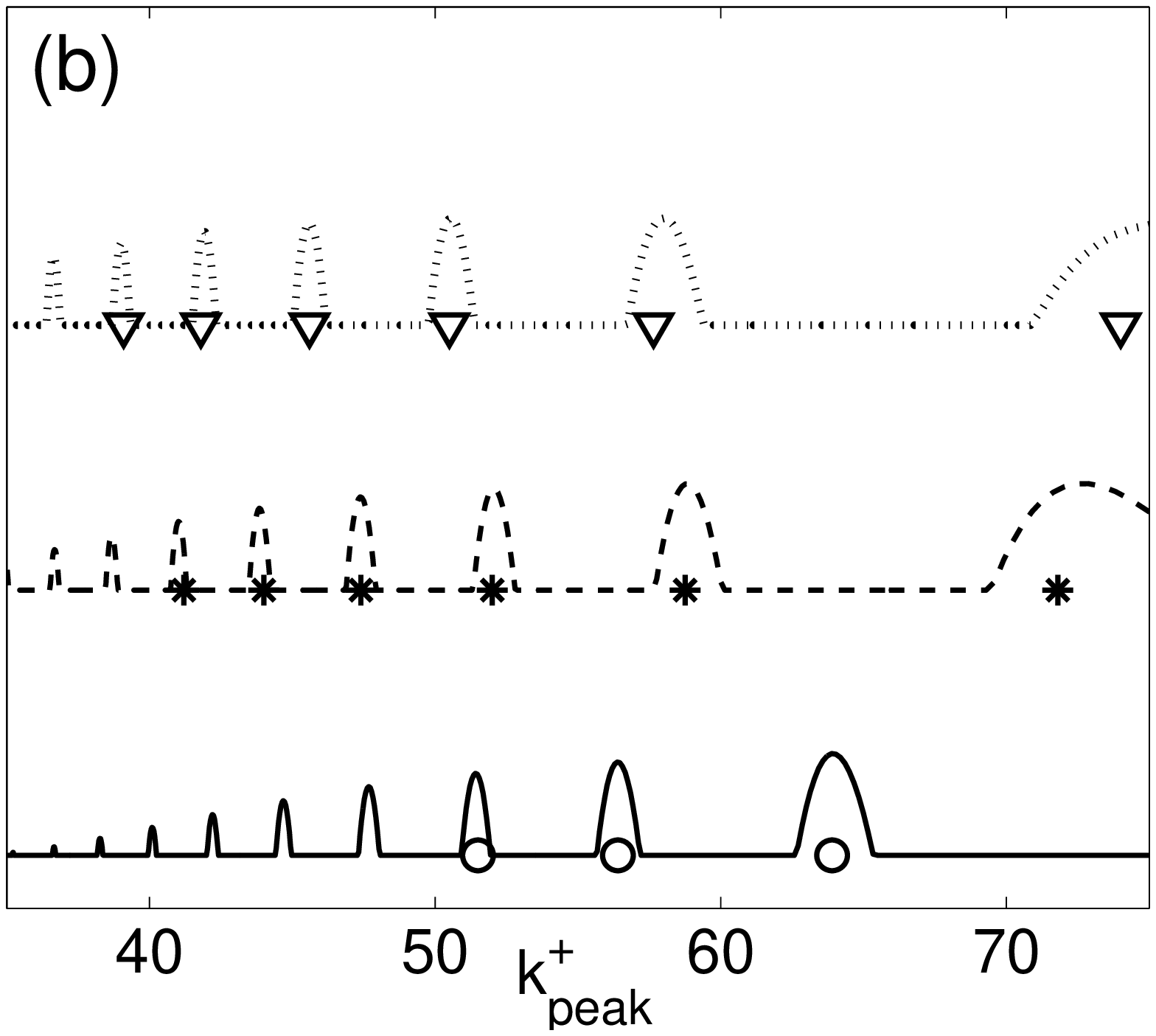}}}
\end{minipage}
 }
\vspace{-1.6cm}
\caption{(a) The l.h.s. of \eqref{esec5_13} versus the wavenumber of unstable
peaks with $k>0$. Solid, dashed, and dotted lines are the analytical expressions
obtained from \eqref{esec5_14} for $C=1,\,1.25,\,1.5$ and the parameters 
stated in the text. Circles, stars, and triangles are the respective numerical
values. \ (b) Line and symbol styles pertain to the same cases as in (a).
Lines are obtained from the analytical expressions for \ 
$\max(|R|,\,1)$, so that the ``bumps" indicate locations of bands of 
unstable modes. Symbols indicate the locations of numerically obtained 
unstable peaks. The data for different values of $C$ are vertically shifted 
for clarity.
}
\label{fig_5_4}
\vspace{1cm}
\end{figure}

Results of Table \ref{table_2} also illustrate feature (iii):
the positive and their respective negative peaks are not symmetric about $\Ksol$.
That is, 
\be
(k^+_{\rm peak}+k^-_{\rm peak})/2 \neq \Ksol.
\label{esec5_13}
\ee
The l.h.s. of this formula is plotted in Fig.~\ref{fig_5_4}(a). The analytical
estimate for this quantity is obtained as follows. Since at any given
time the soliton
occupies only a small part of the computational domain, the unstable mode
is described for the most part by its ``outside of the box" expression \eqref{esec5_07b}.
Along with the expression \eqref{esec5_08} for $\kappa^{\pm}_{\rm out}$ and
the fact that for the unstable modes $\lambda$ is purely real (see \eqref{esec5_09a}
and \eqref{esec5_10}) this implies that $\{p,\,q^*\} \propto \exp[-i(\mu/S)z]$. 
Then from \eqref{esec5_01} it follows that
\be
k^{\pm}_{\rm peak} = \pm K_0 + \Ksol -\mu/S,
\label{esec5_14}
\ee
which confirms \eqref{esec5_13}.

Figure \ref{fig_5_4} demonstrates that the locations of unstable peaks are 
quite accurately predicted by our approximate analysis. 
However, this analysis considerably (by a factor of order two for $C\approx 1$) 
overestimates the instability growth rate. Moreover, as $C$ decreases,
the discrepancy between the analytical and numerically observed growth rates
increases. 

Yet, our analysis easily explains feature (iv), whereby the
instability growth rate scales in inverse proportion to the length of the
computational domain (assuming that it far exceeds the width of the soliton). 
To that end, we will first explain why one typically has
\be
|R|-1 \ll 1 \qquad \mbox{where $|R|>1$},
\label{esec5_15}
\ee
as seen in Fig.~\ref{fig_5_3}. 
For $C$, $A$, $\Ksol$ all of order one, $\mu$ is also of order one; see
\eqref{esec5_04}. (For the specific values $A=\Ksol=|\beta|=1$ used here,
$\mu\ge 1/C$.) Then, even if we conservatively assume $\mu>0$, then from
\eqref{esec5_08} one has $\eta/Q > \sqrt{3}$, and then 
\be
(\mu+\eta+2Q)^2/Q^2 > (\sqrt{3}+2)^2 \approx 14.
\label{esec5_16}
\ee
We stress that this is a conservatively low estimate; in our simulations
the respective values were higher than about 22. 
Relation \eqref{esec5_16} implies that the second term in 
\eqref{esec5_09b} is small. From this one concludes that the ``bumps" of $|R|$
occur where $\Phi^+\approx \pi n$ for some integer $n$ and that
\bsube
\be
|R|-1 \le 2/\left( \big( (\mu+\eta)/Q + 2\big)^2 - 1 \right),
\label{esec5_17a}
\ee
which in view of \eqref{esec5_16} and the note below it we regard as 
as small number. 
Combining this with \eqref{esec5_09a} one obtains
\be
\lambda \approx \sqrt{2}(S/L)\sqrt{|R|-1},
\label{esec5_17b}
\ee
\label{esec5_17}
\esube
which provides the reason behind feature (iv).

\begin{figure}[h]
\vspace{-1.6cm}
\mbox{ 
\begin{minipage}{7cm}
\rotatebox{0}{\resizebox{7cm}{9cm}{\includegraphics[0in,0.5in]
 [8in,10.5in]{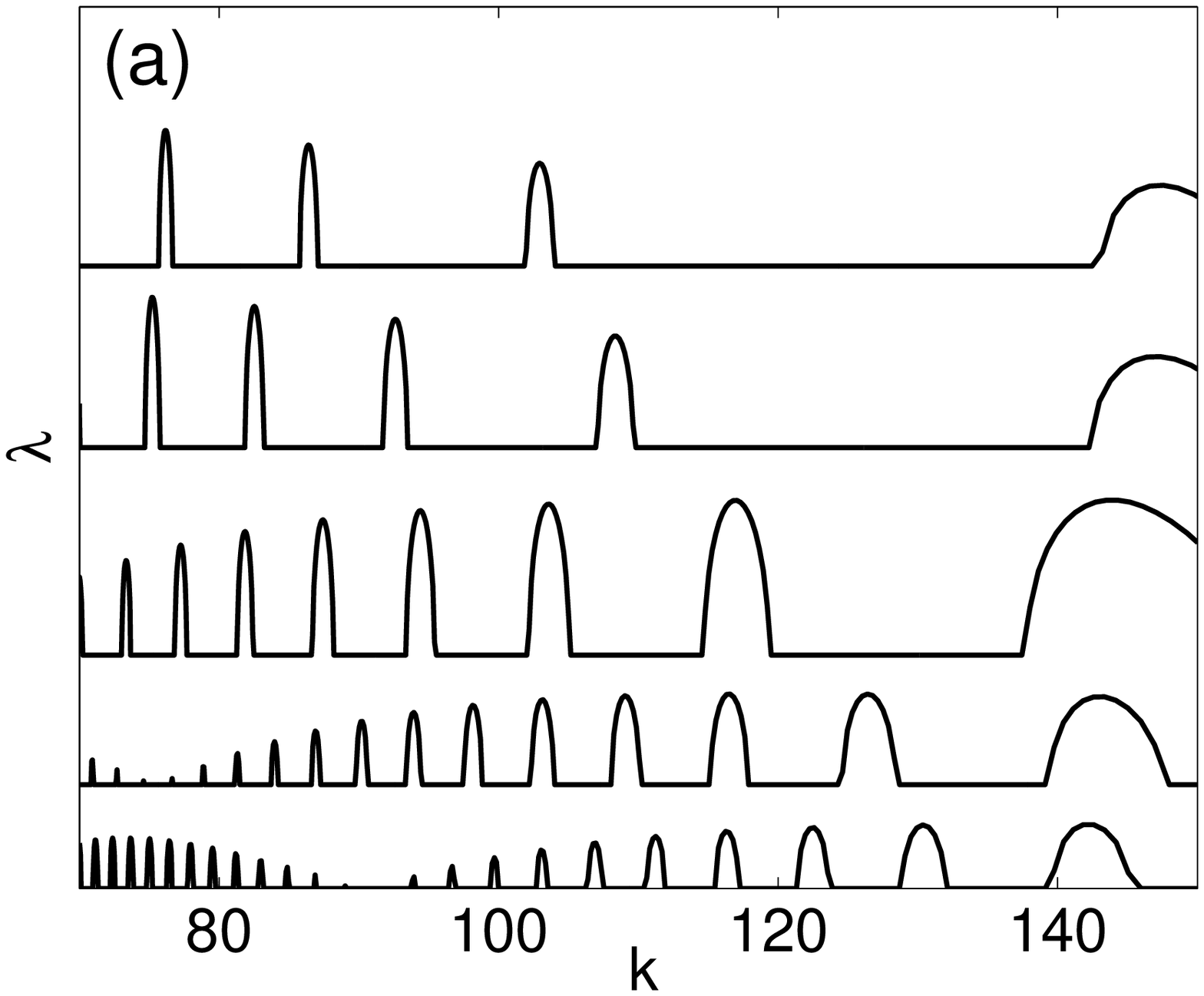}}}
\end{minipage}
\hspace{0.1cm}
\begin{minipage}{7cm}
\rotatebox{0}{\resizebox{7cm}{9cm}{\includegraphics[0in,0.5in]
 [8in,10.5in]{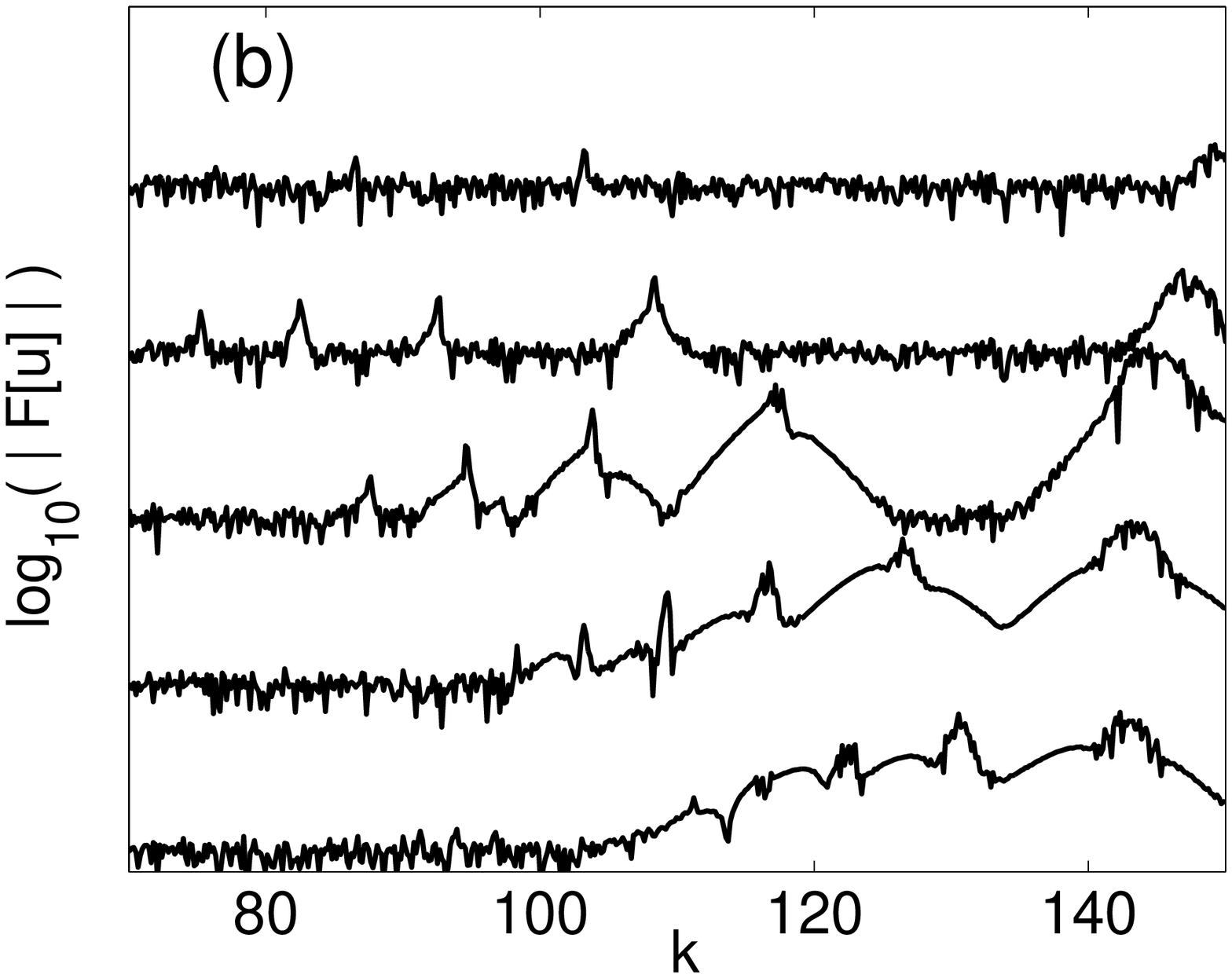}}}
\end{minipage}
 }
\vspace{-1.6cm}
\caption{(a) NI growth rate, computed from \eqref{esec5_17}, as
a function of wavenumber. The parameters are as previously described in the text,
except that $C=1.25$, $\Ksol$ is varied as stated below, and $N=2^{11}$.
(This larger $N$, leading to a smaller $\dx$, is needed to keep the discretization 
error due to
the finite-difference approximation \eqref{e_05} sufficiently small
for the larger values of $\Ksol$.) The curves, from
bottom to top, correspond to $\Ksol=0.31,\,0.47,\,0.94,\,1.57,\,2.04$.
Note that only the higher-$k$ part of the spectrum is shown.
 \ (b) The results of numerical simulations for the same respective
 parameters as in (a).
}
\label{fig_5_5}
\vspace{1cm}
\end{figure}

Formulas \eqref{esec5_17} and \eqref{esec5_04} also explain why
increasing $\Ksol$ (and hence $S=2|\beta|\Ksol$) {\em eventually}
suppresses the NI, which was stated as part of feature (v). 
Below we will present our argument as a crude estimate but will confirm
it with analytical expressions following from our analysis above and
also by results of numerical simulations. For the purpose
of this estimate we will assume that the first two terms on the
r.h.s. of \eqref{esec5_04} approximately cancel each other, and
then $\mu\sim \Ksol^2$. With the same accuracy, from \eqref{esec5_08}
we have $\eta\sim \Ksol^2+2Q$, and then from \eqref{esec5_17} we find
\be
\max\,\lambda \propto \Ksol / ( \Ksol^2/Q + 2 ).
\label{esec5_18}
\ee
This shows that as $\Ksol$ increases, 
the instability growth rate eventually vanishes, although it does 
grow initially as $\Ksol$ increases from zero. 
These conclusions are qualitatively confirmed by Fig.~\ref{fig_5_5}.
As an aside, let us note that the broad ``pedestals" of the unstable peaks
for $\Ksol=0.3$ and $0.5$ seen in Fig.~\ref{fig_5_5}(b)
are reminiscent of the spectrally broad
unstable mode of a standing soliton in Fig.~\ref{fig_2}(a).
This agrees with our remark, made before \eqref{esec5_01}, that
for sufficiently small $\Ksol$ (or $S$) there should be a regime where 
the NI of a moving soliton turns into that of a standing soliton;
as we have stated earlier, our analysis does not capture this regime.

The other part of feature (v) --- that as $C$ decreases, the NI growth rate decreases ---
is explained similarly to the above. 
Indeed, it follows from \eqref{esec5_04} that $\mu$ increases as $C$ decreases, 
which via \eqref{esec5_17}
implies that $\lambda$ decreases. The main difference from \eqref{esec5_18}
here is that this decrease occurs monotonically with $C$.

\begin{figure}[h]
\vspace{-0.6cm}
\hspace*{-0.5cm}
\mbox{ 
\begin{minipage}{7cm}
\rotatebox{0}{\resizebox{7cm}{9cm}{\includegraphics[0in,0.5in]
 [8in,10.5in]{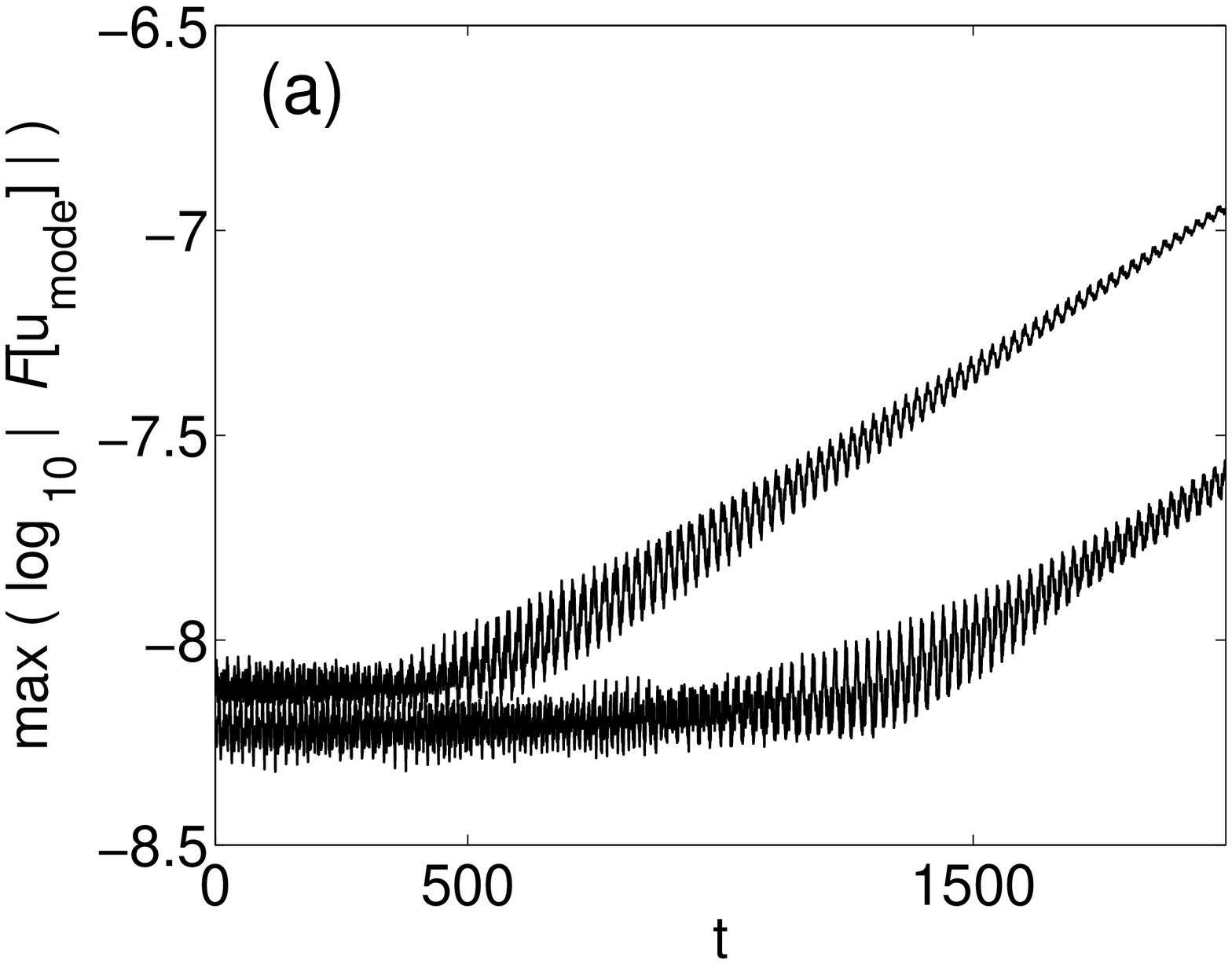}}}
\end{minipage}
\hspace{0.1cm}
\begin{minipage}{7cm}
\rotatebox{0}{\resizebox{7cm}{9cm}{\includegraphics[0in,0.5in]
 [8in,10.5in]{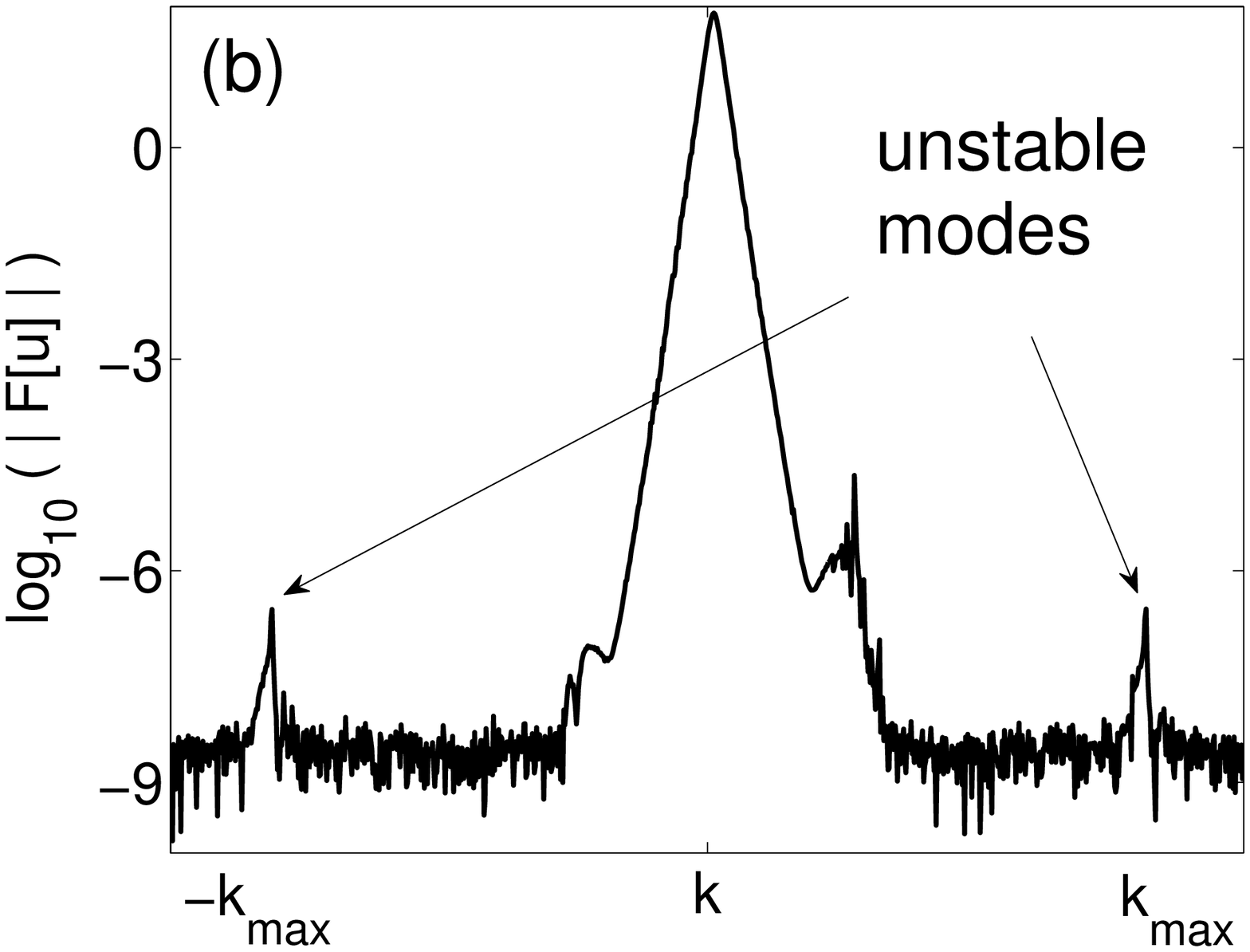}}}
\end{minipage}
 }
\vspace{-1.6cm}
\caption{(a) Evolution of the right peak of the unstable mode for two different
noise realizations in initial condition \eqref{esec5_0}. \ 
(b) Fourier spectrum of the numerical solution corresponding to the smaller
delay time in (a); simulation parameters are the same 
as in Fig.~\ref{fig_5_1}, except that $C=0.7$ and $t=2000$. 
}
\label{fig_5_6}
\vspace{1cm}
\end{figure}

To conclude this section, let us 
note that the ``delayed" NI, first reported in Sec.~5.3, is also observed for
the moving soliton.
Because of estimate \eqref{e4_314}, this phenomenon
is most noticeable for lower values of $C$.
For example, for $C=0.7$, 
NI may become visible around $t=1300$, as shown in Fig.~\ref{fig_5_6}(a).\footnote{
   As we noted before \eqref{esec5_15}, our analysis overestimates the 
   NI growth rate. In particular, it predicts that it should
   be only about two times smaller for $C=0.7$ than for $C=1$,
	 but the numerically observed growth rate for $C=0.7$ is considerably less than that.}
The delay time could be varied by varying the seed of the random
number generator for the background noise in the initial condition \eqref{esec5_0}.
This is illustrated in Fig.~\ref{fig_5_6}(a) and is in agreement with the 
explanation of the ``delayed" NI given in Sec.~5.3.

\section{Numerical instability of oscillating solutions of Eq.~\eqref{e4_301}}
\setcounter{equation}{0}

In the the previous sections we have analyzed the NI on the
background of solitons of the pure and generalized NLS which have stationary shape. 
Here we will extend those studies to the
NI of solutions whose shape varies in time.
As before, we will be concerned with the NI that is weak, i.e. takes a long
time to develop. When the background solution's shape is changing, the development
of weak NI is possible only when those changes are repetitive. Otherwise, the factors
leading to NI will not be able to accumulate coherently, and hence NI would not be
able to occur. Therefore, solutions where weak NI could occur must be periodic or
near-periodic in time. Below we will restrict our attention to such solutions
whose center is not moving; i.e., they extend the standing
solitons considered in Secs.~4 and 5.
We will show that NI on the background of such oscillating pulses is similar
to the ``sluggish" NI reported in Sec.~5.3.

In all simulations reported below, we used $\beta=-1$ and $\gamma=2$.


\subsection{``Sluggish" numerical instability of oscillating pulses}

We began by simulating the NLS \eqref{e_01} with the initial condition
and length of the computational domain given by:
\be
u_0(x)={\rm sech}\,(x)\cdot e^{-(x/3)^4} + 0.2 \cos(2\pi x/L) + \xi(x), 
\qquad L=20,
\label{e7_01}
\ee
instead of \eqref{e_15}. Here the exponential factor was used to ensure zero
(to numerical accuracy), and hence periodic, boundary conditions at this shorter
$L$ than in the previous sections. Near $x=0$, the solution resembles
a soliton, whose amplitude and width oscillate in time; the ``pedestal" outside the pulse
oscillates as well. By varying $C$ around 0.25 (see below), we observed the
``sluggish" NI with all its features described in Sec.~5.3.3. The most notable feature
is still (i): the NI may take a long time to develop; recall 
Figs.~\ref{fig_4_3_1} and \ref{fig_5_6}(a). For example, for $N=2^9$ and $C=0.226$,
it takes $t > 10,000$ for the NI to become distinguishable above the noise floor;
by $t=50,000$ it grows only by half an order of magnitude. At $C=0.265$,
it rises from the noise floor around $t\sim 6,000$ and grows by an order of magnitude
by $t=20,000$.

We have considered possible reasons that could cause ``sluggish" NI in this case.
From Secs.~4 and 5 we have recalled that the unstable mode was found at the ``tails"
of the pulse. Since the ``tails" are being constantly affected by the oscillating ``pedestal",
could that quasi-periodic motion of the ``tails"
have caused growth of unstable modes? We have answered this question to the
negative by showing that qualitatively the same NI is observed for solutions that 
are either exactly or almost
exactly periodic in time. Such solutions were engendered by the following
respective initial conditions:
\be
u_0(x) = 2\,{\rm sech}\,(x) + \xi(x), 
\qquad L=40,
\label{e7_02}
\ee
for the pure NLS \eqref{e_01} and 
\be
u_0(x) = {\rm sech}\,(x)\,
         \left(\, 1 + 0.4(1-2x\,{\rm tanh}\,(x))\;\right) + \xi(x), 
\qquad L=40,
\label{e7_03}
\ee
for the generalized NLS \eqref{e4_301} with $\Pi(x)=1.5\,\exp[-0.2x^2]$. 
Note that initial condition \eqref{e7_02} results in a well-known analytical solution
of the NLS, which is given in Appendix D.
The solution corresponding to \eqref{e7_03} is a sech-like pulse whose amplitude 
oscillates between $1.08$ and $1.44$  almost periodically, with almost no dispersive
radiation being emitted outside the pulse. 
From this numerical evidence we have concluded that it is those oscillations,
rather than just the ``tails" of the soliton,
that cause ``sluggish" NI.
In Appendix D we speculate about a relation between the ``sluggish" NI for an
oscillating pulse and that for a soliton in a wide or tall external potential,
described in Sec.~5.3.
 Unfortunately, unlike in the previous sections, we have not been able
to propose a predictive model of this phenomenon.


\subsection{Estimation of threshold of ``sluggish" numerical instability}

In the absence of such a model, and given a relatively large range of $C$ values 
where ``sluggish" NI of an oscillating pulse is observed, we have considered a
question that may be posed by a researcher interested in avoiding NI
in long-term simulations: \ For what relation between $\dx$ and $\dt$ does NI
not grow above a certain amount (we used `by one order of magnitude')
at a certain simulation time (we used $t=1000$)? In loose terms, what relation
between $\dx$ and $\dt$ gives a ``practical" NI threshold? We address this below.

Our results from Secs.~4 and 5 imply that the {\em exact}
NI threshold should satisfy the relation:
\bsube
\be
\dt = O(\dx) \quad \so \quad C_{\rm thresh,\,exact}={\rm const},
\label{e7_05a}
\ee
given the definition \eqref{e_16} of the parameter in equations \eqref{e_32}
and \eqref{e4_303} for the numerical error.
The only factor that can possibly (and probably only slightly) modify it  is the
dependence of the potentials in \eqref{e_36} and \eqref{e4_309}
on the ``slow" spatial variable $\epsilon \chi$, where $\epsilon=O(\dx)$.
On the contrary, the main result of \cite{Faou_2011} is that NI is guaranteed
not to occur for 
\be
\dt \le O(\dx\,^2) \quad \so \quad C_{\rm thresh,\,exact}=O(\dx\,^2).
\label{e7_05b}
\ee
\label{e7_05}
\esube
These results, again, pertain to the
{\em exact} NI threshold, whereas our question above is about a ``practical"
threshold, which, obviously, is greater.

We have answered that question by numerically simulating initial conditions
\eqref{e7_01}--\eqref{e7_03} and several others, among which we report
on two:
\be
u_0(x) = 2.5\,{\rm sech}\,(x)\cdot e^{-1.2(x/4)^4} + \xi(x), 
\qquad L=15,
\label{e7_06}
\ee
and
\be
u_0(x) = {\rm sech}\,(x)\,
         \left(\, 1 + \varepsilon (1-2x\,{\rm tanh}\,(x))\;\right) + \xi(x), 
\qquad \varepsilon=0.2, \quad L=40;
\label{e7_07}
\ee
both for the pure NLS \eqref{e_01}.
Since the amplitude of the sech-like pulse in \eqref{e7_06} is half-integer,
that initial condition results in a dynamics that is most dissimilar to an ${\mathcal N}$-soliton
solution (for an integer ${\mathcal N}$); the short length of the computational domain  
enhances that dissimilarity. Thus, such a solution represents a rather generic quasi-periodic
(in time), pulse-like  solution of the NLS. 
On the other hand, initial condition \eqref{e7_07} was chosen because 
for $\varepsilon\ll 1$, it results in the soliton of amplitude $1+\varepsilon^2$
plus dispersive radiation of order $O(\varepsilon)$. In other words, the $\varepsilon$-term
only minimally shifts the parameters of the original soliton \cite{KL96_falseinst}.
 The quasi-periodic dynamics here occurs due to the dispersive radiation repeatedly re-entering
the computational domain due to periodic boundary conditions. We had to choose $\varepsilon$
to be not too small since otherwise the ``practical" threshold occurred almost exactly
at the theoretical threshold $C=1$ for the pure soliton \eqref{e_15}.

\begin{figure}[h]
\vspace{-0.6cm}
\centerline{
\begin{minipage}{7cm}
\rotatebox{0}{\resizebox{7cm}{9cm}{\includegraphics[0in,0.5in]
 [8in,10.5in]{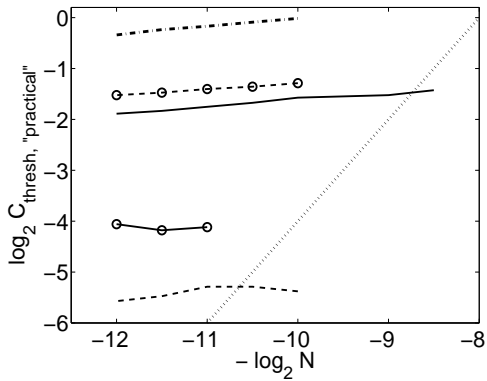}}}
\end{minipage}
 }
\vspace{-1.6cm}
\caption{Dependence of the ``practical" threshold, described before \eqref{e7_05},
on $N^{-1}\propto \dx$. Solid, dashed, and (thicker) dash-dotted lines pertain
to the initial conditions \eqref{e7_01}, \eqref{e7_06}, \eqref{e7_07}, and
solid and dashed lines with circles pertain to initial conditions \eqref{e7_02}, \eqref{e7_03},
respectively. 
Dotted line: the slope predicted by \eqref{e7_05b}.
The smallest $N$ in each case was dictated by ensuring a sufficiently
small discretization error, whereas the largest, $N=2^{12}$, was used to constrain the
computational time. Also, since for lower $N$ the growth rate of ``sluggish" NI is
non-monotonic in $C$ (feature(ii) in Sec.~5.3), we reported the {\em largest} observed
$C_{\rm thresh,\,``practical"}$.
(That is, for some $C$ below that value, NI may grow by more than on order of magnitude
in $t=1000$, but a subsequent increase of $C$ will not necessarily lead to a stronger NI.)
}
\label{fig_7_1}
\vspace{1cm}
\end{figure}

For initial conditions \eqref{e7_01}--\eqref{e7_03}, \eqref{e7_06}, \eqref{e7_07}
the dependence of the ``practical" threshold, as defined above, on $\dx$ is shown
in Fig.~\ref{fig_7_1}. It is seen to be much closer to the dependence \eqref{e7_05a},
predicted in this work, than to \eqref{e7_05b}, predicted in \cite{Faou_2011}. 
The fact that it does not follow \eqref{e7_05a} exactly agrees with
feature (vi) discussed in Sec.~5.3. 
Let us emphasize, again, that all features of the ``sluggish" NI listed there
were also observed for all the cases of the initial conditions considered in this section.

\section{Conclusions}
\setcounter{equation}{0}

\subsection{Summary of results}

The main contribution of this work is the development of the (in)stability analysis of the
fd-SSM beyond the von Neumann (i.e., constant-coefficient) approximation.
Our analysis is valid for spatially-varying, pulse-like background
solutions of the generalized NLS. 
We showed that, as previously for the
s-SSM \cite{ja}, this is done via a modified equation --- \eqref{e_32}, 
\eqref{e4_303}, or \eqref{esec5_03}, ---
derived for the
Fourier modes that approximately satisfy the resonance condition
\be
|\beta| k^2 \dt = \pi.
\label{conc_01}
\ee
Analyzing the (in)stability of the fd-SSM then proceeds similarly to the 
(in)stability analysis of nonlinear waves, i.e., by solving an eigenvalue problem
with a spatially-varying potential. In view of this it is clear that properties
of NI and, in particular, its threshold, 
depend on the simulated solution and thus 
cannot be expected to be universally applicable to all solutions. 
However, our NI analysis {\em does} provide an understanding of  
the mechanism and of generic features of the NI for broad classes of
background solutions. 
Below we summarize such mechanisms and features for
three classes of pulse-like solutions of the generalized NLS.

The first such a class includes, as a prominent representative,
 the standing soliton of the pure NLS \eqref{e_01}. Note that the corresponding 
modified equation for the numerical error, Eq.~\eqref{e_32}, is different 
from the analogous modified equation, \eqref{e_13}, for the s-SSM. 
Their analyses are also qualitatively different, and so are the modes that are
found to cause the instability of these two numerical methods. 
For the s-SSM, these modes are
almost monochromatic (i.e., non-localized) waves $\sim \exp(\pm ikx)$ that ``pass" 
through the soliton very quickly. It is this scattering of those waves 
on the soliton that was shown \cite{ja} to 
lead to their instability. In contrast, for the fd-SSM considered 
in this work, the dominant unstable
modes are stationary relative to the soliton. Moreover, they are 
localized {\em at the sides}, as opposed to the core, of the soliton. 
To our knowledge, such localized modes were not reported before
in studies of instability of nonlinear waves.


It was straightforward to obtain an approximate threshold, \eqref{e_37}
(where $C$ is given by \eqref{e_16}), beyond which NI {\em may} occur. 
Our simulations showed that an NI does indeed occur just slightly
above that threshold. 
In this regard let us note that a qualitatively different expression 
for a bound of the NI threshold, given in \eqref{e7_05b}, was recently proved 
in \cite{Faou_2011} (see Eq.~(2.9) there) by a completely different method.
Our threshold \eqref{e_37}, which satisfies \eqref{e7_05a}, is clearly
greater for $\dx\ll 1$. Also, as we have demonstrated above, it is close
to being sharp. On the downside, it is strictly valid only when the initial condition
is infinitesimally close to the soliton. Indeed, in Sec.~7 we showed numerically
that when the initial deviation from the soliton is not too small, the threshold
value of $C$ may decrease compared to \eqref{e_37}. The NI threshold obtained in 
\cite{Faou_2011} does not require the initial deviation from the soliton to
be infinitesimal.\footnote{
   However, it makes a restrictive assumption of it being an even function:
	 $\tu(-x)=\tu(x)$.}
Yet, as we have demonstrated in Secs.~2, 4, 5, and 7.2, 
it is only a conservative bound and certainly is not sharp.

In Sec.~5 we considered the generalized NLS with an external potential $\Pi(x)$,
Eq.~\eqref{e4_301}, and have shown that NI on the background of {\em its} soliton 
may be similar to that of the standing soliton of the pure NLS \eqref{e_01}.
We have also identified situations when the NI for \eqref{e4_301} can be observed at
different spatial locations, such as minima of $\Pi(x)$ (Sec.~5.1) or the center 
of the soliton (Sec.~5.2). We have not considered the generalized NLS \eqref{e_00}
with a  nonlinearity other than cubic, e.g., saturable. However, we believe that
in that case, the NI follows one of the scenarios described in Secs.~2, 4, 5.1, or
5.2, {\em as long as} the external potential is not wider or substantially
taller than the ``internal" potential created by the soliton itself.

The second class of background solutions, which leads to a noticeably different
NI behavior, are (quasi-)periodic in time solutions, discussed in Sec.~7.
The same kind of NI, which we called ``sluggish", also occurs for stationary
solitons of the generalized NLS \eqref{e4_301} in which the external potential
is either wider or substantially taller (or both) than the ``internal" potential
$\gamma|\Usol(x)|^2$; see Sec.~5.3. The distinguishing feature of the ``sluggish" NI
is that it can remain weak even when $C$ exceeds the NI threshold by several tens
percent. The modes that cause ``sluggish" NI are not localized (see Figs.~\ref{fig_4_3_3}
and \ref{fig_4_3_5}(a,b)), in contrast to the most unstable modes on the background
of the pure NLS soliton (see Fig.~\ref{fig_5}). Yet, they ``hinge" on the pulse's
``tails" (see next paragraph). Other features of the ``sluggish" NI
are listed in Sec.~5.3.

Since the ``sluggish" NI reported in Sec.~5.3 and the NI of the first subclass
of background solutions are described by similar equations, \eqref{e4_309} and
\eqref{e_36}, respectively, they are not unrelated. In fact, the NI of the
standing soliton of the pure NLS also has a ``sluggish" stage, where the most unstable
mode is not localized and the NI growth rate is not a monotonic function of $C$;
see Appendix C.
However, that stage exists only in a narrow interval of $C$ values of about 1\% past
the NI threshold given by \eqref{e_37}, whereas the ``sluggish" NI reported in
Sec.~5.3 exists over several tens percent past the NI threshold.\footnote{
   We mention a possible reason behind this difference in the next subsection
	 when proposing a method of approximate solution of \eqref{e4_309}.}
Therefore, one reason why we have singled out the ``sluggish" NI as a separate 
phenomenon is that it is likely to be noticed in routine simulations, whereas
the behavior described in Appendix C is not. The other reason is that it is
the ``sluggish" NI that is observed for near-soliton and, more generally,
oscillating background solutions.

The third type of background solutions that we considered, in Sec.~6, is the moving soliton
with speed $S=O(1)$ of the pure NLS. In this case, NI develops
in a manner different from that for the other two classes of background solutions.
Namely, the corresponding unstable modes are not localized and also
 are {\em not} ``pinned" to the soliton; see Fig.~\ref{fig_5_1}(b). 
Rather, they and the soliton pass through each other 
(repeatedly, due to the
periodic boundary conditions of the computational domain), 
and the NI is a result of two such waves' interaction
mediated by the soliton. This mechanism is remotely
similar to the NI mechanism for the s-SSM. 
The corresponding Eqs.~\eqref{esec5_03} have the form similar to that of Eqs.~(3.9)
in \cite{ja}. The difference between these two equations is that for the s-SSM, the 
unstable modes pass through the soliton very fast and hence interact with each other
weakly. This is the reason behind the NI of the s-SSM being weak; it is related to
the wavenumbers of the interacting modes, satisfying \eqref{conc_01}, being large,
of order $O(1/\sqrt{\dt})$.
On the other hand, the modes causing the NI of the fd-SSM about the moving soliton
pass through the soliton with speed $S=O(1)$ and, moreover, are almost stationary
relative to one another (their group velocities are close to zero). The weakness of
the fd-SSM does not appear to be related to any physical parameter being small or large;
see the discussion related to estimate \eqref{esec5_16}.

In Sec.~6, we did not mention the NI threshold for the moving soliton.
Let us comment on this issue now.
Formally, from estimate \eqref{esec5_17a}, one may conclude that
such a threshold does not exist (i.e., the fd-SSM is, again, {\em formally},
unconditionally unstable). This is because for arbitrarily small $\dt$ (or,
equivalently, $C$), there are always bands of wavenumbers $k$ where $|R|>1$ and
hence the NI growth rate $\lambda > 0$; see \eqref{esec5_17b} and \eqref{esec5_04}.
Less formally, even thought the {\em continuous} operator $i\partial_z$ in 
\eqref{esec5_06} is sign indefinite and unbounded, the corresponding operator 
on the discrete grid takes on values within the interval $[-k_{\max},\,k_{\max}]$. 
Then, following the reasoning that led to threshold \eqref{e_37},
 one could have obtained a threshold value of $\mu$, and hence $C$, below which
NI of the moving soliton would be guaranteed not to occur.
A simple estimate yields that in this case, one would find $C_{\rm thresh}=O(\dx)$
and hence $\dt_{\rm thresh}=O(\dx\,^{3/2})$. However, such an estimate is of no
practical value. Indeed, we have repeatedly mentioned in Sec.~6 that our analysis
there had overestimated the NI growth rate, and already for $C=0.7$ 
(and $\Ksol\approx 1$),  the NI may take 
$t>1000$ to become just barely visible above the noise floor. 
For  $C=0.5$, it takes several thousand time units to appear
above the noise floor, and for a yet smaller $C$ it will take even longer. 
Thus, it is unlikely that such a weak NI could be significant in simulations.


\subsection{Open problems}

\subsubsection{Analysis of \eqref{e_36} and \eqref{e4_309} in the limit $\epsilon\To 0$}

As we explained in Sec.~3, WKB solution of these eigenproblems would require the
handling of the turning points where a pair of eigenvectors of the non-self-adjoint 
system of linear Schr\"odinger-type equations 
becomes linearly dependent. This appears to have
been a long-standing unsolved problem (see, e.g., \cite{Fulling2,Skorupski}).
Therefore, we find it reasonable to discuss only those aspects of 
\eqref{e_36} and \eqref{e4_309} that do {\em not} require solving that problem.

Let us begin with \eqref{e_36}. Figure \ref{fig_8} in Appendix C suggests that
eigenvalues of the unstable modes at their ``birth" satisfy $|\Lambda|\ll D$.
The smallness of $|\Lambda|/D$ could be used to consider \eqref{e_36} as a 
perturbation of the WKB-solvable system \eqref{e_C1}. This may explain the cascade
of bifurcations that eventually leads to the emergence of a pair of real eigenvalues,
resulting in NI.  It may also give a value of the sharp NI threshold $C_{\rm cr}$, 
defined in Appendix C. It could also be interesting to find out how that cascade
of bifurcations is affected by the size $L$ of the computational domain.
Namely, as $L\To\infty$, will it ``collapse" to a single value of $C$ where
an unstable eigenmode, having once emerged, would persist for all greater
values of $C$, as opposed to the behavior described in Appendix C?

With respect to \eqref{e4_309}, we sketch an approach by which the ``sluggish" NI
of a stationary soliton, described in Sec.~5.3, could be analyzed. As we mentioned
there, the occurrence of ``sluggish" NI requires that the external potential
$\Pi(x)$ be substantially wider or taller than the ``internal" one, $\gamma |\Usol(x)^2|$.
From Fig.~\ref{fig_4_3_5} we notice that in such a case, the eigenmode overlaps
much more with the external than with the ``internal" potential. Therefore,
the $\gamma |\Usol(x)^2|$-term in \eqref{e4_309} can be considered as a small
perturbation. The remaining part of that eigenproblem decouples into two linear
Schr\"odinger equations, which can be solved by the WKB method. Then the solution of 
\eqref{e4_309} could be sought as a perturbation of that solution with $\gamma=0$.
One of the features of ``sluggish" NI, noted in Sec.~5.3, was that the most
unstable mode is delocalized in a rather wide range of $C$ values past the NI
threshold. As for \eqref{e_36}, here it could also be interesting to find out
how this is affected by the size of the computational domain. In other words,
is ``sluggish" NI a finite-$L$ phenomenon or will it persist on the infinite line?

\subsubsection{Numerical instability of oscillating solutions}

Recall that \eqref{e_31} describes the evolution of a high-$k$ numerical error 
with an arbitrary, including oscillatory, background $u_{\rm b}$. 
In Appendix D we argued that the standard technique of splitting the fields
in \eqref{e_31} into slowly and rapidly varying parts does not appear to lead
to an analytically tractable model. An alternative approach could, perhaps,
 be based on proper orthogonal decomposition of the background oscillating solution
into a small number of rapidly diminishing ``principal components". For example,
such a decomposition of solution \eqref{e_D03} into just two principal components
can be found in \cite{Kutz_course}.

\subsubsection{Numerical instability of moving soliton}

We were unable to analytically solve Eqs.~\eqref{esec5_03}
with $\Usol(x) \propto \sech(x)$ and hence had to approximate it by a box profile.
It is unclear whether an analytical solution with \eqref{esec5_03} without such an
approximation is even possible, except, perhaps, in the limit $\mu\To\infty$.
However, that limit is of no practical interest since it corresponds to $C\To 0$,
and we have noted that NI in that case is so weak that it may never be observed
in simulations. Therefore, we do {\em not} propose solving the non-approximated
\eqref{esec5_03} as an open problem.

Instead, we think that obtaining an equation for a high-$k$ numerical error
that could be valid for $0<S<O(1)$ {\em is} an interesting open problem.
Such an equation must somehow account for the Fourier spectrum of the error
having a structure seen in the three bottom curves in Fig.~\ref{fig_5_5}(b):
a narrow peak (i.e., delocalized in $x$) on top of a broad pedestal (i.e.,
variations in $x$ on a scale much shorter than $O(1)$). At the moment we do
not know how to approach that problem. 
If such an equation is obtained, 
its analysis, even only numerical, would be another interesting problem.
Indeed, it would have to exhibit a transition of the unstable mode 
from being ``pinned" to the soliton (for $S=0$) to passing through it 
(for $S=O(1)$).

Even more fundamental seems to be another issue, which lies at the heart of the 
difference between the modified equations of the high-$k$ numerical error
for the standing and moving solitons, i.e., Eqs.~\eqref{e_31} and \eqref{esec5_03},
respectively. This difference stems from that of the spectra of the numerical
error in this two cases: a broad spectrum for the standing soliton (Fig.~\ref{fig_2}(a))
and rather narrow peaks for the moving one (Figs.~\ref{fig_5_1}(a) and \ref{fig_5_6}(b)).
Recall that Eqs.~\eqref{e_31} and \eqref{esec5_03} did not explain those
differences; rather, they were derived {\em based} on the numerically observed
different spectra. Thus, an open question is: 
How can one tell from the form of the background solution of the NLS what the 
spectrum of unstable modes of the fd-SSM should be?

\subsubsection{Miscellaneous}

We believe that analysis of NI of the SSM and related methods (e.g., 
the integrating factor method) is an unexplored area where techniques of
stability analysis of nonlinear waves could be applied. As examples,
let us mention just three broad topics, which were not considered in this work:
 \ (i) NI of s- and fd-SSMs in two and three spatial dimensions;
\ (ii) NI of the generalized NLS \eqref{e4_301} where the external potential
$\Pi(x)$ grows at infinity, e.g., $\Pi(x) \propto x^2$, as in the
Gross--Pitaevskii equation; \ (iii) NI of solitons in long-wave, e.g., 
Korteweg--de Vries, equations.


\section*{Acknowledgement}

I thank Jake Williams for help with numerical simulations at an early stage of this work,
and Eduard Kirr for a useful discussion.
This research was supported in part by NSF grants ECCS-0925706 and
 DMS-1217006.


\section*{Appendix A: 
Modified linearized NLS for fd-SSM with non-periodic boundary conditions}
 \setcounter{section}{9}
 \setcounter{equation}{0}

We consider homogeneous Dirichlet boundary conditions (b.c.),
which are compatible with the standing soliton solution of \eqref{e_01}
in a large computational domain.
Neumann or mixed b.c.
can be treated similarly, and lead to similar results. 

The equation for the dispersive step of the fd-SSM is still given by \eqref{e_05}.
However, now instead of \eqref{e_06} we assume: $u_{n+1}^0=0$, $u_{n+1}^M=0$, where
$m=0$ and $m=M$ are the end points of the spatial grid. Then \eqref{e_05} can be
rewritten as \cite{KincaidCheney}
\be
({\mathcal I}+(i\beta r/2){\mathcal A})\, {\bf u}_{n+1} = 
({\mathcal I}-(i\beta r/2){\mathcal A})\, {\bf \bu},
\label{e_40}
\ee
where: ${\bf \bu}=[\bu^1,\bu^2,\ldots,\bu^{M-1}]^T$, similarly for ${\bf u}_{n+1}$,
${\mathcal I}$ is an $(M-1)\times (M-1)$ identity matrix, and ${\mathcal A}$ is an
$(M-1)\times (M-1)$ tridiagonal matrix with $(-2)$ on the main diagonal and $(+1)$
on the sub- and super-diagonals. 

The starting point of our derivation in Sec.~3, Eq.~\eqref{e_18}, has exacly the 
same form for the case of the Dirichlet b.c., except that $\F$
is replaced with ${\mathcal T}$ --- an expansion over the complete set of the
eigenvectors of ${\mathcal A}$; similarly, $\F^{-1}$ is replaced by ${\mathcal T}^{-1}$.
The exponential in \eqref{e_19} that acts on the $j$th eigenvector is replaced by
\be
e^{iP_j}=\frac{1-i\beta r\lambda_j/2}{1+i\beta r\lambda_j/2},
\label{e_41}
\ee
where $\lambda_j$ is the corresponding eigenvalue \cite{KincaidCheney}:
\be
\lambda_j = -4 \sin^2\big( \pi j/(2M) \big).
\label{e_42}
\ee
Equations \eqref{e_41}, \eqref{e_42} and the middle expression in \eqref{e_19}
coincide provided that we identify:
\be
k=j\pi/(M\dx) = j\pi/L.
\label{e_43}
\ee
However, we are still a step away from proving that the modified linearized NLS for
the Dirichlet b.c. case is the same as that equation for periodic b.c..
This is because $-k^2$, which is the Fourier symbol of the second derivative, is
not the symbol of the second derivative under the transformations ${\mathcal T}$
and ${\mathcal T}^{-1}$. Under those transformation, the required symbol is given 
by \eqref{e_42}. We will now use this observation to supply the last step and show
that the modified linearized NLS for the case of Dirichlet b.c. is indeed the same
as \eqref{e_31}. This follows from \eqref{e_41}--\eqref{e_43} and a calculation that
is similar to \eqref{e_21}:
\bea
e^{iP(k)} & \approx & -\left( 1 + \frac1{i\beta r 
                                         \big(1-\sin^2((k-k_{\max})\dx/2)\big)} \right)
 \nonumber \\
& \approx & -\left( 1 + \frac1{i\beta r} +
 \frac{\sin^2((k-k_{\max})\dx/2)\big)}{i\beta r} \right),
\label{e_44}
\eea
where we have used that $\sin(k_{\max}\dx/2)=1$ and that for highly oscillatory
eigenvectors of ${\mathcal A}$, one has $(k-k_{\max})\dx \ll 1$. The last term
on the r.h.s. of \eqref{e_44} is the desired symbol of the second derivative, and 
then the rest of the derivation is the same as that leading to \eqref{e_28}. From it
one obtains the same modified linearized NLS as \eqref{e_31}. Our numerical simulations
of the NLS using the fd-SSM with zero Dirichlet b.c. confirm this conclusion.


\section*{Appendix B: 
Numerical solution of eigenproblem \eqref{e_36}}
 \addtocounter{section}{1}
 \setcounter{equation}{0}

We work with \eqref{e_36} written in an equivalent form:
\be
\sigma_3 \left( \partial_X^2 + D - V(\epsilon X) 
 \left( \ba{cc} 2 & 1 \\ 1 & 2 \ea \right) -i\Lambda_0 \sigma_3 \,\right)
\vec{\phi} \,=\, i(\Lambda-\Lambda_0) \vec{\phi},
\label{B_01}
\ee
where the reason to include a constant $\Lambda_0$ will be explained later. 
We discretize \eqref{B_01} using Numerov's method, which approximates the equation \ 
$\Phi_{XX}=F(\Phi,X)$ \ by a finite-difference scheme
\be
\Phi^{m+1}-2\Phi^m+\Phi^{m-1}=\frac{\D X\,^2}{12} \big( F^{m+1}+10F^m+F^{m-1} \big)
\label{B_02}
\ee
with accuracy $O(\D X \,^4)$.
Here $\Phi^m\equiv \Phi(X_m)$, $F^m\equiv F(\Phi^m,X_m)$, etc., and
$m=1,\ldots\,,\bar{M}-1$. 
Note that the number of grid points, $\bar{M}$, in the $X$-domain 
is much greater than the number
of grid points, $M$, in the $x$-domain, because $X\propto x/\epsilon$;
see \eqref{e_35}, \eqref{e_27}.
Then, for the discretized solution ${\bf f}_k=[\phi_k^1,\ldots\,,\,\phi_k^{\bar{M}-1}]^T$
($k=1,2$) one obtains:
\be
(-1)^{k-1}\left(\, \left[ \frac1{\D X\,^2}{\mathcal A}_{\rm per} + {\mathcal N}_{\rm per} 
\left\{ D{\mathcal I} - 2{\mathcal V} - i\Lambda_0 (-1)^{k-1} {\mathcal I} \right\} \right] {\bf f}_k
\,-\, {\mathcal N}_{\rm per} {\mathcal V} {\bf f}_{3-k} \, \right) \,=\, 
i(\Lambda - \Lambda_0) {\mathcal N}_{\rm per} {\bf f}_k.
\label{B_03}
\ee
Here all matrices, denoted by script letters, have size $(\bar{M}-1)\times (\bar{M}-1)$;
 ${\mathcal I}$ is defined after \eqref{e_40}; \ ${\mathcal A}_{\rm per}$ is as in \eqref{e_40}
except that its $(1,\bar{M}-1)$th and $(\bar{M}-1,1)$th 
entries equal 1 (to account for the periodic boundary
conditions); \ ${\mathcal N}_{\rm per}$ has a similar structure as ${\mathcal A}_{\rm per}$: \ 
$({\mathcal N}_{\rm per})_{m,m}=10/12$ (see \eqref{B_02}), 
$({\mathcal N}_{\rm per})_{(m-1),m}= ({\mathcal N}_{\rm per})_{m,(m-1)}=1/12$, 
$({\mathcal N}_{\rm per})_{1,(\bar{M}-1)}= ({\mathcal N}_{\rm per})_{(\bar{M}-1),1}=1/12$, 
and the rest of its entries are zero; \ 
and ${\mathcal V}={\rm diag}(V^1,\ldots\,,\,V^{\bar{M}-1})$. 
Next, defining the combined vector and matrices:
$$
\hat{\bf f}=\left[ \ba{c} {\bf f}_1 \\ {\bf f}_2 \ea \right], \quad
\hat{\mathcal A}_{\rm per}= \left( \ba{cc} {\mathcal A}_{\rm per} & {\mathcal O} \\
                                   {\mathcal O} & {\mathcal A}_{\rm per} \ea \right), \quad
\hat{\mathcal N}_{\rm per}= \left( \ba{cc} {\mathcal N}_{\rm per} & {\mathcal O} \\
                                   {\mathcal O} & {\mathcal N}_{\rm per} \ea \right), 
$$
$$
\hat{\mathcal V}= \left( \ba{cc} 2{\mathcal V} & {\mathcal V} \\
                                   {\mathcal V} & 2{\mathcal V} \ea \right), \quad
\hat{\sigma}_3 = \left( \ba{cc} {\mathcal I} & {\mathcal O} \\
                                   {\mathcal O} & -{\mathcal I} \ea \right),
$$
where ${\mathcal O}$ is the $(\bar{M}-1)\times (\bar{M}-1)$ 
zero matrix, one rewrites \eqref{B_03} as:
\be
\hat{\sigma}_3 
\left[ \frac1{\D X\,^2}\hat{\mathcal A}_{\rm per} + D\hat{\mathcal N}_{\rm per} 
 - \hat{\mathcal N}_{\rm per} \hat{\mathcal V} - i\Lambda_0 \hat{\sigma}_3 \hat{\mathcal N}_{\rm per} 
  \right] \hat{\bf f} \,=\, i(\Lambda-\Lambda_0)\hat{\mathcal N}_{\rm per} \hat{\bf f}\,.
\label{B_04}
\ee
This equation has the form of the generalized eigenvalue problem 
${\mathcal G}\hat{\bf f}=\lambda{\mathcal H}\hat{\bf f}$ where
 ${\mathcal H}=\hat{\mathcal N}_{\rm per}$
is a positive definite matrix. This problem can be solved by Matlab's command \verb+eigs+.
As its options, we specified that 108 smallest-magnitude eigenvalues and
the corresponding eigenmodes needed to be computed. Among them, we looked only
at those with complex $\Lambda$.
Beyond the instability threshold there are several such modes.
We visually inspected them and found that the most unstable mode was 
also the most localized and also had a real eigenvalue.

We verified that the eigenvalues did not change to five significant figures whether
 we used $\D X=1/10$
or $1/20$; so we used $\D X=1/10$. Finally, 
for relatively large $D\ge 0.1$,
it was convenient to shift the eigenvalues by some $\Lambda_0$ (found at a previously
considered value of $D$), so that those with largest $\Lambda_R$ would appear
at the beginning of the list, when they are sorted by Matlab according
to their absolute value in ascending order.


\section*{Appendix C: 
``Birth" of localized unstable mode}
\addtocounter{section}{1}
 \setcounter{equation}{0}

The instability growth rates plotted in Fig.~\ref{fig_3} are monotonic functions of 
the parameter $C$. This, however, occurs only when $C$ is sufficiently beyond a critical
value, $C_{\rm cr}$, where the dominant (i.e., with the greatest-$|\Lambda_R|$) unstable 
mode is created. Near $C_{\rm cr}$, which
is slightly above the threshold value given by the r.h.s. of \eqref{e_37}, the evolution
of the greatest-$|\Lambda_R|$ eigenvalue is quite irregular.

Below we present results about this evolution for the dominant unstable mode (shown in
Fig.~\ref{fig_5}) for $L=40$, $N=2^9$ (i.e., $\epsilon=40/1024\approx 0.04$) and the rest of
the parameters being the same as listed in Sec.~2, i.e.: $\beta=-1$, $\gamma=2$, and $A=1$. 
Then,  parameters $C$ and $D$ are related by \eqref{sec4_extra1}: $D=C-1$.
While we have been unable to rigorously establish an analytical expression for $C_{\rm cr}$,
we will present a hypothesis as to what it may be. The main message that we intend to convey
is that the ``birth" of a localized eigenmode of Eq.~\eqref{e_36} occurs via a complex 
sequence of bifurcations, in contrast to a single bifurcation that typically takes place
when an unstable mode of a nonlinear wave is ``born" (see, e.g., \cite{Kapitula98}).

We numerically  observed that eigenmodes of \eqref{e_36} with $\Lambda_R\neq 0$ are 
``born" in two ways. One is  when two imaginary eigenvalues $\pm i\Lambda_I$
 ``collide" at the origin (i.e., $\Lambda_I\To 0$)
and thereby give rise to two real ones.
The other way is a ``collision" of two imaginary eigenvalues $i\Lambda_{I1}$
and $i\Lambda_{I2}$ away from the origin (i.e., $i\Lambda_{I1} \To i\Lambda_{I2} \neq 0$).
In that case  two complex eigenvalues are ``born".\footnote{
   As per the remark after \eqref{e_36}, there is also a pair of eigenvalues with $(-\Lambda_I)$,
   so  a quadruplet of complex eigenvalues actually appears.} 
The very first (i.e., for the smallest $C$)
unstable mode is created in the former way. We will now show that this mode is essentially
non-localized and, moreover, it is {\em not} the mode that eventually becomes the dominant
unstable mode, whose growth rate is plotted in Fig.~\ref{fig_3}. The reason why we still chose
to discuss the former mode while being primarily interested in the latter one, will become clear
as we proceed.

\begin{figure}[h]
\vspace{-1.6cm}
\mbox{ 
\begin{minipage}{7cm}
\rotatebox{0}{\resizebox{7cm}{9cm}{\includegraphics[0in,0.5in]
 [8in,10.5in]{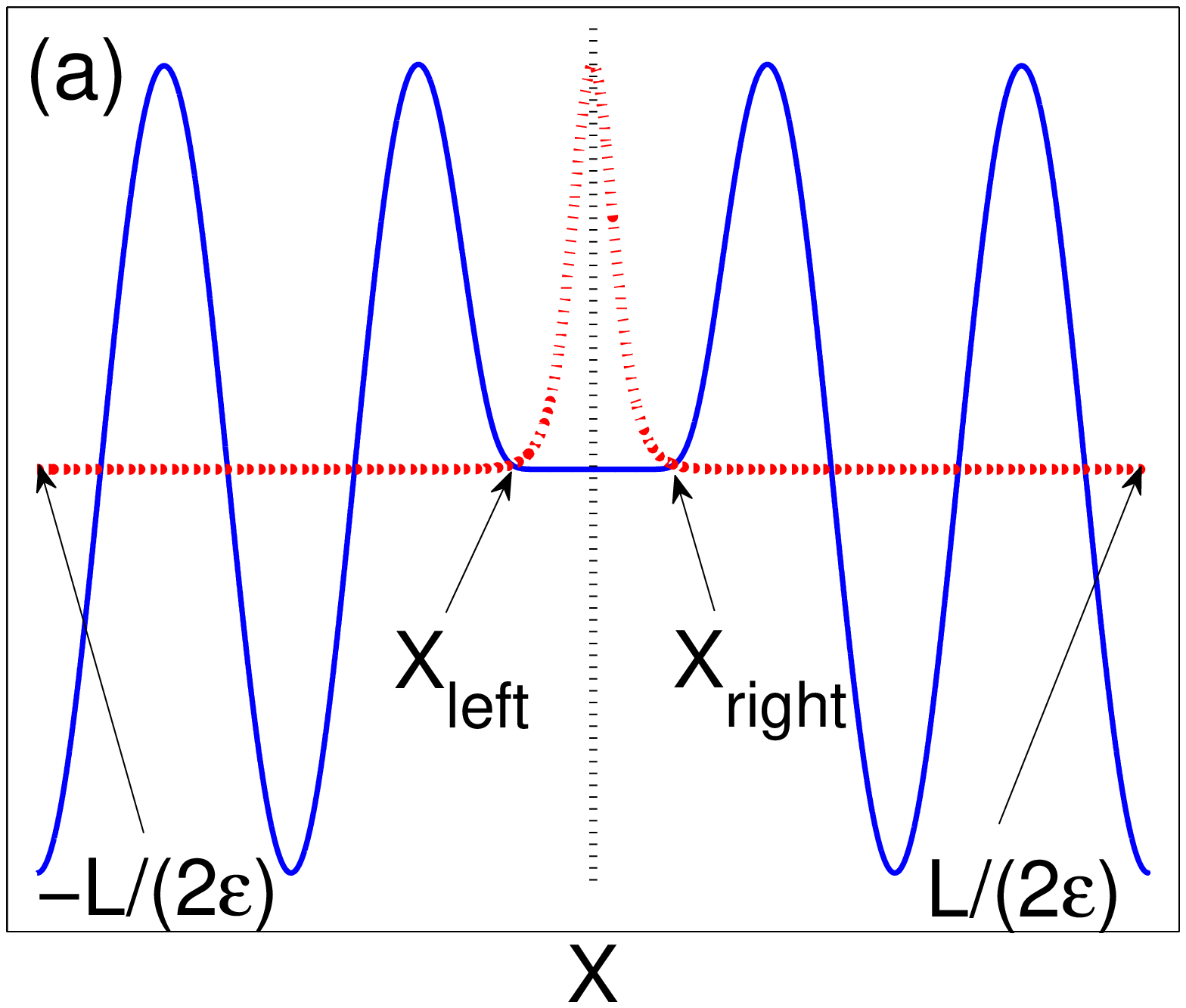}}}
\end{minipage}
 \hspace{0.1cm}
 \begin{minipage}{7cm}
  \rotatebox{0}{\resizebox{7cm}{9cm}{\includegraphics[0in,0.5in]
   [8in,10.5in]{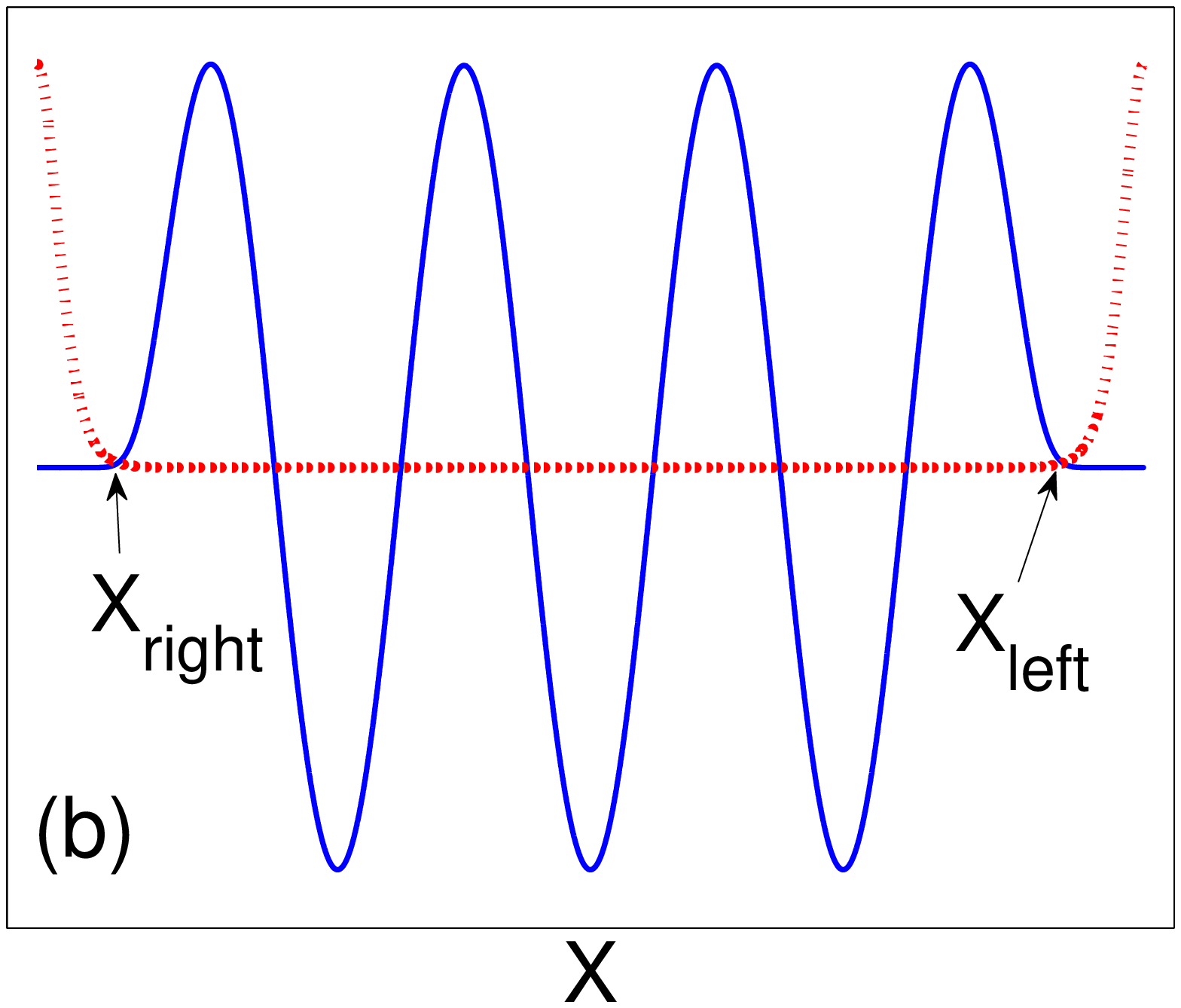}}}
 \end{minipage}
 }
\vspace{-1.6cm}
\caption{ (Color online) \ (a): A solution of \eqref{e_C1} (solid); potential $\sech^2(\epsilon X)$ 
(red dotted). The amplitude of the solution is normalized to that of the potential. \ 
(b): Same as (a), but that panel is ``cut" along the vertical dotted line at the center,
and the resulting halves are interchanged.
}
\label{fig_7}
\end{figure}

For a mode with $\Lambda=0$, Eq.~\eqref{e_36} can be split into two uncoupled 
linear Schr\"odinger equations:
\bsube
\be
\big( \,\partial_X^2 + D - \nu_{\pm}\,V(\epsilon X)\,\big) \phi_{\pm} =0, \qquad 
\phi_{\pm} = \phi_1 \pm \phi_2,
\label{e_C1a}
\ee
where $\nu_-=1$ and $\nu_+=3$. Note that $\phi_{\pm}$ satisfy the periodic boundary conditions:
\be
\phi_{\pm}(-L/(2\epsilon)) \,=\, \phi_{\pm}(L/(2\epsilon)) .
\label{e_C1b}
\ee
\label{e_C1}
\esube
In Fig.~\ref{fig_7}(a) we show an example of a nontrivial solution of \eqref{e_C1}.
In view of the periodic boundary conditions, this figure is equivalent to Fig.~\ref{fig_7}(b).
Recall from Sec.~IV that the eigenmode is exponentially small inside the soliton. Then
the solution shown in Fig.~\ref{fig_7}(b) can be thought of as being localized inside the valley
bounded by the two ``halves" of the potential. Using this observation, one can estimate
the isolated values of $D$ for which one of the equations \eqref{e_C1a}, along with
\eqref{e_C1b}, has a nontrivial solution, by the standard WKB method. 
The condition for the existence
of a mode localized inside the valley of Fig.~\ref{fig_7}(b) is given by the Bohr--Sommerfeld
formula:
\be
\left( \int_{-L/(2\epsilon)}^{X_{\rm left}} + \int_{X_{\rm right}}^{L/(2\epsilon)} \right)
 \, \sqrt{D-\nu V(\epsilon X)} \,dX \,=\, \pi \left(n+\frac12 \right),
\label{e_C2}
\ee
where $\nu$ is either $\nu_-$ or $\nu_+$, $n$ is an integer, and $X_{\rm left,\,right}$ are
the turning points (see Sec.~4), where 
\be
D-\nu V(\epsilon X_{\rm left,\,right}) =0.
\label{e_C3}
\ee
The number of full oscillation periods of the mode inside the valley equals $n$; for example,
in Fig.~\ref{fig_7}, $n=3$.

When $D\ll 1$, the $\sech^2$ potential in \eqref{e_C3} can be approximated by an exponential: \ 
$\sech^2(\epsilon X) \approx 4\exp(-2\epsilon X)$. Then, using \eqref{e_35} and \eqref{e_C3},
we reduce \eqref{e_C2} to
\be
\sqrt{D} 
\int_{X_{\rm right}}^{L/(2\epsilon)} \sqrt{1 - \exp[-2\epsilon(X-X_{\rm right})] } \,dX \,=\,
\frac{\pi}2 \left( n+\frac12 \right),
\label{e_C4}
\ee
with $X_{\rm left}=-X_{\rm right}$ and 
\be
X_{\rm right} = \frac1{2\epsilon} \, \ln \frac{8\nu C \beta^2}{D}\,.
\label{e_C5}
\ee
Neglecting the exponentially small terms of the order
 $O\big( \, \exp[-(L-2\epsilon X_{\rm right})]\,\big)$,
one obtains from \eqref{e_C4}:
\be
\sqrt{D} \left( L - \ln \frac{8\nu C\beta^2}{D} - 2(1-\ln 2) \right) = 
\epsilon\,\pi \left( n+\frac12 \right).
\label{e_C6}
\ee

Note that the WKB condition \eqref{e_C2}, and hence \eqref{e_C6}, is valid when $n$ is sufficiently
large. In particular, it is {\em not} supposed to accurately predict the ``birth" of the first 
unstable mode, where $n=0$. Indeed, Eq.~\eqref{e_C6} predicts that such a mode (for $\nu=1$)
emerges at $D\approx 5.9\cdot 10^{-6}$, while numerically (see Appendix B) it is found at 
$D\approx 1.6\cdot 10^{-5}$. (A similar mode for $\nu=3$ emerges at a slightly higher value
of $D$.) Formula \eqref{e_C6} becomes accurate to the fourth significant figure in $D$ for
$n \gtrsim 20$.

As we noted above, the first unstable mode is {\em not} the one that eventually becomes the 
dominant unstable mode. It disappears already at $D\approx 1.7\cdot 10^{-5}$, and there is
an adjacent interval of $D$ values where all the eigenvalues of \eqref{e_36} are purely imaginary
(i.e., the soliton is numerically stable). 
As $D$ increases, higher-order ``real" (i.e., with $\Lambda_I=0$) modes appear and disappear in a 
similar fashion, as do quadruplets of modes with complex $\Lambda$. In both these types of modes,
$\Lambda_R$ is fairly small: $|\Lambda_R| \lesssim D/10$. There also exist intervals of $D$,
of increasingly small length, where all $\Lambda$'s are purely imaginary. This situation persists
until the dominant unstable mode appears at $D_{\rm cr}\,(=C_{\rm cr}-1)$. This occurs as follows.

\begin{figure}[h]
\vspace*{-0.6cm}
\hspace*{-0.5cm}
\mbox{ 
\begin{minipage}{5.1cm}
\rotatebox{0}{\resizebox{5.1cm}{7cm}{\includegraphics[0in,0.5in]
 [8in,10.5in]{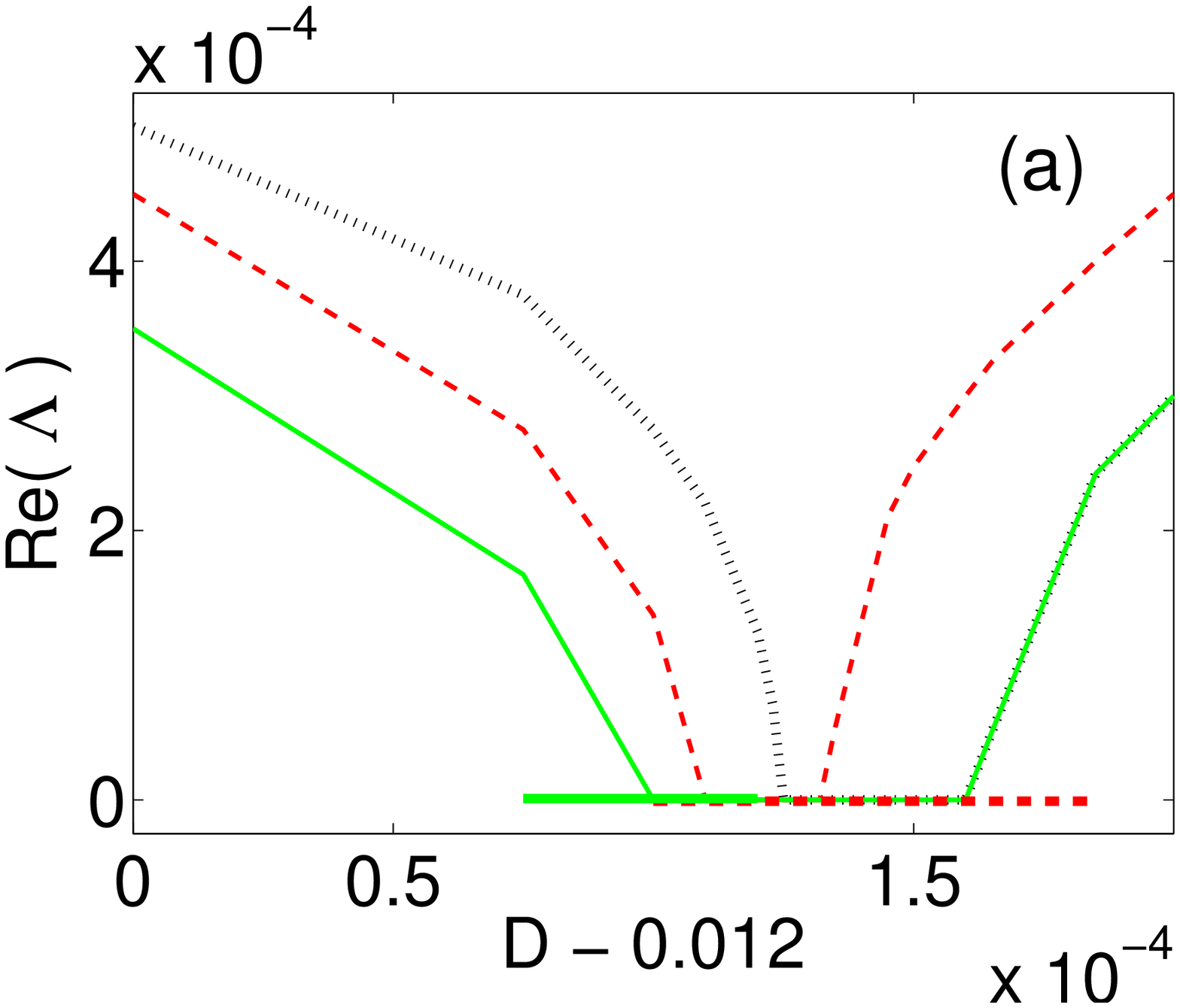}}}
\end{minipage}
\hspace{0.1cm}
\begin{minipage}{5.1cm}
\rotatebox{0}{\resizebox{5.1cm}{7cm}{\includegraphics[0in,0.5in]
 [8in,10.5in]{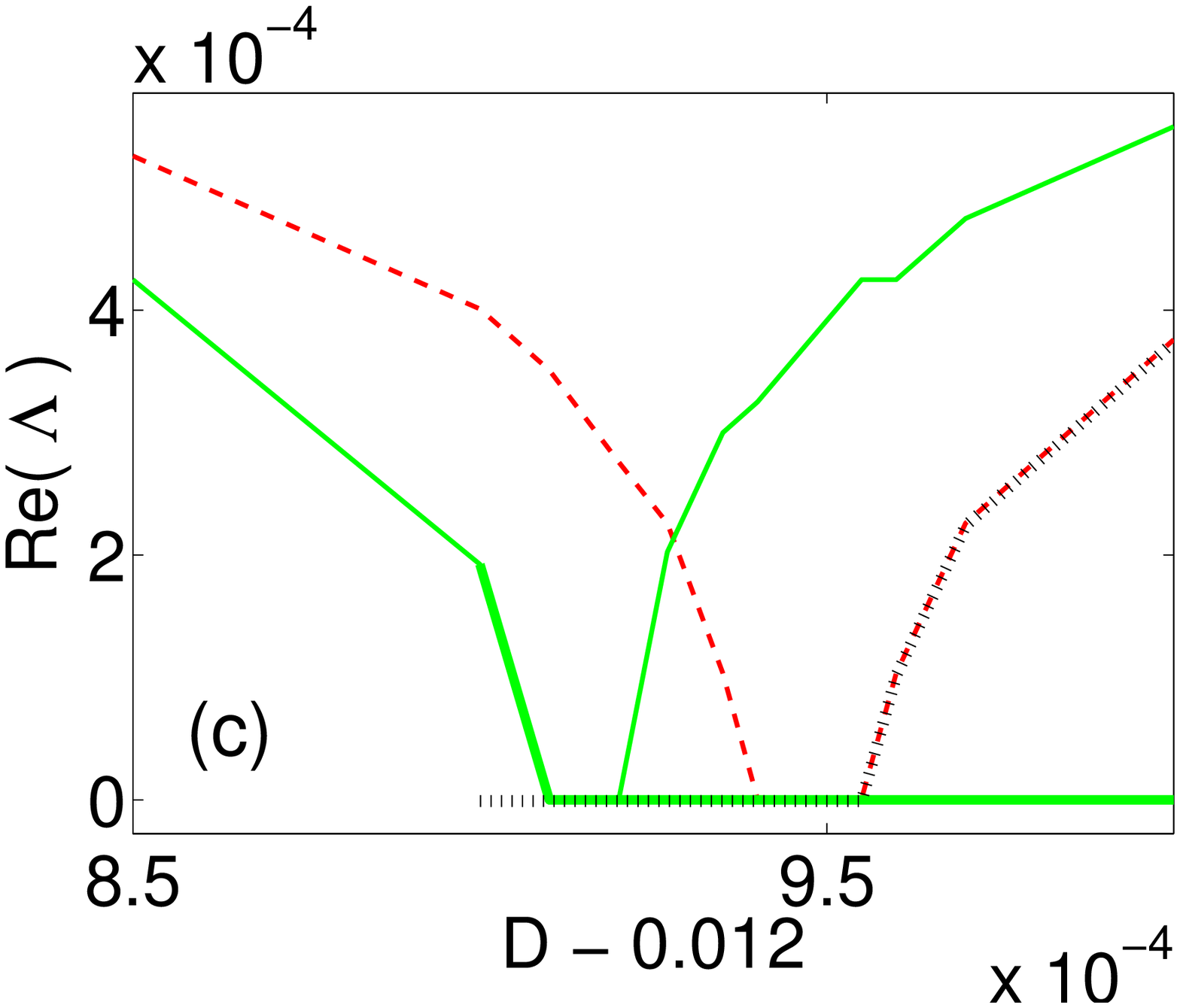}}}
\end{minipage}
\hspace{0.1cm}
\begin{minipage}{5.1cm}
\rotatebox{0}{\resizebox{5.1cm}{7cm}{\includegraphics[0in,0.5in]
 [8in,10.5in]{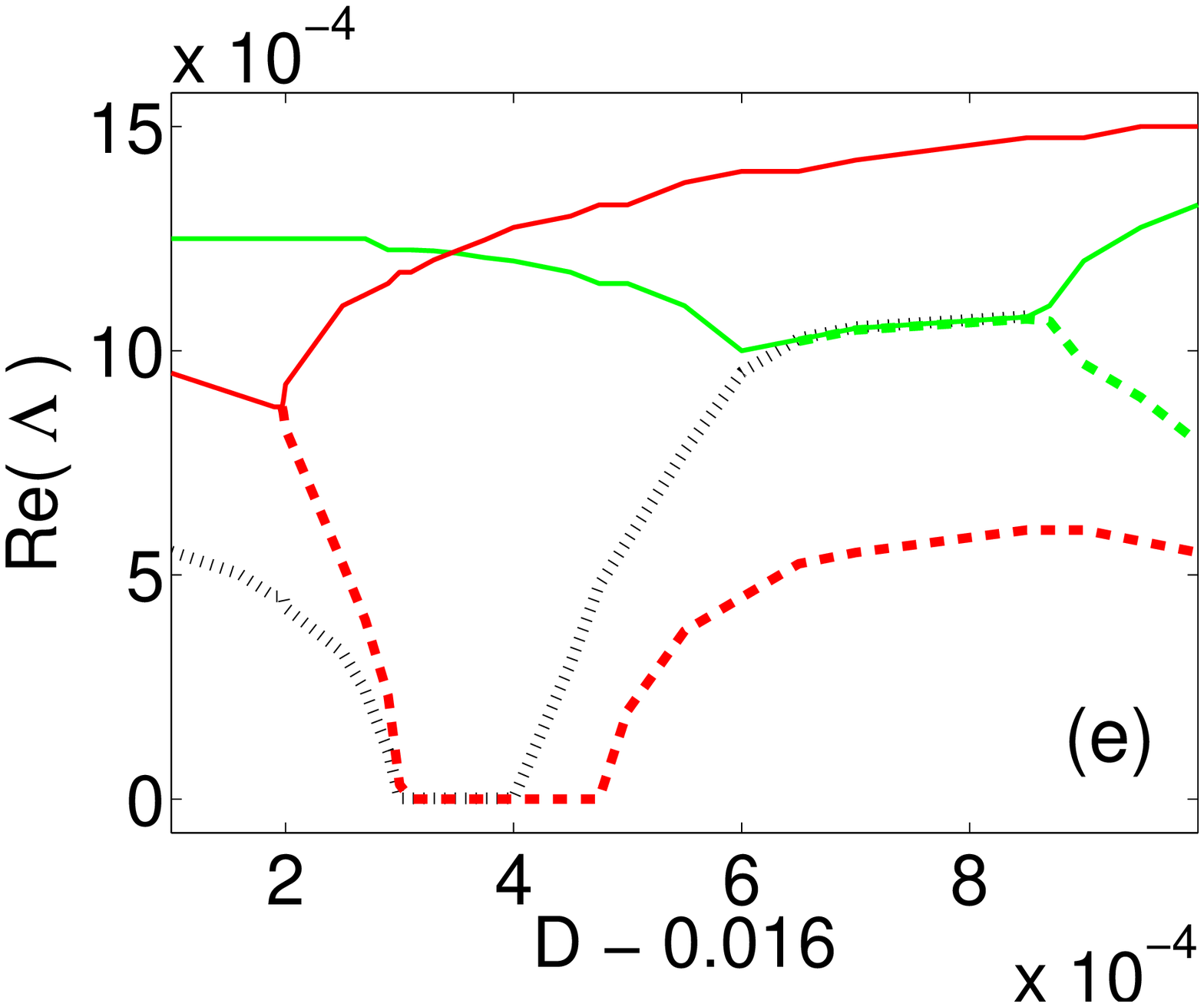}}}
\end{minipage}
 }

\vspace{-2cm}

\hspace*{-0.5cm}
\mbox{ 
\begin{minipage}{5.1cm}
\rotatebox{0}{\resizebox{5.1cm}{7cm}{\includegraphics[0in,0.5in]
 [8in,10.5in]{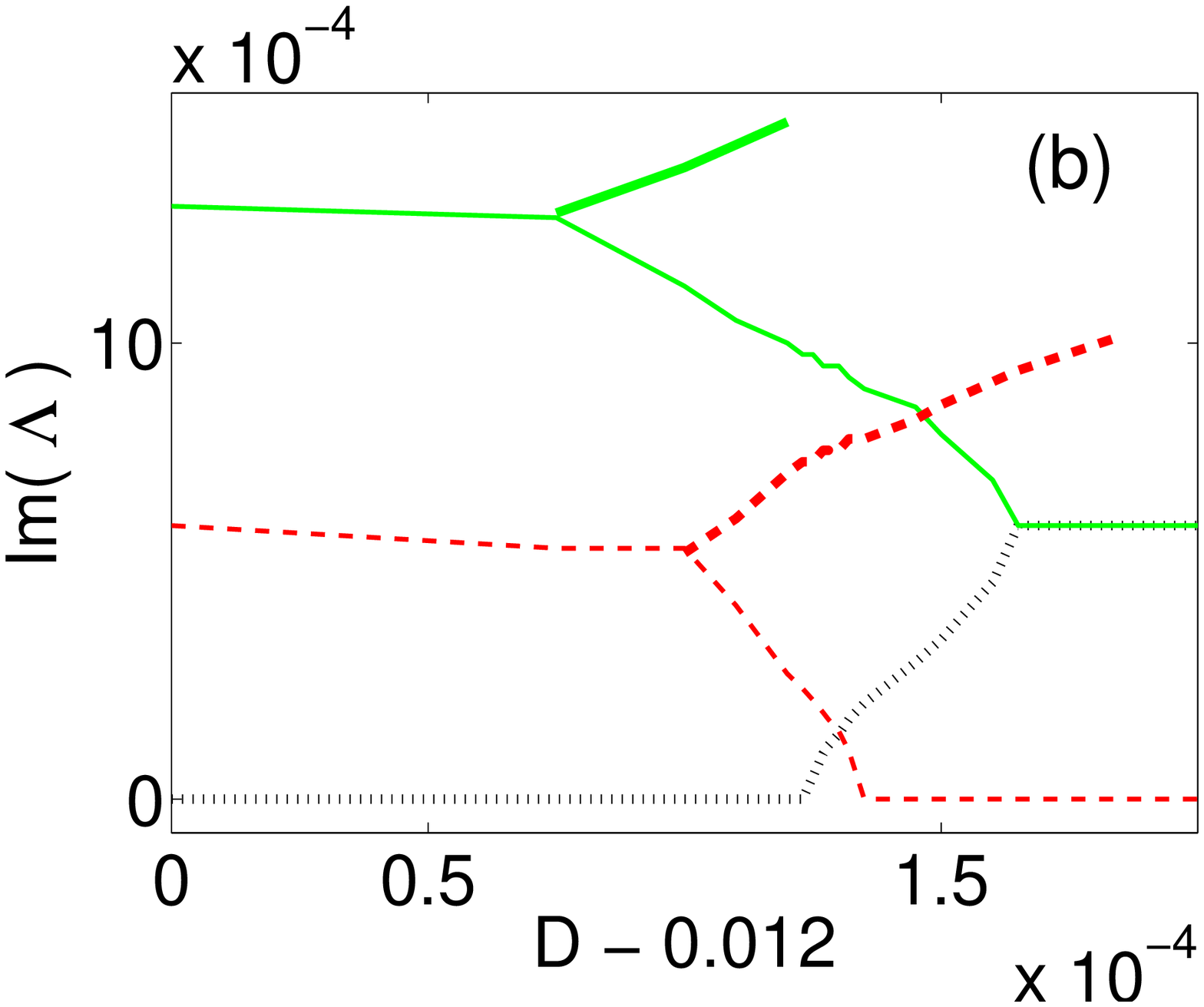}}}
\end{minipage}
\hspace{0.1cm}
\begin{minipage}{5.1cm}
\rotatebox{0}{\resizebox{5.1cm}{7cm}{\includegraphics[0in,0.5in]
 [8in,10.5in]{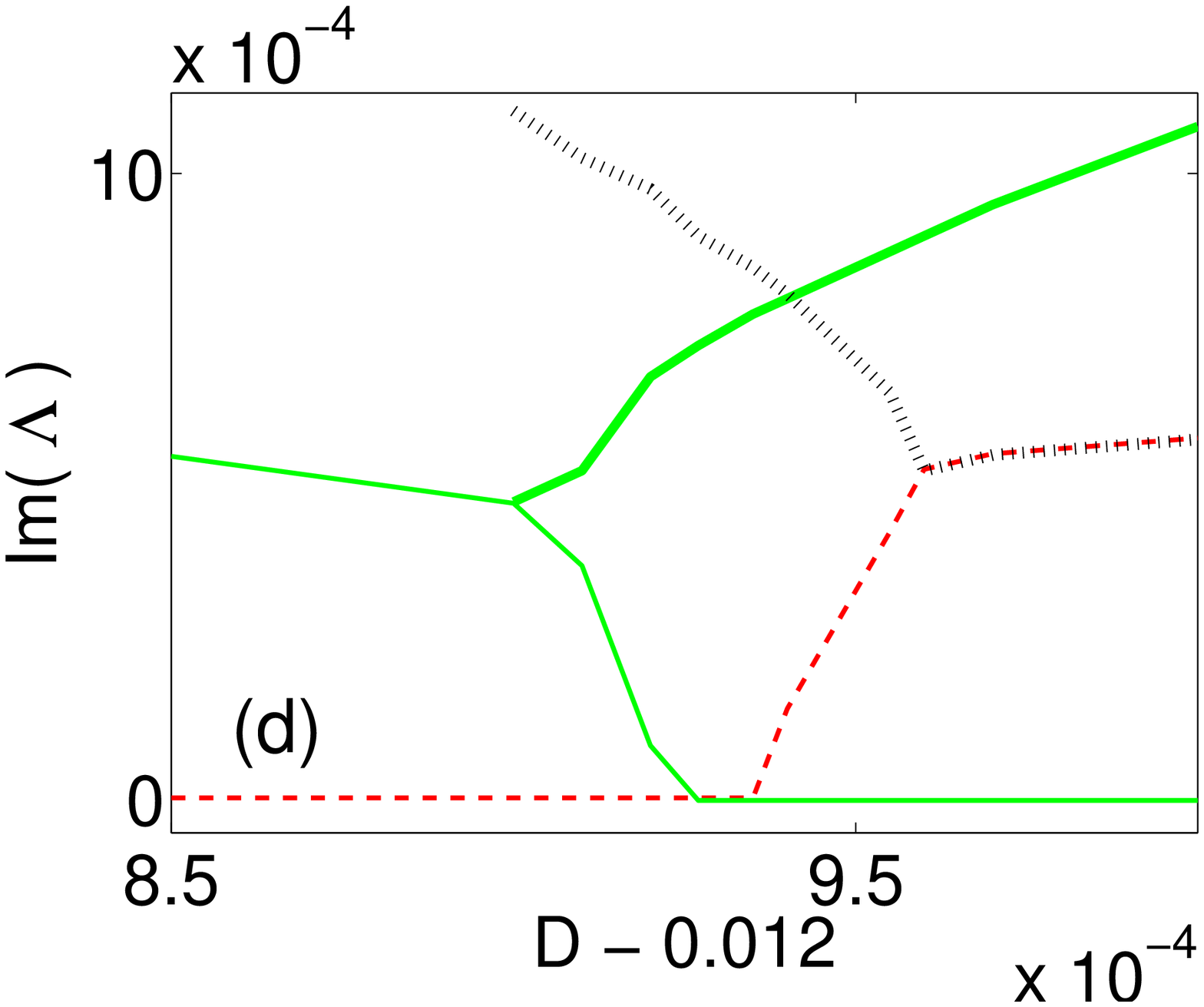}}}
\end{minipage}
\hspace{0.1cm}
\begin{minipage}{5.1cm}
\rotatebox{0}{\resizebox{5.1cm}{7.4cm}{\includegraphics[0in,0.5in]
 [8in,10.5in]{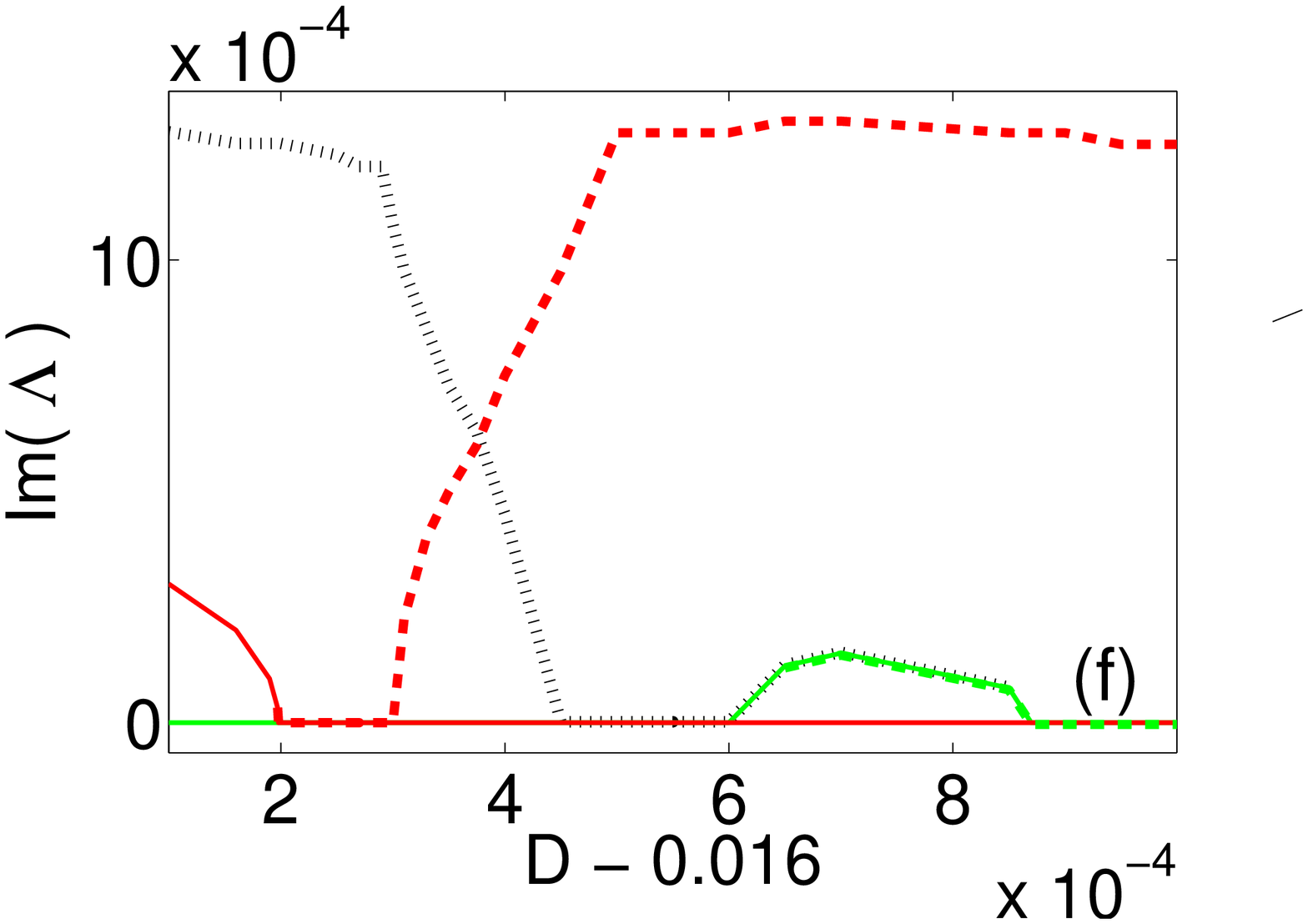}}}
\end{minipage}
 }
\vspace{-1.6cm}
\caption{ (Color online) \ 
Real and imaginary parts of selected modes of \eqref{e_36}, including the
most unstable mode. \ (a) \& (b): $0.01200 \le D \le 0.01220$; \ 
(c) \& (d): $0.01285 \le D \le 0.01300$; \ 
(e) \& (f): $0.01610 \le D \le 0.01700$. Same line colors, styles, and 
widths are used to indicate the same modes within one pair of
panels (e.g., (c) \& (d)). The same line colors/styles/widths in different pairs
of panels (e.g., in (a) \& (b) and (c) \& (d)) do {\em not} imply the same modes.
}
\label{fig_8}
\end{figure}

First, at $D\approx 0.012134$, a ``real" mode appears (see Fig.~\ref{fig_8}(a,b)),
and from this point on there always exists a ``real" mode, even though the particular mode
``born" at $D\approx 0.012134$ disappears later on. Specifically, at $D\approx 0.012928$,
another ``real" mode acquires $\Lambda_R$ greater than that of the mode ``born" 
at $D\approx 0.012134$,
and the latter mode soon disappears (Fig.~\ref{fig_8}(c,d)). A similar switchover between ``real"
modes occurs at least one more time near $D\approx 0.013750$ (not shown). Next, another ``real"
mode is ``born" via a cascade of bifurcations near $D\approx 0.0162$ (Fig.~\ref{fig_8}(e,f)),
and its $\Lambda_R$ crosses that of the previously dominant-$\Lambda_R$ ``real" mode near
$D\approx 0.01635$. At $D=0.0170$, these two dominant ``real" modes have 
$\Lambda_R\approx 1.4\cdot 10^{-3}$
and $1.5\cdot 10^{-3}$ (Fig.~\ref{fig_8}(e)). Finally, these two modes gradually approach each
other while crossing at least once more near $D=0.01725$. At $D=0.023$ and beyond, their eigenvalues
are the same to five significant figures. Thus, remarkably, the dominant unstable mode 
eventually becomes doubly degenerate. We verified that such a degeneracy also occurs
 for the higher-order localized unstable modes, shown in Fig.~\ref{fig_6}.

To conclude, we present a hypothesis as to why the value $C_{\rm cr}$, where
a ``real" mode appears permanently (see above), is near $C= 1.013$. 
Let us interpret \eqref{e_C6} in a way that the $n$ on its r.h.s. is not necessarily
an integer, but
a continuous function of the parameter $D$. For those values of $D$ when $n$ {\em is} an integer,
a mode with a real $\Lambda$ either appears or disappears at the origin $\Lambda=0$. 
Evaluating $n$ at the values
of $D$ listed in the previous paragraph in connection with Figs.~\ref{fig_8}(a)--(d), one finds:
\bsube
\bea
\underline{{\rm at} \;\;D=0.012134}: & & n|_{\nu=1}\approx 29.02, \qquad n|_{\nu=1}-n|_{\nu=3}\approx 0.99; 
\label{e_C7a} \\
\underline{{\rm at} \;\;D=0.012928}: & & n|_{\nu=1}\approx30.02, \qquad n|_{\nu=1}-n|_{\nu=3}\approx 1.02; 
\label{e_C7b} \\
\underline{{\rm at} \;\;D=0.013750}: & & n|_{\nu=1}\approx31.04, \qquad n|_{\nu=1}-n|_{\nu=3}\approx 1.05.
\label{e_C7c} 
\eea
\label{e_C7}
\esube
That is, both $n|_{\nu=1}$ and $n|_{\nu=3}$ are simultaneously very close to integers.
At $D=0.012928$, one of the ``real" modes has not yet disappeared while the next 
one has appeared (Fig.~\ref{fig_8}(c)). From \eqref{e_C7} we observe that at this value of $D$,
the difference $(n|_{\nu=1}-n|_{\nu=3})$ exceeds $1$ for the first time. Thus, we hypothesize
that $C_{\rm cr}\equiv 1+D_{\rm cr}$ 
is found from the condition that $(n|_{\nu=1}-n|_{\nu=3})$ exceeds $1$ 
for the first time. Verification of this hypothesis requires a deeper analytical insight than
we have at the moment. Moreover, finding a value of $C$ past which the dominant real eigenvalue 
increases monotonically (as seen in Fig.~\ref{fig_8}(e)) is also an open question.


\section*{Appendix D: 
Possible reason behind ``sluggish" numerical instability of oscillating pulse}
\addtocounter{section}{1}
 \setcounter{equation}{0}

Here we will give an argument in favor of a relation between mechanisms of
``sluggish" NI for a stationary pulse in a potential (Sec.~5.3) and for an
oscillating pulse (Sec.~7). The key observation here is that the (approximate)
period of the oscillations, which is $O(1)$, is much smaller than the characteristic
time over which NI develops, which is $O(100)$ and greater. 
This allows one to invoke well-known techniques
of analysis of the evolution with rapidly oscillating perturbations in
solving Eq.~\eqref{e_31}. Specifically, we split the background 
$u_{\rm b}^2$ and the solution $\tilde{w}$ into slowly (`s') and rapidly (`r') 
varying parts:
\bsube
\be
u_{\rm b}^2 = \left( \varrho_s + \varrho_r \right)\,
 e^{2i[\vartheta_s + \vartheta_r]},
\label{e_D01a}
\ee
where all the variables are real-valued and depend on $x$ and $t$,
and 
\be
\tw = \left( \zeta_s + \zeta_r \right)\,  e^{i[\vartheta_s + \vartheta_r]},
\label{e_D01b}
\ee
\label{e_D01}
\esube
where now $\zeta_{s,r}$ may be complex. 
In what follows we will denote $\partial_t\vartheta_{s,r}\equiv \dot{\vartheta}_{s,r}$
and $\partial_{\chi}\vartheta_{s,r}\equiv \vartheta\,'_{s,r}$. 
Substitution of \eqref{e_D01} into 
\eqref{e_31} and separation of the slow and fast parts yields:
\bsube
\bea
i \zeta_{s,\,t} & \hspace*{-0.6cm} & 
   + \left( \Gamma \big(1+(\vartheta_s')^2 - i\vartheta_s''  \big)  -\dot{\vartheta}_s \right) 
	   \zeta_s - 2i\Gamma \vartheta_s' \zeta_{s,\,\chi} -\Gamma \zeta_{s,\,\chi\chi} +
		 \gamma \varrho_s (2\zeta_s + \zeta_s^*) 
		 \nonumber \\
 & \hspace*{-0.6cm}  & + 
 \left\langle \left( \Gamma \big(\,2\vartheta_r' \vartheta_s' - i\vartheta_r''  \big) 
  -\dot{\vartheta}_r  \right) \zeta_r - 2i\Gamma \vartheta_r' \zeta_{r,\,\chi} 
	+ \gamma \varrho_r (2\zeta_r + \zeta_r^*) 
	 \right\rangle \;=\; 0; 
\label{e_D02a}
\eea
\bea
i \zeta_{r,\,t} & \hspace*{-0.6cm} & 
   + \left( \Gamma \big(1+(\vartheta_s')^2 - i\vartheta_s''  \big)  -\dot{\vartheta}_s \right) 
	   \zeta_r - 2i\Gamma \vartheta_s' \zeta_{r,\,\chi} -\Gamma \zeta_{r,\,\chi\chi} +
		 \gamma \varrho_s (2\zeta_r + \zeta_r^*) 
		\nonumber \\ 
 & \hspace*{-0.6cm}  & + 
 \left\{ 
 \left( \Gamma \big(\,2\vartheta_r' \vartheta_s'  - i\vartheta_r''  \big) 
  -\dot{\vartheta}_r  \right) \zeta_s - 2i\Gamma \vartheta_r' \zeta_{s,\,\chi} 
	+ \gamma \varrho_r (2\zeta_s + \zeta_s^*) 
	\right\} 
	\;=\; 0. 
\label{e_D02b}
\eea
\label{e_D02}
\esube
Here $\Gamma\equiv 1/(C|\beta|)$ and $\langle\ldots\rangle$ denotes averaging over time;
as usual, we have omitted terms that oscillate faster than (e.g., twice as fast as)
$\vartheta_r$ and $\zeta_r$. 
The standard method is to solve \eqref{e_D02b} for $\zeta_r$ and substitute the result
in the second line of \eqref{e_D02a}. Solving \eqref{e_D02b} analytically requires
 the assumption $|\zeta_r| \ll |\zeta_s|$, which, upon some calculations, reduces the
terms in the second line of \eqref{e_D02a} to $\langle \vartheta_r\varrho_r\rangle\zeta_s^*$,
etc.. We numerically computed $\langle \vartheta_r\varrho_r\rangle$ for some of the
solutions considered in Sec.~7 and found that it is considerably smaller than, e.g., 
$\varrho_s$. Therefore, we believe that 
the occurrence of ``sluggish" NI for oscillating pulses is primarily related to
terms in the first line of \eqref{e_D02a}.

\begin{figure}[h]
\vspace*{-0.6cm}
\hspace*{-0.5cm}
\mbox{ 
\begin{minipage}{5.1cm}
\rotatebox{0}{\resizebox{5.1cm}{7cm}{\includegraphics[0in,0.5in]
 [8in,10.5in]{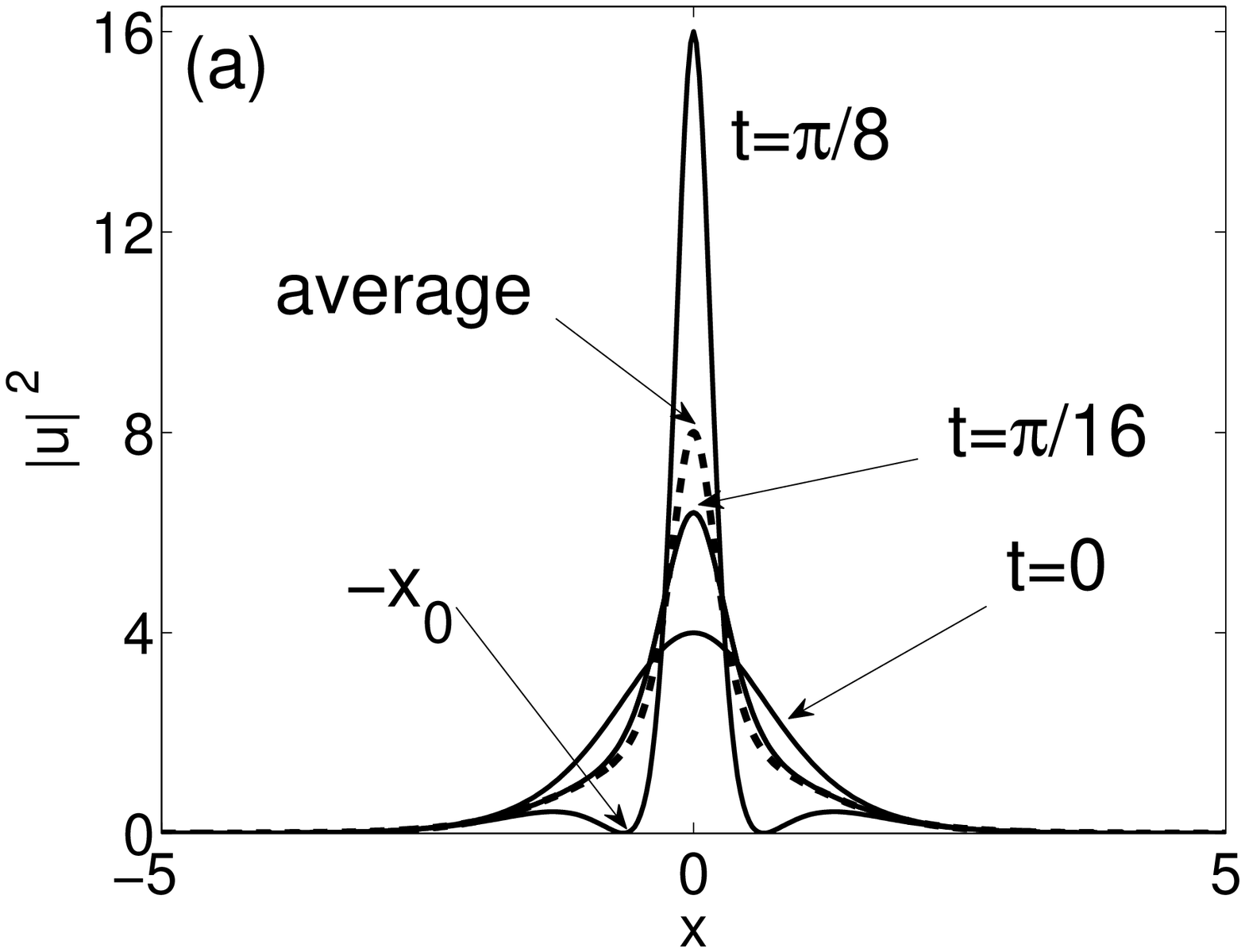}}}
\end{minipage}
\hspace{0.1cm}
\begin{minipage}{5.1cm}
\rotatebox{0}{\resizebox{5.1cm}{7cm}{\includegraphics[0in,0.5in]
 [8in,10.5in]{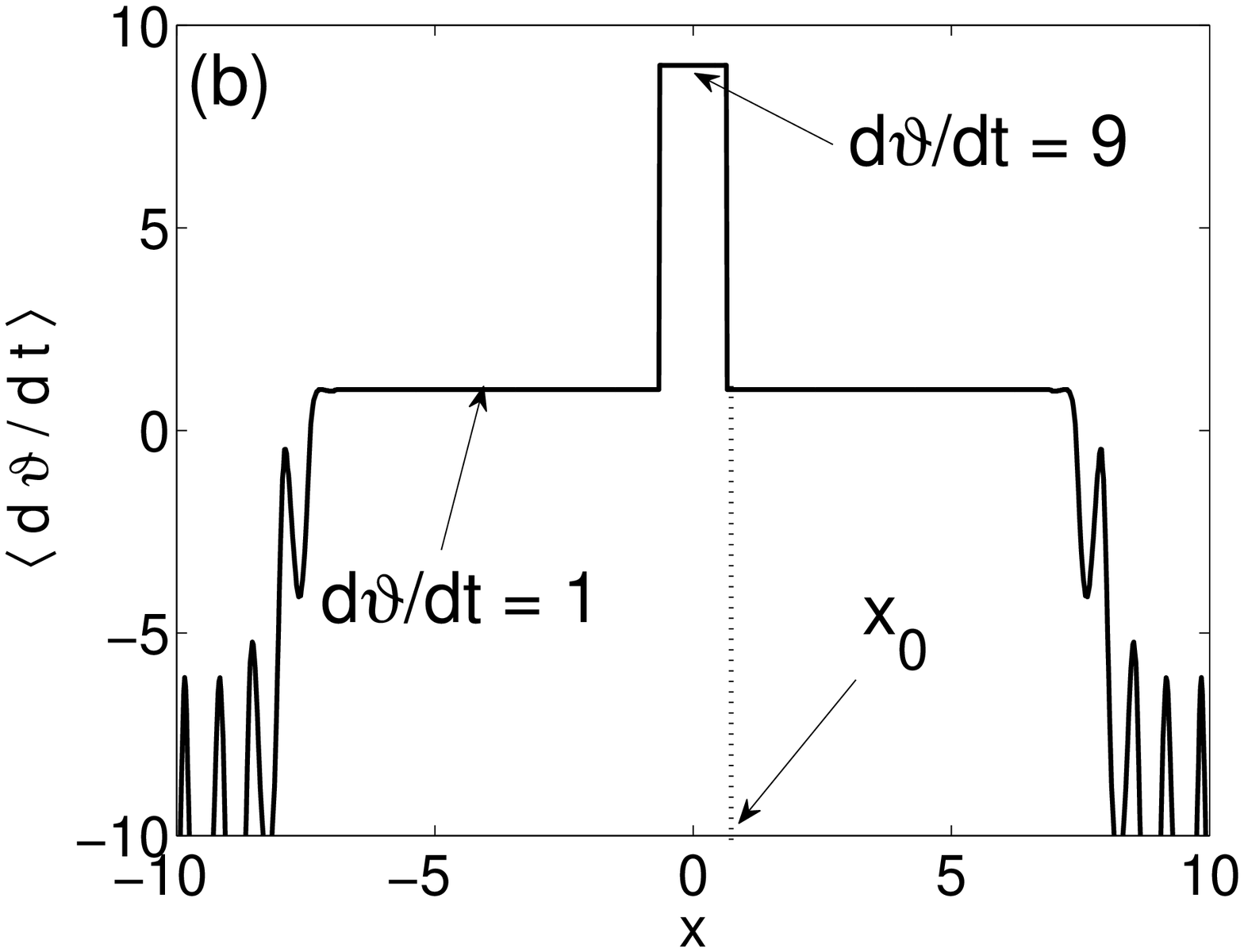}}}
\end{minipage}
\hspace{0.1cm}
\begin{minipage}{5.1cm}
\rotatebox{0}{\resizebox{5.1cm}{7.4cm}{\includegraphics[0in,0.5in]
 [8in,10.5in]{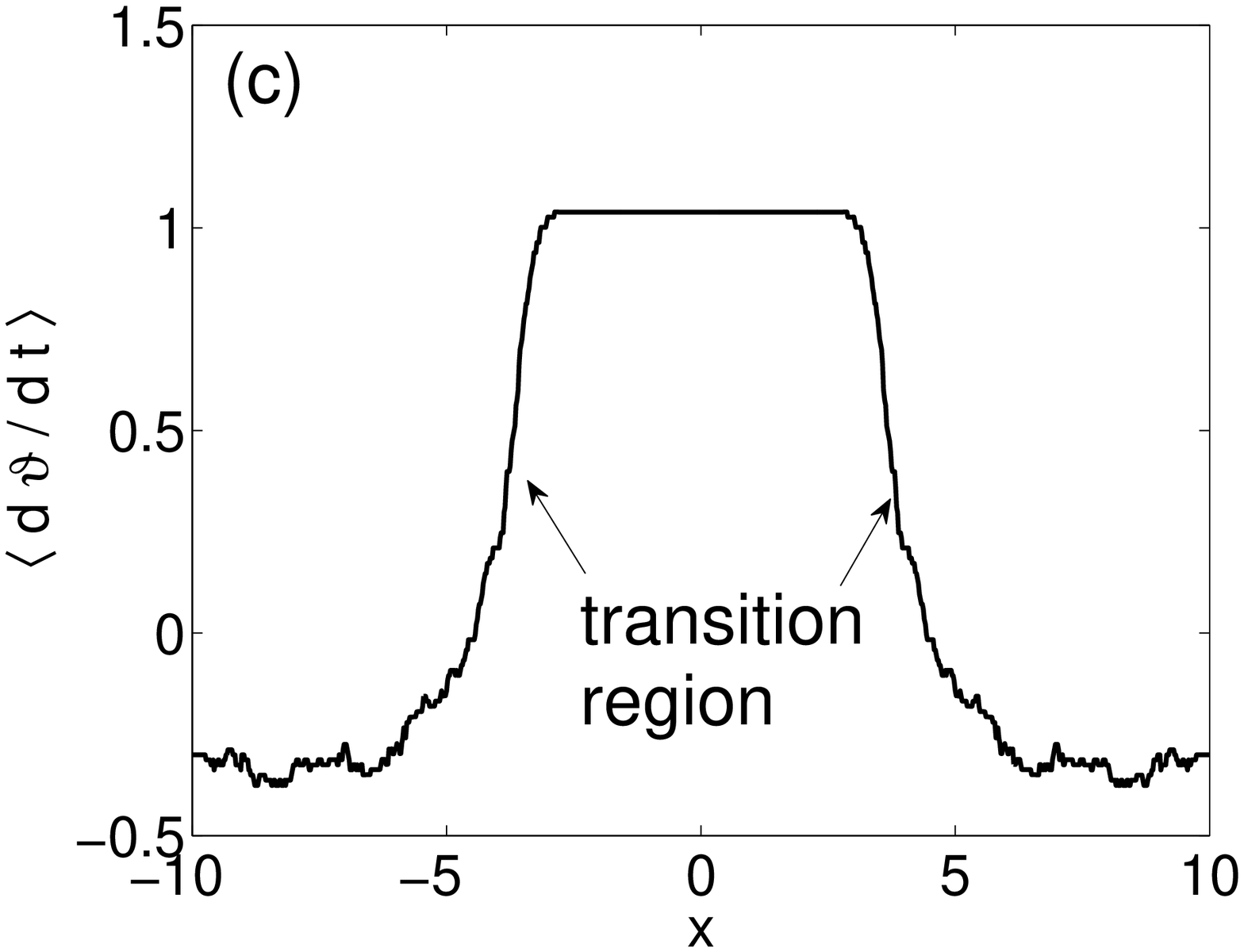}}}
\end{minipage}
 }
\vspace{-1.6cm}
\caption{ (a) \ Snapshots and the average value of $|u|^2$ from \eqref{e_D03}. \ 
(b) \ $ \dot\vartheta_s $ of the same solution. \ 
In both panels, $x_0={\rm arcsinh}(1/\sqrt{2})$ is where $u(x_0,\pi/8)=0$ and hence the phase
has a discontinuity. \ 
(c) \ $ \dot\vartheta_s $ of the solution generated by initial
condition \eqref{e7_07}.
}
\label{fig_appD_1}
\end{figure}

We will argue this point using the solution generated by initial condition \eqref{e7_02}
as an example. The analytical form of this solution is
\be
u(x,t) = \frac{\displaystyle 4e^{it} \left( \cosh(3x)+3e^{8it}\cosh(x)\right)}
              {\cosh(4x) + 4\cosh(2x) + 3\cos(8t)};
\label{e_D03}
\ee
its snapshots are shown in Fig.~\ref{fig_appD_1}(a). In Fig.~\ref{fig_appD_1}(b) we
show $\dot\vartheta_s$, computed as $\langle \dot\vartheta \rangle$. At the far ``tails"
of the pulse ($|x|>7$) the graph is irregular since the numerical error dominates
over the solution \eqref{e_D03}. The important feature to note in Fig.~\ref{fig_appD_1}(b)
is that $\dot\vartheta_s$ is {\em piecewise constant} where $|u|^2$ is essentially nonzero.
Thus, the corresponding term in the first line of \eqref{e_D02a} creates an effective
potential, which is one reason why NI in this case may be related to that described in
Sec.~5.3. We emphasize that the piecewise-constant shape of $\dot\vartheta_s$ is
common for all time-periodic (or almost periodic) solutions that we have simulated, 
e.g., that generated by initial condition \eqref{e7_03}. For quasi-periodic 
oscillating pulses, the transition region connecting one constant value of $\dot\vartheta_s$
to another or to the outside region is smooth rather than abrupt; see Fig.~\ref{fig_appD_1}(c).

However,
the reason presented in the previous paragraph may not be the only one behind
the development of ``sluggish" NI. Two other terms in the same equation,
$\big((\vartheta_s')^2 - i\vartheta_s''\big)$, create an additional, {\em time-dependent},
effective potential. Indeed, while $\vartheta_s'=0$ wherever $\dot\vartheta_s={\rm const}$
{\em with respect to $x$}, it is nonzero in the transitional region. Since in that region 
$\dot\vartheta_s={\rm const}$ {\em with respect to $t$}, then  $\vartheta_s' \propto t$ there.
Let us note that while $\partial_{x}\vartheta_s \sim \delta(x-x_0)$, where $x_0$
is the location of the sharp transition region (see Fig.~\ref{fig_appD_1}(b)),
$\vartheta_s'\equiv\partial_{\chi}\vartheta_s=\epsilon \partial_{x}\vartheta_s$ 
is a function varying on the scale $O(1)$ in $\chi$. 
Unfortunately, analytical solution of even simplest equations with 
time-dependent potential, like
\be
iw_t + w_{xx} \pm t\,\delta(x)\,w=0,
\label{e_D04}
\ee
does not appear to be possible. It is, therefore, an open problem to relate the
development of ``sluggish" NI of an oscillating pulse to some reduced model 
described by an equation of the form
\be
iw_t + w_{xx} + \big(\Pi_1(x)w + \Pi_2(x)w^*\big)=0
\label{e_D05}
\ee
for some {\em time-independent} $\Pi_1$ and $\Pi_2$. Such a model would generalize
that considered in Secs.~3--5 and would be more amenable to standard methods of
eigenvalue analysis than the time-dependent model discussed above.


\begin{thebibliography}{99}

\bibitem{HardinTappert_73}
R.H.~Hardin and F.D.~Tappert, \ 
Applications of the split-step Fourier method to the numerical soltion of nonlinear
and variable coefficient wave equations, \ 
SIAM Review (Chronicle) {\bf 15}, 423 (1973).

\bibitem{AblowitzTaha_84}
T.~Taha and M.~Ablowitz, \ 
Analytical and numerical aspects of certain nonlinear evolution equations.
II. Numerical, Nonlinear Schrodinger equation, \ 
J.~Comp.~Phys. {\bf 55}, 203--230 (1984).  

\bibitem{Agrawal_book}
G.P.~Agrawal, \ 
{\em Nonlinear fiber optics, 3rd Ed.} (Academic Press, San Diego, 2001).

\bibitem{BaoWang}
W.~Bao and H.~Wang, \ 
An efficient and spectrally accurate numerical method for computing 
dynamics of rotating Bose--Einstein condensates, 
J.~Comput.~Phys. {\bf 217}, 612--626 (2006).  


\bibitem{Bandrauk93}
A.D.~Bandrauk and H.~Shen, \ 
Exponential split operator methods for solving coupled time-dependent Schr\"odinger equations, 
J.~Chem.~Phys. {\bf 99}, 1185--1193  (1993).   


\bibitem{Bandrauk07}
E.~Lorin, S.~Chelkowski,  A.~Bandrauk, \ 
A numerical Maxwell--Schr\"odinger model for intense laser--matter interaction and propagation, \ 
Comput.~Phys.~Commun. {\bf 177}, 908--932 (2007).  

%
\bibitem{SSMReactDiff_99}
D.~Lanser and J.G.~Verwer, \ 
Analysis of operator splitting for advection--reaction--diffusion problems
from air pollution modelling, \ 
J.~Comp.~Appl.~Math.  {\bf 111}, 201--206 (1999).  

%
\bibitem{SSMHydrology_04}
J.~Carrayroua, R.~Mos\'e, and P.~Behra, \ 
Operator-splitting procedures for reactive transport and comparison of mass balance errors, \ 
J.~Contamin.~Hydrology {\bf 68}, 239--268  (2004).  

%
\bibitem{SSMConvDiff_09}
A.~Chertock, A.~Kurganov, and G.~Petrova, \ 
Fast explicit operator splitting method for convection--diffusion equations,
Int.~J.~Numer.~Meth.~Fluids {\bf 59}, 309--332 (2009).  



\bibitem{KGE_2012}
F.~Blumenthal, H.~Bauke, \ 
A stability analysis of a real space split operator method for the Klein--Gordon equation, 
J.~Comput.~Phys. {\bf 231}, 454--464 (2012).   

%
\bibitem{Strang}
G.~Strang, \ 
On the construction and comparison of difference schemes, \ 
SIAM J.~Numer.~Anal. {\bf 5} 506--517  (1968).   

\bibitem{Yevick91}
M. Glassner, D. Yevick, and B. Hermansson, \
High-order generalized propagation techniques, 
J.~Opt.~Soc.~Am.~B {\bf 8}, 413--415 (1991).  

\bibitem{WH}
J.A.C.~Weideman and B.M.~Herbst, \ 
Split-step methods for the solution of the nonlinear Schr\"odinger equation, 
SIAM J.~Numer.~Anal. {\bf 23}, 485--507 (1986).  

\bibitem{SelAreasCommun_1997}
 A.~Carena, V.~Curri, R.~Gaudino, P.~Poggiolini, and S.~Benedetto, \ 
 A time-domain optical transmission system simulation package accounting for 
 nonlinear and polarization-related effects in fiber, \ 
 IEEE J.~Sel.~Areas Commun. {\bf 15}, 751--764 (1997).   

\bibitem{JCP1999}
Q.~Chang, E.~Jia, and W.~Sun, \ 
Difference schemes for solving the generalized nonlinear Schr\"odinger equation, 
J.~Comp.~Phys. {\bf 148}, 397--415 (1999). 

\bibitem{Faou_2011}
D.~Bambusi, E.~Faou, and B.~Gr\'ebert, \ 
Existence and stability of solitons for fully discrete approximations of
the nonlinear Schr\"odinger equation,
Numer.~Math.  {\bf 123}, 461--492 (2013).  

\bibitem{CPC_2013}
X.~Antoine, W.~Bao, and C.~Besse, \ 
Computational methods for the dynamics of the nonlinear 
Schr\"odinger/Gross--Pitaevskii equation, 
Comp.~Phys.~Commun. {\bf 184}, 2621--2633 (2013).

\bibitem{AOP_2009}
G.~Li, 
Recent advances in coherent optical communication,
Adv.~Opt.~Photon. {\bf 1}, 279--307 (2009).  
%


\bibitem{Kenkre_1986}
V.M.~Kenkre and D.K.~Campbell, \ 
Self-trapping on a dimer: Time-dependent solutions of a discrete
nonlinear Schr\"odinger equation, \ 
Phys.~Rev.~B {\bf 34}, 4959--4961 (1986). 

\bibitem{Aceves_2014}
A.B.~Aceves and J.-G.~Caputo, \ 
Mode dynamics in nonuniform waveguide arrays: a graph Laplacian approach, \ 
J.~Opt. {\bf 16}, 035202 (2014).


\bibitem{vonNeumannRichtmeyer}
J.~Von Neumann and R.D.~Richtmeyer, \ 
A method for the numerical calculation of hydrodynamic shocks, \ 
J.~Appl.~Phys. {\bf 21}, 232--237  (1950).   

\bibitem{Trefethen_book}
L.N.~Trefethen, \ 
{\em Spectral methods in Matlab} \ 
(SIAM, Philadelphia, 2001), Chap.~10.

\bibitem{ja}
T.I.~Lakoba, \ 
Instability analysis of the split-step Fourier method on the 
background of a soliton of the nonlinear Schr\"odinger equation,
Num.~Meth.~Part.~Diff.~Eqs. {\bf 28}, 641--669 (2012).  
 
\bibitem{Gauckler_2010}
L.~Gauckler and C.~Lubich, \ 
Splitting integrators for Nonlinear Schr\"odinger equations over long times,
Found.~Comput.~Math. {\bf 10}, 275--302 (2010). 

\bibitem{Kaup90}
D.J.~Kaup, \ 
 Perturbation theory for solitons in optical fibers, 
Phys.~Rev.~A {\bf 42}, 5689--5694 (1990).  

\bibitem{Tran92}
H.T.~Tran, \ 
Stability of dark solitons: Linear analysis, \ 
Phys.~Rev.~A {\bf 46}, 7319--7321 (1992).  

\bibitem{Peli98}
D.E.~Pelinovsky, Yu.S.~Kivshar, and V.V.~Afanasjev, \ 
 Internal modes of envelope solitons, \ 
Physica D {\bf 116}, 121--142 (1998).  

\bibitem{RefLA}
I.M.~Gelfand, \ 
{\em Lectures on linear algebra} 
(Interscience Publishers, New York, 1961), Sec.~15.

\bibitem{WKB_system}
G.~Chen and J.~Zhou, \ 
{\em Vibration and damping in distributed systems, vol. II (WKB and wave methods,
vizualization and experimentation)} \  (CRC Press, Boca Raton, 1993), Sec.~1.2.

\bibitem{Fulling2}
S.A.~Fulling, \ 
Adiabatic expansions of solutions of coupled secondorder linear differential equations. II, \ 
J.~Math.~Phys. {\bf 20}, 1202--1209 (1979).  

\bibitem{Skorupski}
A.A.~Skorupski, \ 
Phase integral approximation for coupled ODEs of the Schr\"odinger type, \ 
J.~Math.~Phys. {\bf 49}, 053523 (2008).

\bibitem{gPetviashvili}
T.I.~Lakoba, J.~Yang, \ 
A generalized Petviashvili iteration method for scalar and vector Hamiltonian 
 equations with arbitrary form of nonlinearity, \ 
J.~Comp.~Phys. {\bf 226}, 1668--1692 (2007). 

\bibitem{Segev_1994}
M.~Segev, G.C.~Valley, B.~Crosignani, P.~DiPorto, A.~Yariv, \
Steady-state spatial screening solitons in photorefractive materials
with external applied field, \ 
Phys.~Rev.~Lett. {\bf 73}, 3211--3214 (1994). 

\bibitem{Ascher}
U.~Ascher, \ 
Surprising computations, \ 
Appl.~Numer.~Math. {\bf 62}, 1276--1288 (2012).  

\bibitem{jaJOSAB}
T.I.~Lakoba, \ 
Instability of the split-step method for a signal with nonzero central frequency, \ 
J.~Opt.~Soc.~Am.~B {\bf 30}, 3260--3271 (2013).  

\bibitem{KL96_falseinst}
D.J.~Kaup, T.I.~Lakoba, \ 
Variational method: How it can generate false instabilities, \ 
J.~Math.~Phys. {\bf 37}, 3442--3462 (1996).  

\bibitem{Kutz_course}
J.N.~Kutz, {\em Computational Methods for Data Analysis}, 
online course: \verb+http://courses.washington.edu/amath582/+; Sec.~18.

\bibitem{KincaidCheney}
D.R.~Kincaid, E.W.~Cheney, \ 
{\em Numerical analysis: Mathematics of scientific computing, 3rd Ed.}
(Brooks/Cole, Pacific Crove, CA, 2002), Sec.~9.1.

\bibitem{Kapitula98}
T.~Kapitula and B.~Sandstede, \ 
Instability mechanism for bright solitary-wave solutions to the
cubic-quintic Ginzburg--Landau equation
J.~Opt.~Soc.~Am.~B {\bf 15}, 2757--2762 (1998).  



\end{thebibliography}
\end{document}